\documentclass[24pt,notitlepage]{revtex4-1}
\usepackage{graphicx}

\usepackage{amsmath}
\usepackage{amsfonts}
\usepackage{amssymb}
\usepackage{slashbox}
\usepackage{epstopdf}

\font\msytw=msbm9 scaled\magstep1 
scaled\magstep1

\let\a=\alpha \let\b=\beta  \let\g=\gamma  \let\d=\delta \let\e=\varepsilon
\let\z=\zeta  \let\h=\eta   \let\th=\theta \let\k=\kappa \let\l=\lambda
\let\m=\mu    \let\n=\nu    \let\x=\xi     \let\p=\pi    \let\r=\rho
\let\s=\sigma    \let\f=\varphi \let\c=\chi
   \let\o=\omega
\let\G=\Gamma \let\D=\Delta  \let\L=\Lambda 
         
\let\O=\Omega 

 \def\VV{{\cal V}}
\def\CC{{\cal C}} \def\WW{{\cal W}}
\def\NN{{\cal N}} \def\BBB{{\cal B}}
\def\RR{{\cal R}}\def\LL{{\cal L}}  \def\OO{{\cal O}}
\def\DD{{\cal D}} \def\SS{{\cal S}}

\def\qq{{\bf q}} \def\pp{{\bf p}}
 \def\xx{{\bf x}} \def\yy{{\bf y}} 
\def\kk{{\bf k}}

\def\dd{{\boldsymbol{\delta}}}

\def\ddd{\boldsymbol{\d}}

\def\nn{\nonumber}

\def\os{\overset}

\def\RRR{\hbox{\msytw R}}

 \def\ZZZ{\hbox{\msytw Z}}

\def\\{\hfill\break}
\def\={:=}
\let\io=\infty
\let\0=\noindent
\def\media#1{{\langle#1\rangle}}
\let\dpr=\partial

\def\const{{\rm const}}
\def\tende#1{\,\vtop{\ialign{##\crcr\rightarrowfill\crcr\noalign{\kern-1pt
    \nointerlineskip} \hskip3.pt${\scriptstyle #1}$\hskip3.pt\crcr}}\,}
\def\otto{\,{\kern-1.truept\leftarrow\kern-5.truept\to\kern-1.truept}\,}

\def\to{\rightarrow}

\def\qed{\hfill\raise1pt\hbox{\vrule height5pt width5pt depth0pt}}

\def\ul#1{{\underline#1}}
\def\lis{\overline}
\def\V#1{{\bf#1}}
\def\be{\begin{equation}}
\def\ee{\end{equation}}
\def\bp{\begin{pmatrix}}
\def\ep{\end{pmatrix}}
\def\bea{\begin{eqnarray}}
\def\eea{\end{eqnarray}}
\def\nn{\nonumber}
\def\pref#1{(\ref{#1})}

\def\Tr{\mathrm{Tr}}
\def\eu{\mathrm{e}}

\begin{document}
\title{Lattice quantum electrodynamics for graphene}
\author{A. Giuliani} \affiliation{Universit\`a di Roma Tre, L.go
S. L. Murialdo 1, 00146 Roma - Italy}
\author{V. Mastropietro}
\affiliation{Universit\`a di Roma Tor Vergata, V.le della Ricerca
Scientifica, 00133 Roma - Italy}
\author{M. Porta}
\affiliation{Institut f\"ur Theoretische Physik, ETH
H\"onggerberg, CH-8093, Z\"urich - Switzerland}
\begin{abstract}
The effects of gauge interactions in graphene have been analyzed up
to now in terms of effective models of Dirac fermions. However, in
several cases lattice effects play an important role and need to be taken consistently into
account. In this paper we introduce and analyze a lattice gauge theory model for graphene,
which describes tight binding electrons hopping on the
honeycomb lattice and interacting with a three-dimensional quantum
$U(1)$ gauge field. We perform an exact Renormalization Group analysis, which leads to
a renormalized expansion that is finite at all orders. The flow of the effective parameters is
controlled thanks to Ward Identities
and a careful analysis of the discrete lattice symmetry properties of the model.
We show that the Fermi velocity increases up to the speed of light and Lorentz
invariance spontaneously emerges in the infrared. The interaction
produces critical exponents in the response functions; this
removes the degeneracy present in the non interacting case and
allow us to identify the dominant excitations. Finally we add mass
terms to the Hamiltonian and derive by a variational argument the
correspondent gap equations, which have an anomalous non-BCS form,
due to the non trivial effects of the interaction.
\end{abstract}

\maketitle

\renewcommand{\thesection}{\arabic{section}}

{\bf keywords}: Graphene, Lattice Gauge Theory, honeycomb lattice,
Kekul\'e mass generation, Ward Identities, Renormalization Group,
Critical exponents

\tableofcontents

\section{Introduction}\label{sec1}
\setcounter{equation}{0}
\renewcommand{\theequation}{\ref{sec1}.\arabic{equation}}

Graphene, a two dimensional crystal of carbon atoms that has been
recently experimentally realized  \cite{N1,N2}, has highly unusual
electronic properties. The lattice in graphene has a honeycomb
shape and the corresponding energy bands intersect at two points, close to
which the effective dispersion relation is approximately
conical and similar to a ``relativistic" one. The low energy excitations
of the half-filled system consist of hole-particle pairs created
close to the tips of these two cones. These quasi-particles behave like
two-dimensional (2D) massless Dirac fermions: for this reason the infrared (IR)
properties of the system can be understood in terms of a model of
2D Dirac particles in the continuum \cite{W,S}. As a result, a number of
concepts and features of high energy physics have a correspondence and can be
observed at much lower energies in this crystal.

The description in terms of Dirac fermions is quite accurate in
the free case, and it is also helpful in the presence of
many-body interactions (see, e.g., \cite{C} for a review), as
it allows to translate and adopt a number of powerful methods from
the realm of quantum field theory (QFT) to that of graphene. However, taking
the effective description too seriously has some
drawbacks, since a model of interacting 2D Dirac fermions in the continuum
has spurious {\it ultraviolet divergences} due to the linear bands.
In order to make the continuum theory well-defined, ad hoc {\it regularizations} must
be introduced to cure the short distance singularities, which are obviously
absent in the tight binding model, where the honeycomb lattice acts naturally
as an ultraviolet (UV) cut-off. It is unfortunate that
the computation within the Dirac model of certain physical
observables, such as the conductivity, is sensitive to the specific choice
of the regularization scheme, a fact that makes the comparison with experiments
difficult or inaccurate, see, e.g., \cite{He3}.

These ambiguities make an approximation-free analysis of the effects of the
lattice and of the non-linear bands in graphene highly desirable. In this paper, we
study a lattice gauge model for graphene that describes
electrons hopping on the honeycomb lattice and weakly interacting with a
three-dimensional (3D) quantum $U(1)$ gauge field. Our model has two
independent parameters: the bare Fermi velocity $v$ and the electric charge $e$
(we use units such that the reduced Planck constant $\hbar$ and the speed of
light $c$ are equal to 1). The analysis is performed by using exact
Renormalization Group (RG) methods, which allow us to express the physical
observables in terms of renormalized expansions in the electric charge
with {\it finite} coefficients at all orders, uniformly in the volume
and in the temperature (more precisely, we show that the coefficient
multiplying $e^{2n}$ in the renormalized expansion grows at most as $n!$).
Our main physical predictions are the following.
\begin{enumerate}
\item Thanks to the validity of exact lattice {\it Ward
Identities}, the gauge field remains massless and the IR behavior
of the system is characterized by a {\it line of fixed points}
(i.e., the effective charge has vanishing beta function).
Correspondingly, the physical parameters are strongly renormalized
by the interaction. In particular, {\it the Fermi velocity
increases up to the speed of light} and Lorentz invariance
spontaneously emerges in the infrared. This is proved by fully
taking into account the discrete lattice symmetries, which are
used to exclude the presence of dangerous extra marginal or
relevant terms in the RG flow. \item {\it The wave function
renormalization diverges at the Fermi points with an anomalous
exponent}. This last properties strongly resembles one of the
crucial features of one-dimensional {\it Luttinger liquids}; in
this sense, the model considered in this paper is one of the very
few established examples of Luttinger liquid behavior in {\it two}
dimensions (it has been suggested that also the Hubbard model on
the square lattice close to half filling is a Luttinger liquid,
but this is still an unproven fact). \item The response functions
have an {\it anomalous behavior} expressed in terms of non trivial
scaling exponents. The difference between the interacting and
non-interacting exponents is small at small coupling. In
particular, the response functions associated to fermionic
bilinears, which decay as $r^{-4}$ at large distances in the
non-interacting case, remain integrable even in the presence of
weak interactions; therefore, magnetic, phonon or superconducting
susceptibilities are finite and {\it no evidence for quantum
instabilities} is found, in agreement with the fact that the fixed
point is close to the trivial one at weak coupling. \item On the
other hand, the interaction removes the degeneracy in the decay
exponents of the response functions: some exponent increase and
some other decrease and this depresses or enhances the effects of
local perturbations associated to specific fermionic bilinears.
This gives us a criterium to extrapolate the qualitative behavior
of the system at larger values of the electric charge and decide
what are the favorite {\it quantum instabilities at intermediate
to strong coupling}. An explicit computation of the anomalous
exponents shows that the {\it dominant excitations} correspond to:
(i) {\it Kekul\'e distortions}, associated to a dimerized Peierls'
pattern, (ii) {\it charge-density waves} associated to an
excess/deficit of the electron density on the two sublattices of
the honeycomb lattice, (iii) {\it Ne\'el antiferromagnetism}, (iv)
the {\it Haldane circulating currents} \cite{Ha}. In all these
cases, the logarithmic singularity at zero transferred momentum in
the first derivative of the response function in momentum space is
changed into a power law singularity with an anomalous exponent.
The singularity in the other responses is either absent or weaker
than in the four cases mentioned above. \item If we add a symmetry
breaking field coupled to a Kekul\'e distortion the induced energy
gap in the spectrum is dramatically amplified by the interaction:
the ratio between the energy gap and the amplitude $\D_0$ of the
external field diverges as $\D_0\to 0$ with an anomalous power
law. \item The effect of the electronic repulsion on the
Peierls-Kekul\'e instability, usually neglected, is evaluated by
deriving an exact non-BCS gap equation, from which evidence is
found that the gauge interaction facilitates the spontaneous
distortion of the lattice and the gap generation. Similar
conclusions can be drawn for the mass terms corresponding to the
other dominant excitations.
\end{enumerate}

The exact Wilsonian RG methods we use are based on ideas borrowed
from constructive QFT and have been introduced in the context of
interacting Fermi systems in \cite{BG} (and then extended in
\cite{Sh,Po}) at the beginning of the 90s, and since then
successfully applied to various problems in solid state physics,
see, e.g., \cite{Mbook} for an updated review. In particular, they
appear to be very well suited to analyze the properties of
graphene without any Dirac approximation, large-N approximations
or unphysical regularization schemes. Such methods have been first
applied in \cite{GM} to the Hubbard model on the honeycomb
lattice, where we proved in a full {\it non-perturbative} fashion
the analyticity of the ground state of the half-filled system; we
used the same methods to rigorously establish the universality of
the optical conductivity of graphene with weak short range
interactions \cite{GMP3,GMP3long}, an issue that still needs to be
fully understood in the case of electromagnetic interactions.
Technically, the analysis in this paper extends the one in
\cite{GMP1}, the main differences being that here we fully exploit
the lattice symmetries and, besides computing the reduced density
matrices, we construct the response functions, we compute the
critical exponents and discuss the effects of external symmetry
breaking fields. Most of the results in this paper were announced
in \cite{GMP2}.

Let us add a comment on the range of applicability of our theory.
Our analysis is based on resummations of perturbation theory in $\a$,
where  $\a=e^2/(4\p\e\hbar v)$ is the effective fine structure
constant of graphene, which unfortunately is not small: e.g., in
suspended graphene, $\a\simeq 2$, which makes graphene an intrinsically
{\it strongly coupled} problem, apriori not accessible to approaches based
on power series expansions. However, one needs to take into account that
the effective Fermi velocity is considerably increased by the interactions
(recent experiments \cite{E} can observe a factor three amplification of the
velocity close to the Fermi points
and even a larger enhancement is expected at lower energies), an effect that
goes in the direction of decreasing the effective fine structure constant.
Therefore, our theory is valid close to the IR
fixed point, with effective parameters $e$ and $v$ ($v$ close to the speed of
light) that should be thought
as being obtained by the (non-perturbative) integration of the ``first few IR
scales'', possibly by using numerical methods, like those of \cite{DL1,DL2}.
Let us also remark that since we predict the emergence of anomalous critical
exponents, our final results can be easily extrapolated to intermediate
coupling, which would not be the case if the apparent logarithmic divergences
emerging in perturbation theory were not correctly resummed at all orders.

The rest of the paper is organized as follows. In Section \ref{sec2}, we define the model, we
present in detail our main results and compare them with existing literature.
In Sections \ref{sec2.2} to \ref{appC} we discuss the proof of our results: in Section \ref{sec2.2} we
discuss the functional integral representation of our model and derive the relevant
Ward Identities (WIs); in Section \ref{sec3} we describe our RG scheme to compute the functional integral;
in Section \ref{secWI} we show how to use WIs to prove the vanishing of the photon mass, the
vanishing of the beta function for the effective charge, and how to control the flow of the
renormalization parameters (Fermi velocity, wave function renormalization, vertex functions);
in Sections \ref{sec3ac} and \ref{secoth} we compute the response functions associated to several
fermionic bilinears; in Sections \ref{seckek} and \ref{appC} we explain how to modify the general RG scheme
to include the effects of an external field coupled to local order parameters and how to derive
the non-BCS equation for the gap. A number of more technical aspects of the proof are deferred to
the Appendices: in Appendix \ref{app1b} we derive the functional integral representation, we derive the WIs
and we explicitly show the
equivalence between the Feynman and Coulomb gauges; in Appendices \ref{app1} and \ref{app3} we analyze the
symmetry properties of the theory and use them to identify the symmetry structure of the relevant
and marginal kernels; in Appendix \ref{app2d} we perform the lowest order computations of the beta
function and of the critical exponents; finally, for completeness, in Appendix \ref{WIcheck} we check at lowest
order the cancellation of the photon mass and of the charge beta function, which follows
from the general gauge invariance of the model.

\section{A lattice gauge theory for graphene}\label{sec2}
\setcounter{equation}{0}
\renewcommand{\theequation}{\ref{sec2}.\arabic{equation}}

In this section we introduce the model, define the main quantities of interest and present
our main results. A comparison with existing literature is also presented.

\subsection{The model}\label{sec2_model}

We let $\L =
\{n_{1}\vec l_{1} + n_{2}\vec l_{2}: n_{i} = 0,\ldots , L-1\}$ be
a periodic triangular lattice of period $L$, with basis vectors
$\vec l_{1} = \frac{1}{2}(3,\sqrt{3})$, $\vec l_{2} =
\frac{1}{2}(3,-\sqrt{3})$. We denote by $\L_{A} = \L$ and
$\L_{B} = \L + \vec \d_{i}$ the $A$- and $B$- sublattices of the
honeycomb lattice, with $\vec \d_{i}$ the nearest neighbors
vectors defined as:
\be \vec\d_{1} = (1,0)\;,\qquad \vec\d_{2} =
\frac{1}{2}(-1,\sqrt{3})\;, \qquad \vec\d_{3} =
\frac{1}{2}(-1,-\sqrt{3})\;.\label{1.1} \ee
\begin{figure}[hbtp]
\centering
\includegraphics[width=.3\textwidth]{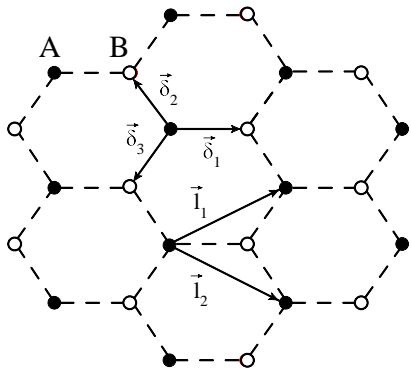}
\caption{The honeycomb lattice of graphene.}\label{figlat}
\end{figure}
We introduce creation and annihilation fermionic operators for
electron sitting at the sites of the $A$- and $B$- sublattices
with spin index $\s=\uparrow\downarrow$ as
\bea &&a^{\pm}_{\vec x,\s} = L^{-2}\sum_{\vec k\in\BBB_{L}}
e^{\pm i \vec k\vec x}\hat a^{\pm}_{\vec k,\s}\;,\hskip1.3truecm \vec x\in \L_{A}\;,\nn\\
&&b^{\pm}_{\vec x,\s} = L^{-2}\sum_{\vec k\in\BBB_{L}} e^{\pm
i\vec k(\vec x - \vec\d_{1})}\hat b^{\pm}_{\vec k,\s}\;,\qquad
\vec x\in \L_{B}\;,\label{1.2} \eea
where $\BBB_{L} = \{\vec k = n_{1}\vec G_{1}/L + n_{2}\vec
G_{2}/L: 0\leq n_{i} < L\}$, with $\vec G_{1,2} =
\frac{2\pi}{3}(1,\pm \sqrt{3})$, is the first Brillouin zone; note
that in the thermodynamic limit $L^{-2}\sum_{\vec k\in
\BBB_{L}}\rightarrow |\BBB|^{-1}\int_{\BBB}d\vec k$, with $\BBB =
\{\vec k = \xi_{1}\vec G_{1} + \xi_{2}\vec G_{2}: \xi_{i}\in
[0,1)\}$ and $|\BBB| = 8\pi^2/(3\sqrt{3})$. The operators
$a^{\pm}_{\vec x,\s}$, $b^{\pm}_{\vec x,\s}$ satisfy the canonical
anticommutation rules, and are periodic over $\L$; their Fourier
transforms are normalized in such a way that, if $\vec k, \vec
k'\in\BBB_L$,
\be \{\hat a^{+}_{\vec k,\s},\hat a^{+}_{\vec k',\s'}\} =\{\hat a^{-}_{\vec k,\s},\hat a^{-}_{\vec k',\s'}\}=
\{\hat b^{+}_{\vec k,\s},\hat b^{+}_{\vec k',\s'}\} =\{\hat b^{-}_{\vec k,\s},\hat b^{-}_{\vec k',\s'}\}=0\;,
\qquad \{\hat a^{-}_{\vec k,\s},\hat a^{+}_{\vec k',\s'}\} =
\{\hat b^{-}_{\vec k,\s},\hat b^{+}_{\vec k',\s'}\} =L^2\d_{\vec k,\vec k'}\d_{\s,\s'}\;.
\label{1.3} \ee
Definition Eq.(\ref{1.2}) implies that $\hat a^{\pm}_{\vec k,\s}$,
$\hat b^{\pm}_{\vec k,\s}$ are periodic over the reciprocal
lattice $\L^{*}$.

We also introduce a quantized photon field living in the 3D continuum. Let
$\bar\SS_{L,L'} = \SS_{L}\times [-\frac{L'}2, \frac{L'}2)$ be a subset of $\mathbb{R}^3$ with periodic
boundary conditions and $\SS_{L} = \{\vec x= L\xi_{1}\vec l_{1} + L\xi_{2}\vec l_{2}: \xi_{i}\in
[0,1)\}$; let also $\bar \DD_{L,L'} = \DD_{L}\times \frac{2\p}{L'}\mathbb{Z}$
be the corresponding dual momentum space, with $\DD_{L} = \{ \vec p = n_{1}\vec G_{1}/L + n_{2}\vec
G_{2}/L: n_{i} \in \ZZZ \}$ (the honeycomb lattice can be thought as being contained in the section
$\SS_L\times{0}$ at $z=0$ of $\bar\SS_L$). For all $p = (\vec p, p_3)\in
\bar\DD_{L,L'}$ we introduce bosonic creation and annihilation
operators $\hat c^{\pm}_{p,r}$, with helicity index $r=1,2$; they satisfy the
commutation relation
\be \big[\hat c^{+}_{p, r}, \hat c^{+}_{p',r'}\big] =\big[\hat c^{-}_{p, r}, \hat c^{-}_{p',r'}\big] =0\;,
\qquad \big[\hat c^{-}_{p, r}, \hat c^{+}_{p',r'}\big] =L'|\SS_{L}| \d_{p,p'}\d_{r,r'}\;.\label{1.4} \ee
Let $A(x) = (\vec A(x),A_{3}(x))$ be the quantized vector potential on $\bar\SS_L$
defined as:
\be A(x) = \frac{1}{L'|\SS_{L}|} \sum_{p\in\bar\DD_{L,L'}}\,\sum_{r=1,2}
\sqrt{\frac{\chi(|p|)}{2|p|}}\e_{p,r} \Big(\hat c^{-}_{p, r}e^{-ip x} +\hat
c^{+}_{p,r}e^{ipx}\Big)\;,\label{1.5}\ee
where $\e_{p,r}\in\mathbb{R}^3$ are polarization vectors satisfying the conditions
\be \e_{p,r}\cdot \e_{p,r'} = \d_{r,r'}\;, \qquad
\e_{p,r}\cdot p =0\;,\label{1.5a} \ee
which reflect the choice of the {\it Coulomb gauge}. Note that, in the thermodynamic limit
$|\bar\SS_{L}|^{-1}\sum_{p \in \bar\DD_{L}}
\rightarrow (2\p)^3\int_{\mathbb{R}^{3}}\,dp$. Moreover, the function $\chi(|p|)$ acts as an
UV cutoff function: more exactly $\chi(t)$ is a smooth compact support function
equal to $1$ for $0\leq t\leq \frac{1}{2}$ and to $0$ for $t\geq 1$. The photon UV cutoff is chosen to be on the same scale as the
inverse lattice spacing. We expect that this cutoff should be removable, but this
is not our main concern here.

The interacting electron-photon system we are interested in is described at half filling and
in the Coulomb gauge by the following grandcanonical Hamiltonian:
\be H_{\L} = H^{h}_{\L} + H^{f}_{\L} +
V_{\L}\;,\label{1.7} \ee
where the first term is the (gauge-invariant) hopping term, the
second represents the field energy, and the third is
the Coulomb interaction, namely:
\bea H^{h}_{\L}&=& -t\sum_{\substack{\vec x\in \L_{A} \\
j=1,2,3}} \sum_{\s= \uparrow\downarrow} \Big(e^{ie\int_0^1ds\vec A(\vec x+s\vec\d_j,0)\cdot\vec\d_j}
a^{+}_{\vec x,\s} b^{-}_{\vec x+\vec\d_{j},\s} + e^{-ie\int_0^1ds\vec A(\vec x+s\vec\d_j,0)\cdot\vec\d_j}
b^{+}_{\vec x + \vec \d_{j},\s}a^{-}_{\vec x,\s}\Big)\;,\label{1.8}\\
H^{f}_{\L} &=& \frac{1}{L'|\SS_{L}|}\sum_{\substack{p\in\bar\DD_{L,L'}\\r=1,2}}
|p| \hat c^{+}_{p,r}\hat c^{-}_{p,r}\;,\qquad \qquad
V_{\L} = \frac{e^2}{2}\sum_{\vec x,\vec
y\in\L_{A}\cup\L_{B}}( n_{\vec x} - 1
)\varphi\big(\vec x - \vec y\big) ( n_{\vec y} -
1 )\;,\nn \eea
where $t>0$ is the hopping strength, $e$ is the electric charge,
and $n_{\vec x}$ is equal to $\sum_{\s = \uparrow\downarrow}
a^{+}_{\vec x,\s}a^{-}_{\vec x,\s}$ or to $\sum_{\s = \uparrow\downarrow}b^{+}_{\vec
x,\s}b^{-}_{\vec x,\s}$ depending on whether $\vec x \in
\L_{A},\L_{B}$, respectively; moreover, $\f(\vec x)$ is a regularized periodic version of the
3D Coulomb potential:
\be \varphi(\vec x) = \frac{1}{L'|\SS_{L}|}\sum_{p\in\bar \DD_{L,L'}} \frac{\chi(|p|)}{p^2}
e^{-i \vec p\,\cdot\vec x}\;,\label{1.6} \ee
where we remind the reader that $p=(\vec p,p_3)$. Note that the electron-photon interaction is induced both by
the complex hopping rate
$t\exp\{ie\int_0^1ds\vec A(\vec x+s\vec\d_j)\cdot\vec\d_j\}$ in $H^h_\L$ and by the
static Coulomb interaction $V_\L$; the combination of the two describes the retarded
electromagnetic interaction mediated by 3D photons. If the electric charge $e$ is $0$, then
the Hamiltonian decouples into a sum of two quadratic terms:
\be H_\L\big|_{e=0}=-t\sum_{\substack{\vec x\in\L \\
j=1,2,3}}\sum_{\s=\uparrow\downarrow}\Big(a^{+}_{\vec x,\s}
b^{-}_{\vec x +\vec \d_j,\s}+ b^{+}_{\vec x +\vec
\d_j,\s}a^{-}_{\vec x,\s}\Big)+\frac{1}{L'|\SS_{L}|}\sum_{\substack{p\in\bar\DD_{L,L'}\\r=1,2}}
|p| \hat c^{+}_{p,r}\hat c^{-}_{p,r}=:H^0_\L+H^f_\L\;,\label{1a}\ee
which can be explicitly diagonalized; the corresponding correlation functions can be
computed exactly, via the Wick rule, in terms of the electron and photon propagators, which read
as follows. Let $\Psi^+_{\xx,\s}=(a^+_{\xx,\s}, b^+_{\xx,+\dd_1,\s})$
and $\Psi^-_{\xx,\s}=\begin{pmatrix}a^-_{\xx,\s}\\ b^-_{\xx+\dd_1,\s}\end{pmatrix}$ be row and
column spinors, with $\dd_1 = (0,\vec\d_1)$, $\xx = (x_0,\vec x)$,
$x_0\in[0,\b)$ an imaginary time, $\b>0$ the inverse temperature,
and $a^{\pm}_{\xx,\s}=e^{H_{\L} x_0}a^{\pm}_{\vec x,\s}e^{-H_{\L} x_0}$,
$b^{\pm}_{\xx+\dd_1,\s}=e^{H_{\L} x_0}b^{\pm}_{\vec x+\vec\d_1,\s}e^{-H_{\L} x_0}$
the imaginary time evolved of the creation/annihilation operators.
Then the {\it free electron propagator} is \cite{GM}
\be S^{\b,L}_{0}(\xx) := \media{{\bf T}
\{\Psi^{-}_{\xx,\s}\Psi^{+}_{\V0,\s}\}}_{\b,L}\big|_{e=0}=
\frac{1}{\b L^{2}}\sum_{\kk \in\BBB_{\b,L}} e^{-i\kk\xx}\hat S_0(\kk)\;,\qquad \hat S_0(\kk):=
\frac{1}{k_0^2 + v^2 |\O(\vec
k)|^2}\begin{pmatrix} i k_0 & -v \O^{*}(\vec k) \\ -v \O(\vec
k) & i k_0 \end{pmatrix}
\;,\label{free1.1}
\ee
where $\media{\cdot}_{\b,L}$ denotes the average
with respect to $e^{-\b H^0_{\L}(t)}$, ${\bf T}$ is the
fermionic time ordering, $\BBB_{\b,L}:=\frac{2\p}{\b}(\mathbb{Z}+\frac12)\times\BBB_L$,
$v = \frac{3}{2}t$ is the {\it bare Fermi velocity}
and $\O(\vec k) = \frac{2}{3}\sum_{j=1}^{3} e^{i \vec k(\vec \d_j
- \vec\d_1)}$ the {\it complex dispersion relation}.
The function $\O(\vec k)$ is vanishing if and only if $\vec k =
\vec p_{F}^{\pm}$, where $\vec p_{F}^{\pm}=\Big(\frac{2\pi}{3},\,\pm\frac{2\pi}{3\sqrt{3}}\Big)$
are the two {\it Fermi points}, close to which $\O(\vec k' + \vec p_{F}^{\pm}) = i k'_{1} \pm k'_{2}
+ O(|\vec k'|^2)$. Therefore, setting $\pp_{F}^{\o} = (0,\vec
p_{F}^{\o})$, $\kk' = \kk - \pp_{F}^{\o}$, $\o=\pm$, the propagator in momentum space reads
\be \hat S_{0}(\kk' + \pp_{F}^{\o}) = -\frac{1}{Z}\begin{pmatrix}
i k_0 & v(-i k'_{1} + \o k'_{2})\\  v(i
k'_{1} + \o k'_{2})& i k_{0}
\end{pmatrix}^{-1}\Big(1+O(|\vec k'|^2)\Big)\;,\label{free5} \ee
where $Z=1$ is the {\it bare wave function renormalization}.
Eq.(\ref{free5}) has the form of the propagator for {\it massless Dirac fermions} in $2+1$
dimensions.

Similarly, defining, for $\xx=(x_0,\vec x)$,
$\vec A_{\xx}=e^{H^f_\L x_0}\vec A(\vec x,0)\,e^{-H^f_{\L}x_0}$, the in-plane {\it free photon propagator} is,
for $i,j\in\{1,2\}$,
\be w^{\b,L,L'}_{ij}(\xx):=\media{{\bf T}\big[(A_{\xx})_i(A_{\V0})_j\big]}_{\b,L}\big|_{e=0}=
\frac1{L'|\SS_L|}\sum_{\pp\in\DD_{\b,L}}\,
\sum_{p_3\in \frac{2\p}{L'}\mathbb{Z}}
e^{-i\pp\xx}\frac{\chi(|p|)}{\pp^{2} + p_3^2}\Big(\d_{ij} - \frac{p_{i}p_j}{|\vec p|^2+p_3^3}\Big)\;,
\label{1.14}\ee
where $\DD_{\b,L}:=\frac{2\p}{\b}\mathbb{Z}\times\DD_L$.
In the limit $L'\to\infty$,
\be w^{\b,L}_{ij}(\xx)=
\frac1{|\SS_L|}\sum_{\pp\in\DD_{\b,L}}e^{-i\pp\xx}\hat w^{(C)}_{ij}(\pp)\;,\qquad
\hat w^{(C)}_{ij}(\pp):=\int_{\mathbb{R}}\frac{dp_3}{2\p}\frac{\chi(|p|)}{\pp^2+p_3^2}
\Big(\d_{ij} - \frac{p_{i}p_j}{|\vec p|^2+p_3^2}\Big)\;,\label{1.15} \ee
where the apex $(C)$ reminds the choice of the Coulomb gauge. Note that the IR singularity of the in-plane
photon propagator in momentum space is $\sim |\pp|^{-1}$,
rather than the usual $\sim |\pp|^{-2}$ of standard quantum electrodynamics (QED). Therefore,
interacting graphene at low energies is similar to a gas of massless 2D Dirac particles interacting
via a modified $|\pp|^{-1}$ photon propagator.

\subsection{Response functions}

\begin{figure}[hbtp]
\centering
\includegraphics[width=.85\textwidth]{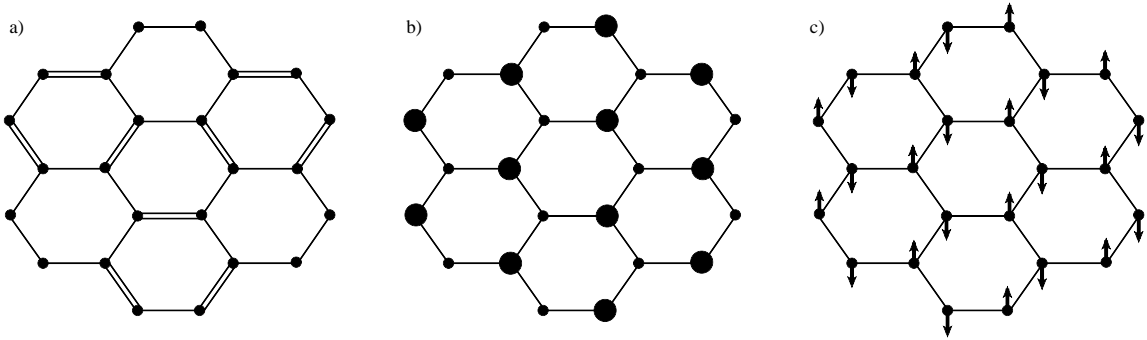}\caption{$a)$ The {\it Kekul\' e distortion}; the double and
single bonds correspond to higher and smaller hopping rates,
respectively. $b)$ The {\it charge density wave instability}; big
dots correspond to a charge excess, while small dots correspond to
a charge deficit. $c)$ The {\it antiferromagnetic instability};
the arrows represent the spins of the electrons, and they have to
be understood as lying on the axis orthogonal to the honeycomb
lattice. }\label{KCAF}
\end{figure}

Our goal is to understand the behavior of the system in the presence of a non-zero electron-photon coupling.
We will be mainly concerned with the computation of the interacting correlations, in particular
of the interacting electronic propagator and response functions. The latter are particularly relevant from
a physical point of view, since we can read from their long distance behavior (or, equivalently, from
their singularities in momentum space) the tendency of the system to develop quantum instabilities
associated to several putative local order parameters, in the same spirit as \cite{So}.
For illustrative purposes, we restrict our attention to
the response functions associated to the following bond fermionic bilinears:
\be \begin{array}{lll} &\zeta^K_{\xx,j} =
\sum_{\s}\Big(e^{ ie\int_{0}^{1}ds\,\vec\d_j\vec A_{\xx+s\dd_j}}
a^{+}_{\xx,\s}b^{-}_{\xx+\dd_j,\s} + c.c. \Big)
&\qquad {\rm (lattice\ distortion)}
\\
&\zeta^{CDW}_{\xx,j}=\sum_\s\Big(a^+_{\xx,\s}a^-_{\xx,\s}-b^+_{\xx+\dd_j,\s}b^-_{\xx+\dd_j,\s}\Big)
&\qquad
{\rm (staggered\ density)}
\\
&\z^{AF}_{\xx,j}=\sum_\s\s\Big(a^+_{\xx,\s} a^-_{\xx,\s}-b^+_{\xx+\dd_j,\s} b^-_{\xx+\dd_j,\s}\Big)
&\qquad{\rm (staggered\ magnetization)}\\
& \zeta^{D}_{\xx,j}=\sum_\s\Big(a^+_{\xx,\s}a^-_{\xx,\s}+b^+_{\xx+\dd_j,\s}b^-_{\xx+\dd_j,\s}\Big)
&\qquad {\rm (bond\ density)}\\
&\zeta^{J}_{\xx,j}=\sum_{\s}\Big(ie^{ ie\int_{0}^{1}ds\,\vec\d_j\vec A_{\xx+s\dd_j}}
a^+_{\xx,\s}b^-_{\xx+\dd_j,\s}+c.c. \Big)
&\qquad {\rm (bond\ current)}
\\
&\z^H_{\xx,j}=\sum_\s
\Big(ie^{ ie\int_{0}^{1}ds\,\vec m_j\vec A_{\xx+s{\bf m}_j}}
a^+_{\xx,\s}a^-_{\xx+{\bf m}_j,\s}-ie^{-ie\int_{0}^{1}ds\,\vec m_j\vec A_{\xx+s{\bf m}_j}}&
b^+_{\xx+\dd_j,\s}b^-_{\xx+\dd_j+{\bf m}_{j},\s}+c.c.\Big)\\
&&\qquad{\rm (Haldane\ circulating\ currents)}
\end{array}\label{bil6}
\ee
where in the last line ${\bf m}_1=\dd_2-\dd_3$,
${\bf m}_2=\dd_3-\dd_1$ and ${\bf m}_3=\dd_1-\dd_2$ indicate next to nearest neighbor vectors.
The corresponding response functions are defined as:
\be R^{(a)}_{ij}(\xx-\yy)=\media{\z^{a}_{\xx,i};\z^{a}_{\yy,j}}=
\lim_{\b\to\infty}\lim_{L\to\infty}\media{\z^{a}_{\xx,i};\z^{a}_{\yy,j}}_{\b,L}\;,\label{res1}\ee
where the semicolon in $\media{\cdot;\cdot}$ indicates truncated
expectation: $\media{A;B}:=\media{AB}-\media{A}\media{B}$. From the
long distance behavior of $R^{(a)}_{ij}(\xx)$ we can read the possible emergence
of long range order. For instance, if
the lattice distortion response function behaved as
$R^{(K)}_{ij}(\xx)\simeq C \cos(\vec p_F^+(\vec
x-\vec\d_i+\vec\d_j))$ for some $C\neq 0$ asymptotically as
$|\xx|\to\infty$, this would signal the spontaneous emergence of a
Peierls' instability in the form of the dimerized Kekul\'e pattern
of Fig.\ref{KCAF}a, which is one of the possible distortion
patterns of graphene \cite{HCM,FL}. In fact, $\z^{K}_{\xx,j}$ is
the local order parameter coupled in the Hamiltonian to the
hopping strength, whose mean value heuristically represents the
intensity of the hopping from a given site to its nearest
neighbor. Having $C\neq 0$ means that far away bonds are
strongly correlated and that the joint distribution of their
hopping rates is non uniform, but rather oscillating with a cosine
dependence. With reference to Fig.\ref{KCAF}a, this oscillation
can be heuristically understood by associating a factor
proportional to $1$ (to $-\frac12$) to the double (single) bonds
and by averaging over the three equiprobable configurations
obtained by rotating Fig.\ref{KCAF}a by $0,\frac{2\p}3,
\frac{4\p}3$. In momentum space, this would correspond to the
appearance of a delta singularity (i.e., of a ``condensate") in
$\hat R^{(K)}_{ij}(\pp)$ at $\pp= \pp_F^\pm$. Similarly, a
condensate in the $\pp=\V0$ mode of $\hat R^{(CDW)}_{ij}(\pp)$,
$\hat R^{(AF)}_{ij}(\pp)$ or $\hat R^{(H)}_{ij}(\pp)$ would signal
the spontaneous emergence of a staggered density pattern ({\it
charge density wave}), of a staggered magnetization pattern ({\it
Ne\'el order}) as in Fig.\ref{KCAF}b and Fig.\ref{KCAF}c,
respectively, or of the specific pattern of circulating currents discussed in
\cite{Ha}. Even in the
absence of condensation, that is of delta-like singularities in
$\hat R_{ij}^{(a)}(\pp)$, the possible loss of regularity in $\hat
R^{(a)}_{ij}(\pp)$ due to the interaction can be interpreted as a
tendency of the system to develop quasi-long range order in the
corresponding channel.
%

\subsection{Critical behavior and anomalous exponents}\label{sec2.1.1c}
%
We are now ready to state our main results. As shown in the following, after systematic resummations of
perturbation theory, we are able to express the observables of our theory
as renormalized series in the electric charge, with finite (and explicitly bounded) coefficients
at all orders; the coefficient of $e^{2n}$ grows at most as $n!$, a behavior compatible with
Borel summability of the theory at weak enough coupling strength.

In particular, in the limit $\b,L\to\infty$, the interacting two-point function in the Feynman gauge
(see Section \ref{sec2.2}) is given by:
\bea &&S(\xx)=\media{{\bf T}(\psi^-_{\xx,\s}\psi^+_{\V0,\s})}=\int_{-\infty}^{\infty}\frac{dk_0}{(2\p)}
\int_{\BBB}\frac{d\vec k}{|\BBB|}e^{-i\pp\xx}\hat S(\kk)\;,\nn\\
&&\hat S(\kk' + \pp_{F}^{\o}) = - \frac{1}{Z(\kk')}\begin{pmatrix}
i k_0 & v(\kk')(-i k'_1 + \o k'_2) \\ v(\kk')(i k'_1 + \o k'_2) &
i k_0 \end{pmatrix}^{-1}\big(1 + B(\kk')\big)\label{res1z} \eea
where $Z(\kk')$ and $v(\kk')$ are the interacting wave function renormalization and Fermi velocity,
while $B(\kk)$ is a subdominant term, vanishing at the Fermi points $\kk'=\V0$. The interacting propagator
close to the Fermi points has a structure very reminiscent of the free propagator, Eq.(\ref{free5}). However,
$Z(\kk')$ and $v(\kk')$ are {\it strongly renormalized} by the interaction:
\be Z(\kk') \simeq |\kk'|^{-\eta}\;,\qquad 1-v(\kk') \simeq (1 - v)|\kk'|^{\tilde\eta}\;,\label{Zv}\ee
where
\be \eta = \frac{e^2}{12\pi^2} + O(e^4)\;,\qquad \tilde\eta =
\frac{2 e^2}{5\pi^2} + O(e^4)\;,\label{res2} \ee
are anomalous critical exponents, well defined at all orders in renormalized perturbation theory
(the $O(e^4)$ remainders in Eq.(\ref{res2}) are written below in terms of a series in $e^{2n}$, with
coefficients growing at most as $n!$). Eqs.(\ref{res1z})-(\ref{res2}) are very reminiscent of the
IR behavior of the 2-point function of a Luttinger liquid \cite{ML65,So,Ha2}, and are consistent with
results obtained in a model of interacting Dirac fermions in the continuum \cite{V1}.

On the contrary, the interacting 2-point function for the photon has the same IR singularity $\sim|\pp|^{-1}$
as the free case; this means that the gauge field remains {\it
massless} (no screening).

Regarding the response functions associated to fermionic bilinears,
we find that in general the interaction changes their decay exponents
at large distances as compared to the non-interacting case, where all the responses decay as
$\sim r^{-4}$ at large distances. The presence of the interaction makes these exponents non
trivial functions of $e$ (anomalous dimensions). In particular we prove that
\bea
&&R^{(K)}_{ij}(\xx)=\frac{27}{8\p^2} A_K\frac{\cos\big(\vec p_F^{+}(\vec x-\vec\d_i+\vec\d_j)\big)}
{|\xx|^{4-\xi^{(K)}}}
+r^{(K)}_{ij}(\xx)\;,\label{RK}\\
&&R^{(CDW)}_{ij}(\xx)= \frac{27}{8\p^2}A_{CDW}
\frac{1}{|\xx|^{4-\xi^{(CDW)}}}
+r^{(CDW)}_{ij}(\xx)\;,\label{RCDW}\\
&&R^{(AF)}_{jj'}(\xx)=\frac{27}{8\p^2}A_{AF}\frac{1}{|\xx|^{4-\xi^{(AF)}}}
+r^{(AF)}_{ij}(\xx)\;,\label{RAF}\\
&&R^{(H)}_{jj'}(\xx)=\frac{81}{8\p^2}A_{H}\frac{1}{|\xx|^{4-\xi^{(H)}}}
+r^{(H)}_{ij}(\xx)\;,\label{H}\eea
where
\be \xi^{(K)}= \frac{4 e^2}{3\pi^2} + O(e^4);\quad \xi^{(CDW)}=
\frac{4 e^2}{3\pi^2} + O(e^4);\quad \xi^{(AF)}= \frac{4
e^2}{3\pi^2} + O(e^4) ;\quad \xi^{(H)}= \frac{4 e^2}{3\pi^2} +
O(e^4) \label{expKCAF}\ee
and $A_\#=A_\#(v,e)$ are constants that are equal to 1 at the free
Dirac point, i.e., $A_\#\big|_{v=1, e=0}=1$ (in particular, for
$v$ close to 1, $A_\#=1+O(1-v)+O(e^2)$). Moreover, the correction
terms $r^{(a)}_{ij}(\xx)$ are subdominant contributions, decaying
at infinity faster than $|\xx|^{-4+\x^{(a)}}$; they include both the effects coming from
the irrelevant terms in a RG sense and the effects proportional to
$1-v(\kk')$ (see Eq.(\ref{Zv})) coming from the Lorentz symmetry breaking terms.
From Eqs.(\ref{RK})--(\ref{RAF}), we see that the decay of the
interacting responses in the K, CDW, AF, H channels is slower than
the corresponding non-interacting functions; i.e., the responses
to K, CDW, AF, H are strongly enhanced by the interaction. On the
contrary, all other responses decay at infinity faster than
$|\xx|^{-4+(\const.)e^4}$, i.e., if $a=D,J,$ (and similar bounds are valid for other observables
like the Cooper pairs, see Section \ref{secoth} below)
\be |R^{(a)}_{ij}(\xx)|\le
\frac{C}{|\xx|^{4-Ce^4}}\;,\label{oth_dec}\ee
for some constant $C>0$. These results can be naturally extrapolated to larger values of the electric charge,
in which case they suggest that the lattice distortion, staggered density, staggered magnetic order
and the ``Haldane circulating currents"
are the dominant quantum instabilities at intermediate to strong coupling strength.

As we noticed, quantum instabilities are also signaled by divergences
in the Fourier transform of the response functions. Even in the
presence of non trivial exponent, the power law decay remains
integrable at weak coupling: therefore, no divergence is found in
the Fourier transform of the response function.
On the other hand,
\be
\partial_\pp \hat R^{(K)}_{ij}(\pp'+\pp_F^\pm)\sim
|\pp'|^{-\xi^{(K)}}\quad{\rm and}\quad \partial_\pp \hat
R^{(a)}_{ij}(\pp)\sim |\pp|^{-\xi^{(a)}}\;, \quad {\rm with}\quad a=CDW, AF, H\;,\ee
so that the singularity in the first derivative of the response functions in
momentum space, which appeared as a discontinuity or at most
as a logarithmic divergence in the non interacting case, is enhanced and
turned into a power law singularity by the interaction for these four
responses. The singularity at different momenta or for other
response functions is either weaker or absent. Also in this respect, {\it the conclusion is that the
system shows a tendency towards Kekul\'e, charge density wave, Ne\'el ordering} or to the formation of the {\it Haldane gap}.

\subsection{Mass renormalization and gap equation}\label{sec2.1.1}

The enhancement of the response functions suggests that the effects of small external staggered
fields coupled to the $K, CDW, AF,H$ local order parameters are dramatically enhanced by the interactions.
This is in fact the case.
Let us, for instance, add an external staggered
field coupled to the lattice distortion local order parameter, with the same cosine dependence as
the long distance decay of $R^{(K)}_{ij}(\xx)$, see Eq.(\ref{RK}). Physically, this can be interpreted as a fixed distortion
of the lattice into a Kekul\'e pattern as in Fig.\ref{KCAF}a. In fact, if we allow
distortions of the honeycomb lattice, the hopping becomes a
function of the bond length $\ell_{\vec x,j}$ that, for small
deformations, can be approximated by the linear function $t_{\vec
x,j}=t+g (\ell_{\vec x,j}-\bar\ell)=:t+\phi_{\vec x,j}$, where $\bar\ell$ is the
equilibrium length of the bonds and $\phi_{\vec x,j}$ plays the role of a classical phonon field.
If a Kekul\'e distortion of amplitude proportional to $\D_0/g$ is present, then
$\phi_{\vec x,j}=
\frac{\D_0}3\, 2\cos(\vec p_F^{\,+}(\vec x-
\vec\d_j+\vec\d_{j_0}))$ for some $j_0\in\{1,2,3\}$
and the Hamiltonian becomes
\be H^{\D_0}_{\L}=H_{\L}-\frac{\D_0}{3}\sum_{j=1,2,3}\sum_{\substack{\vec x\in\L\\ \o=\pm}}
e^{i\vec p_F^{\,\o}(\vec x-\vec \d_j+\vec \d_{j_0})}\zeta^{K}_{\vec x,j}\;.\label{hd0} \ee
If the electron-photon coupling $e$ is equal to $0$, then the fermionic 2-point function has
a mass proportional to $\D_0$. If we switch on the interaction, the mass (i.e.,
the decay constant describing the exponential decay of the 2-point function at large distances)
becomes
\be \D=\D_0^{1/(1+ \h_\D)}\;,\qquad
\h_\D=\frac{2e^2}{3\pi^2}+\cdots\;,\label{aa} \ee
that is, the Kekul\'e mass is strongly amplified by the interaction (note that the ratio
between the interacting and bare masses diverges with an anomalous exponent
as $\D_0\to 0$). This phenomenon is very reminiscent of the
spontaneous mass generation phenomenon in QFT, the
main difference being that while in a truly relativistic theory in the continuum
the flow of the mass can be studied also in
the UV region and the bare mass can be let to zero with the UV cutoff (still keeping the
same dressed mass at fixed IR scale), see \cite{GGMS},
here the lattice acts as a fixed UV cut-off, so that the mass is amplified but not spontaneously
generated. Similar arguments can be repeated for the other dominant excitations.

Let us finally discuss a possible mechanism for the spontaneous generation of a Kekul\'e instability in our
model. Rather then fixing the distortion pattern $\phi_{\vec x,j}$ once and for all,
we can let $\ul\phi=\{\phi_{\vec x,j}\}_{\vec x\in\L}^{j=1,2,3}$ be a classical field to be fixed
self-consistently, in such a way that the total energy in the Born-Oppenheimer approximation is minimal, i.e.,
\be \ul\phi={\rm argmin}\Big\{
E_0(\ul\phi)+
\frac{\k}{2g^2}\sum_{\substack{\vec x\in\L\\ j=1,2,3}}
\phi^2_{\vec x,j}\Big\}\;,\label{var_eq}\ee
where $E_0(\ul\phi)$ is the ground state energy of
$H_\L^{\phi}=H_\L-\sum_{\substack{\vec x\in\L\\ j=1,2,3}}\phi_{\vec x,j}\z^{(K)}_{\vec x,j}$.
We find that the Kekul\'e distortion pattern
\be \phi^{(j_0)}_{\vec x,j}=\phi_0+\frac23\D_0
\cos\big(\vec p_F^{\,+}(\vec x-\vec \d_j+\vec \d_{j_0})\big)\label{phi0*}\ee
is a stationary point
of the total energy, provided that $\phi_0=c_0 g^2/\k+\cdots$ for
a suitable constant $c_0$ and that $\D_0$ satisfies the following
non-BCS gap equation:
\be\D_0\simeq 6 \frac{g^2}{\k}\!\!\!\int\limits_{\D\lesssim|\kk'|\lesssim 1}
\!\!\!\!\!d\kk'\frac{ Z^{-1}(\kk')\D(\kk')}
{k_0^2+v^2(\kk')|\O(\vec k'+ \vec p_F^\o)|^2+|\D(\kk')|^2}\;,\label{nonsimpl}\ee
where $\D=\D_0^{1/(1+\h_\D)}$ and, for
$\D\lesssim|\kk'| \ll1$, $Z(\kk') \sim |\kk'|^{-\h}$, $v(\kk')\sim1-(1-v)|\kk'|^{\tilde\h}$ and
$\D(\kk')\sim \D_0\, |\kk'|^{-\h_\D}$. Our gap equation has the same qualitative
properties of the simpler equation:
\be 1=g^2\int_\D^1d\r\frac{\r^{\h-\h_\D}}{1-(1-v)\r^{\tilde\h}}\label{simpl3}\ee
from which it is apparent that at small $e$, the equation admits a non trivial solution
only for $g$ larger than a critical coupling $g_c$;
remarkably, $g_c\sim\sqrt{v}$, with $v$ the free Fermi velocity,
even though the effective Fermi velocity tends to the speed of
light. Therefore, at weak coupling, the prediction for $g_c$ is
qualitatively the same as in the free case \cite{HCM}; this can be
easily checked by noting that the Fermi velocity $v(\kk')$ is sensibly
different from $v$ only for momentum scales exponentially small in $v/e^2$.
See Section \ref{appC} for more comments about this point.

Even more interestingly, the value of $g_c$ decreases as $e$ increases; i.e., interactions
facilitate the formation of a Kekul\'e pattern. Eqs.(\ref{nonsimpl})-(\ref{simpl3}) can be naturally
extrapolated to intermediate coupling: if in such a regime $\h_\D-\h=
\frac{7e^2}{12\p^2}+\cdots$ exceeds $1$, then the integrand in the
r.h.s. of the gap equation diverges as $\D\to 0$, a fact that guarantees the existence
of a non-trivial solution {\em for arbitrarily small} $g$.
In other words, the larger the electron-photon interaction, the easier
is to form a Kekul\'e patterned state; it is even possible that at intermediate coupling $g_c=0$,
which would imply a spontaneous generation of the Peierls'-Kekul\'e instability.
Note the non BCS-like form of the gap,
similar to the one appearing in certain Luttinger superconductors
\cite{M1}. A similar analysis can be repeated for the gap generated by the
staggered density, the magnetization or the Haldane mass.

\subsection{A comparison with existing literature}

Before we enter the technical part of our work, let us conclude
this expository section by a comparison of our model and our
results with existing literature. We do not pretend to give a full
account of the rapidly expanding literature on the effects of
interactions in graphene; several excellent reviews already
exists, like \cite{C,KUC2}, which we refer to for extensive
bibliography. Here we focus on the difference between the
approaches and results based on the effective models of Dirac gas
in the continuum, which is the most popular and widely studied
model of graphene, and ours, which is based on a tight binding
lattice model.

{\it Short range interactions}. Graphene with electron-electron
screened interactions has been studied in terms of an effective
model of 2D massless Dirac fermions interacting with a local
quartic potential: in the weak coupling regime this interaction is
irrelevant in the Renormalization Group sense \cite{V2}, while at
strong coupling analyses based on large $N$ expansions found some
evidence for quantum critical points \cite{H,SS,Son,HJR,Sac}. This
model requires an UV regularization and some observables, like the
conductivity, appear to be sensitive to the specific UV
regularization scheme used: different results are found
\cite{He1,He2} depending on whether momentum or dimensional
regularizations is chosen.

A more realistic model for graphene with short range interactions is a tight
binding model that keeps into full account lattice effects, such as
the half-filled {\it Hubbard model on the honeycomb lattice}.
Formally, the Hubbard model reduces to the continuum Dirac gas in
the limit as the lattice spacing goes to zero; in this sense the
latter can be thought as a scaling limit approximation of the
former. One important advantage of the lattice model as compared to the
continuum one is that within the former no ambiguities arise
in the computation of the conductivity.
In particular,  all the interaction correction to the conductivity
{\it exactly cancel out} in the optical limit \cite{GMP3},
in agreement with experimental results, \cite{Na}, and in
disagreement with the Dirac model with momentum regularization.

{\it Gauge-invariant electromagnetic interactions}. In the early paper
\cite{V1} (written much before the actual realization of
graphene) an effective model for interacting of graphene was
proposed, in which massless Dirac fermions in the 2D continuum
are coupled to a quantum 3D photon field, with the fermionic propagation
speed much smaller than the speed of light. The main result of
\cite{V1}, based on second order perturbation theory (and,
therefore, valid in the weak coupling regime), is that at low
energies the Fermi velocity tends to the speed of light and the
wave function renormalization diverges with an anomalous exponent.
No computation of the response function exponents was performed.

In \cite{GMP1}, we revisited the model proposed in \cite{V1} and,
rather than using dimensional regularization as in \cite{V1}, we
used an UV momentum cut-off; this is a much more natural choice,
since the Dirac continuum model should be though of as an
effective model emerging in a Wilsonian RG after the integration
of the high energy degrees of freedom. Using this model, we
extended the results in \cite{V1} at all orders, but still we
computed neither the response functions nor the gap equation. On
the other hand, the momentum cutoffs, necessary to avoid spurious
UV divergences, break gauge invariance and this imposed the
introduction of counterterms in order to keep the photon mass
vanishing and in order to have one (rather than three) effective
charge \cite{GMP1}.

In the more realistic model considered in this paper the electrons
live on a honeycomb lattice and interact with a quantum photon
field living in the 3D continuum. The fact that gauge invariance
is not broken has the effect that no unphysical counterterms need
to be introduced. The critical exponents and the gap equation have
been computed here for the first time. Moreover, lattice gauge
invariance prevents the generation of several potentially
dangerous marginal and relevant terms.
As in the case of
Hubbard interactions, the conductivity computed in the continuum
model show an unphysical dependence from the conductivity
\cite{He3}, while considering the model and the formalism
introduced in this model will resolve such ambiguities.

{\it Static Coulomb interactions.} The most popular model used to
describe graphene with unscreened electromagnetic interactions is
a Dirac gas with static density-density Coulomb interactions; this
is an apparently sensible approximation, since the (bare)
propagation speed of quasi-particles in graphene is much smaller
than the speed of light, so that retardation effects should be
negligible at least in a wide range of energy scales. However, in
the weak coupling regime, a second order RG analysis predicts an
unbounded growth of the Fermi velocity in the IR and the vanishing
of the effective charge at the Fermi points \cite{V3}. Therefore,
at low energy scales the Fermi velocity becomes comparable with
the speed of light and the model with static interactions loses
its significance, a fact that can be seen as a dramatic manifestation of
its ``uncompletness", see \cite{Voz}. This is also consistent with
the fact that the theory with static Coulomb interactions does not
appear to be renormalizable at all orders, see
\cite[p.1425]{GMP2}. In conclusion, retardation effects are
important to understand the nature of the IR fixed point of the
theory.

The effective Dirac model with Coulomb interactions has been also
extensively analyzed in the strong coupling regime. It has been
argued that, at large enough coupling,  an excitonic gap
spontaneously opens \cite{GGMS,K1, K2}, by a mechanism similar to
mass generation in QED$_{2+1}$ \cite{MI}: in these works, the gap
equation is derived by a self-consistence argument; the
corresponding solution is shown to be momentum-dependent and
vanishing in the limit of high momenta. However, these findings
rely on several approximations, in particular: in \cite{GGMS,K1}
the vertex, wave function and velocity renormalizations are
neglected; in \cite{K2} the renormalization of the velocity is
taken into account, but the corresponding flow is IR-unbounded and
UV-cutoff dependent. In our work, the gap equation is obtained by
using an {\it exact} energy optimization problem and fully takes
into account the renormalization of all relevant and marginal
operators; moreover, since we do not neglect effects of the
honeycomb lattice, our results are free from the ambiguities
related to the presence of the spurious UV divergences typical of
the Dirac approximation.

In \cite{GSC,HJR} a systematic classifications of the possible
interaction and mass terms allowed by symmetry is performed in the
Dirac model with Coulomb interactions; in the present paper
a similar analysis is carried out, with the difference that the lattice discrete
symmetries rather than the continuum symmetries are taken into account.

RG analyses based on large $N$ expansions \cite{H, Son, HJR,DS}
and Quantum Montecarlo analyses \cite{DL1,DL2} in the presence of
Coulomb interactions have identified critical exponents for the
response functions and found evidence for the presence of
excitonic phase transitions (like CDW and Kekul\'e instabilities).
Again, these results are in qualitative agreement with our finding
that the effective Kekul\'e mass (or the CDW, AF, H mass) grows at
low momenta with an anomalous power law, although the model and
the method are quite different (expansion in the charge and
retarded interactions versus $1/N$ expansions and instantaneous
interactions). A strong coupling expansion for a lattice
gauge theory for graphene has been also performed in \cite{A,A2} and
evidence for spontaneous Kekul\'e mass generation was found. It is
worth stressing that the lattice used for the Quantum Montecarlo
analysis in \cite{DL1,DL2} is a {\it square} lattice, rather than the original {\it
honeycomb} lattice; it would be interesting to repeat a similar
analysis for the more realistic honeycomb lattice gauge theory
introduced in the present paper.

Let us finally note that the Peierls-Kekul\'e instability was
first discussed in the non-interacting case in \cite{HCM}, as a
key ingredient for the emergence of electron fractionalization
without the breaking of time-reversal symmetry (on this issue, see
also \cite{JP,CHJMPS}). Our gap equation generalizes the one of
\cite{HCM} to the interacting case.

\section{Functional integral representation and Ward Identities}\label{sec2.2}
\setcounter{equation}{0}
\renewcommand{\theequation}{\ref{sec2.2}.\arabic{equation}}

The correlation functions introduced in the previous section can be conveniently expressed in
terms of a functional integral. We introduce the generating functional of correlations in
the $\x$-gauge  with infrared cutoff on the photon propagator as
\be e^{\WW^{\x,h^*}(\Phi,J,\l)}=\int P(d\Psi)P^{\x,h^*}(dA)e^{\VV(\Psi,A+J)+\BBB(\Psi,A+J,\Phi)+(\l,\Psi)}\;,
\label{gen}\ee
where:
\begin{enumerate}
\item $\Psi=\{\Psi_{\xx,\s}^\pm\}$ are two-components Grassmann fields, the components being denoted by
$\Psi^\pm_{\xx,\s,\r}$, $\r=1,2$,
the first corresponding to the $a^\pm$-fields, the second to the $b^\pm$-fields.
Moreover, $A=\{A_{\m,\xx}\}$, with $\m=0,1,2$, are real fields. The convention on
the Fourier transform of the fields that we use is the following:
\be \Psi^\pm_{\xx,\s,\r}=\frac1{\b L^2}\sum_{\kk\in\BBB_{\b,L}}e^{\pm i\kk\xx}\hat \Psi_{\kk,\s,\r}\;,
\qquad A_{\m,\xx}=\frac1{\b|\SS_L|}\sum_{\pp\in\DD_{\b,L}}e^{- i\pp\xx}\hat A_{\m,\pp}\;.\label{fou}\ee
\item If $n_\m=\d_{\m,0}$, $\m=0,1,2$, and
\be \hat w^{\x,h^*}_{\m\n}(\pp)=
\int_{\mathbb{R}}\frac{dp_3}{2\p}\frac{\c(|p|)-\c(2^{-h^*}|\pp|)}{\pp^2+p_3^2}
\Big(\d_{\m\n}-\x\frac{p_\m p_\n-p_0(p_\m n_\n+p_\n n_\m)}{|\vec p|^2+p_3^2}\Big)\label{pho_fey7}\ee
is the photon propagator in the $\x$-gauge (the Coulomb gauge corresponding to $\x=1$ and
the {\it Feynman gauge} corresponding to $\x=0$) with infrared cutoff at momenta of the order $2^{h^*}$,
$h^*<0$, then $P(d\Psi)$ and $P^{\x}(d\vec A)$ are the gaussian ``measures" associated to the propagators
$\hat S_0(\kk)$ and $\hat w^{\x,h^*}(\pp)$, respectively:
\bea &&
P(d\Psi) = \frac1{\NN_{\Psi}}\prod_{\kk\in\BBB_{\b,L}}
\prod_{\substack{\r=1,2\\ \s=\uparrow,\downarrow}}d\hat\Psi^+_{\kk,\s,\r}
d\hat\Psi^-_{\kk,\s,\r}\exp\Big\{-
\frac1{\b L^2}\sum_{\substack{\kk\in\BBB_{\b,L}\\ \s=\uparrow\downarrow}}
\hat\Psi^{+}_{\kk,\s}\big[\hat S_0(\kk)\big]^{-1}\hat\Psi^{-}_{\kk,\s}\Big\}\;,\label{Pdpsi}\\
&& P^{\x,h^*}(d A) =\frac1{\NN_{\x}}\prod_{\pp\in\DD^+_{\b,L}}\prod_{ \m=0,1,2}
d{\rm Re}A_{\m,\pp}d{\rm Im}A_{\m,\pp} \exp\Big\{-\frac1{2\b|\SS_L|}
\sum_{\substack{\pp\in\DD_{\b,L}\\ \m,\n=0,1,2}}^*\hat A_{\m,\pp}\big[\hat w^{\x,h^*}(\pp)]_{\m\n}^{-1}\hat
A_{\n,-\pp}\Big\}\;,\label{PdA}\eea
where $\NN_{\Psi}$, $\NN_{\x}$ two normalization factors; in the second line, the product over $\pp$
runs over the subset $\DD^+_{\b,L}$ of $\DD_{\b,L}$ such that $\pp>0$ (here $\pp>0$ means that
either $p_0>0$, or $p_0=0$ and $p_1>0$, or $p_0=p_1=0$ and $p_2>0$)
and $\pp$ is in the support of $\c(|\pp|)$. Similarly, the $*$ in the sum at exponent indicates that
the summation runs over the momenta in the support of $\c(|\pp|)$.
\end{enumerate}
{\bf Remark.}
Note that, by the very definition
of the cutoff function $\c$ (see the lines following Eq.(\ref{1.5a})), the modes $\hat A_{\m,\pp}$
associated to momenta close to $\pm(\pp_F^+-\pp_F^-)$
and to their images over the dual lattice $\L^*$ (i.e., to the points
$(0,\pm(\vec p_F^+-\vec p_F^-)+n_1\vec G_1+n_2\vec G_2)$, with
$n_1,n_2\in\mathbb{Z}$) are vanishing. In other words, the UV cutoff on the photon field
is chosen so small that {\it umklapp processes are suppressed}.
\begin{enumerate}
\setcounter{enumi}{2}
\item The interaction is
\bea \VV(\Psi,A) &=&\!
t\!\sum_{\substack{\s=\uparrow,\downarrow\\ j=1,2,3}}\!\int d\xx\Big[\big(e^{ie\int_0^1ds\,
\vec\d_j\vec A_{\xx+s\dd_j}}-1\big)
\Psi^{+}_{\xx,\s,1}\Psi^{-}_{\xx + \dd_{j} - \dd_1,\s,2} +
\big(e^{-ie\int_0^1ds\, \vec\d_j\vec A_{\xx+s\dd_j}}-1\big)\Psi^{+}_{\xx+\dd_j-\dd_1,\s,2}
\Psi^{-}_{\xx,\s,1}\Big] \nn\\
& -& ie\sum_{\s=\uparrow\downarrow}\int d\xx\big(
A_{0,\xx}\Psi^{+}_{\xx,\s,1}\Psi^{-}_{\xx,\s,1} +
A_{0,\xx+\dd_1}\Psi^{+}_{\xx,\s,2}\Psi^{-}_{\xx,\s,2}
\big)\;,\label{1.15uy} \eea
where $\int d\xx$ is a shorthand for $\int_{-\b/2}^{\b/2}dx_0\sum_{\vec x\in\L}$.
The interaction can be equivalently rewritten in momentum space, see Eq.(\ref{intmom}).
\item The external sources are
\be \BBB(\Psi,A,\Phi) = \int d\xx\,\sum_{a}\sum_{j=1,2,3}
\Phi^{a}_{j,\xx}\zeta^{a}_{j,\xx}\;, \qquad (\l,\Psi) = \int
d\xx\,\sum_{\s=\uparrow\downarrow}(
\l^{+}_{\xx,\s}\Psi^{-}_{\xx,\s} +
\Psi^{+}_{\xx,\s}\l^{-}_{\xx,\s})\;,\label{1.16} \ee
where the sum over the index $a$ runs over the choices $a=K,CDW,AF,D,J,C$, and
$\zeta^{a}_{i,\xx}$ are given by the same expressions Eqs.(\ref{bil6}) after the replacement of the
femionic and bosonic operators by the Grassmann and real fields $\Psi$ and $A$. The external fields
$\l^\pm_{\xx,\s}$ are Grassmann (two-components) fields, while $\Phi^a_{j,\xx}$ and $J_{\m,\xx}$ are real
fields, the first defined on the lattice, the second in the continuum with the same UV cutoff
as the photon field (therefore, as discussed in the previous remark, the UV cutoff on the
$J$ field is chosen so small that the modes $\hat J_{\m,\pp}$ with $\pp$ sufficiently close to
$\pm(\pp_F^+-\pp_F^-)$ and to their images over $\L^*$ are vanishing).
The source term can be equivalently rewritten in momentum space, see Eq.(\ref{intmom}).
\end{enumerate}

The response functions introduced in Section \ref{sec2} can be written as
functional derivatives of the generating function,
\be R^{(a)}_{ij}(\xx-\yy) =\lim_{\b\to\infty}\lim_{L\to\infty}
\lim_{h^*\to-\io}\frac{\partial^{2}}{\partial\Phi^{a}_{i,\xx}\partial\Phi^{a}_{j,\yy}}
\WW^{\x,h^*}(\Phi,0,0)\big|_{\Phi=0} \label{1.17} \ee
that, remarkably, are {\it independent} of the choice of the gauge, $\x\in[0,1]$
A sketch of the
proof of the functional integral representation of the observables of our theory, as well
as a proof of the independence of Eq.(\ref{1.17}) on the specific choice of the gauge, is
given in Appendix \ref{app1b}. Since the response functions are independent of $\x$, from now
on we choose to evaluate them in the Feynman gauge, i.e., $\x=0$.

Similarly, the {\it Schwinger functions} in the $\xi$-gauge at finite volume and finite temperature
are defined as the $h^*\to-\infty$ limit of
\bea && S^{\x}_{n,m;\ul\e,\ul\s,\ul\r,\ul\m}(\xx_1,\ldots,\xx_n;\yy_1,\ldots,\yy_m)
=\frac{\partial^{n+m}\WW^{\xi,h^*}(0,J,\l)}{\partial
\l^{\e_1}_{\xx_1,\s_1,\r_1} \cdots \partial\l^{\e_n}_{\xx_n,\s_n,\r_n}
\partial J_{\m_1,\xx_{n+1}}\cdots \partial J_{\m_m,\xx_{n+m}}}\Big|_{\l=J=0}\;.\label{1.17b}
\eea
Contrary to the response functions, the Schwinger functions (at least the way they
are defined in Eq.(\ref{1.17b})) are not gauge invariant and, therefore, they depend on the specific
choice of the gauge. Nevertheless, we decide to compute them in the Feynman gauge, which is
technically the simplest where to perform computations.
Although the Schwinger functions 
in the Feynman
gauge do not have an obvious Hamiltonian counterpart, we believe that they are a source of
valuable information on the behavior of the system. In particular, as we will see,
the 2-point functions and the vertex function, defined as
\bea && \big[\hat S^{\x,h^*}_{2,0}(\kk)\big]_{\r\r'}= \b L^2\frac{\dpr^2\WW^{\x,h^*}(0,0,\l)}
{\dpr\hat \l^-_{\kk,\s,\r'}\dpr\hat \l^+_{\kk,\s,\r}}\;,\qquad \qquad
\hat S^{\x,h^*}_{0,2;(\m,\n)}(\pp)=\b |\SS_L|\frac{\dpr^2\WW^{\x,h^*}(0,J,0)}
{\dpr\hat J^-_{\m,\pp}\dpr\hat J_{\n,-\pp}}\;,\nn\\
&&\big[\hat S^{\x,h^*}_{2,1;\m}(\kk,\pp)\big]_{\r\r'}= \b^2 L^2|\SS_L|\frac{\dpr^3\WW^{\x,h^*}(0,J,\l)}
{\dpr\hat J_{\m,\pp}\dpr\hat \l^-_{\kk,\s,\r'}\dpr\hat \l^+_{\kk+\pp,\s,\r}}\;,
\label{schw9}\eea
will play a crucial role in the study of the flow of
the effective couplings (and, therefore, of the response function themselves).

The independence of the gauge invariant observables on the specific choice of $\x$ is strictly related to
the gauge invariance of the generating functional $\WW^{\x,h^*}$ with respect to $U(1)$ gauge transformations.
Namely, for all $\x\in[0,1]$ and $h^*<0$, we have:
\be 0 = \frac{\partial}{\partial\hat \a_{\pp}}
\WW^{\xi,h^{*}}(\Phi,J + \partial\a,\l \eu^{ i e \a})
\Big|_{\a=0}\;,\label{1.23} \ee
see Appendix \ref{app1b} for a proof. By taking derivatives with respect to the external fields
in Eq.(\ref{1.23}), we can generates infinitely many identities between correlations,
also known as {\it Ward identities}. In fact, Eq.(\ref{1.23}) is equivalent to
\be \sum_{\m=0}^2p_{\m}\frac{\partial\WW^{\xi,h^{*}}(\Phi,J,\l)}
{\partial \hat J_{\m,\pp}} = \frac{e}{\b
|\SS_{L}|}\sum_{\substack{\kk\in\BBB_{\b,L}\\\s=\uparrow\downarrow}}
\Big[\hat\l^{+}_{\kk+\pp,\s}\G_0(\pp)\frac{\dpr\WW^{\xi,h^{*}}(\Phi,J,\l)}{\partial{\hat \l^{+}_{\kk,\s}}}+
\frac{\partial\WW^{\xi,h^{*}}(\Phi,J,\l)}{\dpr\hat\l^{-}_{\kk+\pp,\s}}
\G_0(\pp)\hat\l^{-}_{\kk,\s} \Big]\;,\label{1.23a}\ee
with $\G_0(\pp)=\bp 1 & 0\\ 0& e^{-i\pp\dd_1}\ep$. Taking, e.g., one derivative with respect to
$J$ or two derivatives with respect to $\l$, we find:
\be \sum_{\m=0}^2 p_\m \hat S^{\x,h^*}_{0,2;(\m,\n)}(\pp)=0\;,\qquad
\sum_{\m=0}^2 p_\m \hat S^{\x,h^*}_{2,1;\m}(\kk,\pp)=e\Big(\G_0(\pp)\hat S_{2,0}^{\x,h^*}(\kk)-
\hat S_{2,0}^{\x,h^*}(\kk+\pp)\G_0(\pp)\Big)\;,\label{WI87}\ee
which will be used below to deduce that the dressed photon mass is zero and the dressed electric charge is
close to the bare one within $O(e^3)$. Note the crucial fact that Eqs.(\ref{1.23})--(\ref{WI87}) are valid
at finite volume, finite temperature and {\it for any value of $h^*$}.

\section{Renormalization Group analysis}\label{sec3}
\setcounter{equation}{0}
\renewcommand{\theequation}{\ref{sec3}.\arabic{equation}}

In this Section we start the evaluation of the generating functional Eq.(\ref{gen}).
In the following, for simplicity, we set the external fermionic and bosonic fields to 0, $J=\l=0$,
and $h^*=-\io$ (dropping the $h^*$ index in the formulas). The effect of the external fields $J$ and $\l$ has
been discussed several times in the literature, see, e.g., \cite[Appendix B]{GMP1}.
The presence of a finite  bosonic infrared cutoff on scale $2^{h^*}$ will be discussed at the beginning of the
next section. Moreover, we set all the external fields $\Phi^{(a)}$
{\it but the one coupled to the Kekul\'e distortion} to zero.
The presence of the external field $\Phi^{(K)}$ will be discussed in detail, for illustrative purposes.
The effect of its addition is non trivial, particularly because of the new marginal terms it can generate.
The effect of other external fields $\Phi^{(a)}$, $a\neq K$, can be studied along the same lines and will be
discussed in Section \ref{secoth}.

As mentioned above, from now on we will work in the Feynman gauge, $\x=0$, in which case, the bosonic propagator
simply reads
\be \hat w(\pp)\d_{\m\n}:=  \hat w^{\x}_{\m\n}(\pp)\Big|_{\x=0}=
\int_{\mathbb{R}}\frac{dp_3}{2\p}\frac{\c(|p|)}{\pp^2+p_3^2}\d_{\m\n}\label{pho_fey17}\ee
The bosonic propagator $\hat w(\pp)$ is singular at $\pp=\V0$, while the fermionic
propagator is singular at the two Fermi points
$\kk=\pp_F^{\pm}$. The first step of the RG analysis consists in rewriting both the fermionic and the bosonic
propagators as sums of two propagators, one supported close to the singularity (infrared propagator) and one
in the complementary region (ultraviolet propagator), that is,
\be \hat S_0(\kk) = \hat g^{(\leq 0)}(\kk) + \hat g^{(1)}(\kk)\;,\qquad
\hat w(\pp) = \hat w^{(\leq 0)}(\pp) + \hat w^{(1)}(\pp)\;.\label{2.1}\ee
where
\be g^{(\leq 0)}(\kk) = \sum_{\o=\pm} \chi(|\kk-\pp_F^{\o}|)\hat S_0(\kk) =:
\sum_{\o=\pm}\hat g^{(\le 0)}_{\o}(\kk-\pp_F^\o)\;,
\qquad \hat w^{(\leq 0)}(\pp) = \chi(|\pp|)\int_{\mathbb{R}}\frac{dp_3}{2\p}\frac{1}{\pp^2+p_3^2}=
\frac{\c(|\pp|)}{2|\pp|}\;.\label{prople0}\ee
Note that in the first formula the support functions
$\chi(|\kk-\pp_F^{+}|)$ and $\chi(|\kk-\pp_F^{-}|)$ have disjoint supports.
Correspondingly, we rewrite the Gaussian measures as
\be
P(d\Psi) = \big[\prod_{\o=\pm} P(d\Psi_{\o}^{(\leq 0)})\big]P(d\Psi^{(1)})\,,\qquad P^\x(dA)\Big|_{\x=0} =
P(dA^{(\leq 0)})P(dA^{(1)})\;,\label{addpri}\ee
where $\hat \Psi^{(\leq 0)}_{\o}$, $\hat \Psi^{(1)}$, $\hat A^{(\leq 0)}$ and $\hat A^{(1)}$ have propagators
given by $\hat g_{\o}^{(\leq 0)}$, $\hat g^{(1)}$, $\hat w^{(\leq 0)}$ and $\hat w^{(1)}$, respectively.
The fields $\hat \Psi^{(\leq 0)}_{\o}$ are called {\it quasi-particle} fields, and the index $\o=\pm$ is
called quasi-particle or valley index.

Using this decomposition of the fermionic and bosonic fields,
the generating functional $\WW(\Phi)=\WW^{\x,-\io}(\Phi,0,0)\big|_{\x=0}$ at $J=\l=0$ can be rewritten as
\bea  e^{\WW(\Phi)}&=&\int
P(d\Psi^{(\le 0)})P(dA^{(\le 0)})\int P(d\Psi^{(1)})P(dA^{(1)})
e^{\VV(\Psi^{(\le 0)}+\Psi^{(1)},A^{(\le 0)}+A^{(1)})+\BBB(\Psi^{(\le 0)}+
\Psi^{(1)},A^{(\le 0)}+A^{(1)},\Phi)}=\nn\\
&=& e^{-\b L^{2} F_{0} + S^{(\geq0)}(\Phi)} \int
P(d\Psi^{(\le 0)})P(dA^{(\le 0)})e^{\VV^{(0)}(\Psi^{(\le 0)},A^{(\le 0)})+
\BBB^{(0)}(\Psi^{(\le 0)},A^{(\le 0)},\Phi)}\;,\label{1.16d}\eea
where the expression in the second line is obtained by an explicit integration of the UV
degrees of freedom, which is very simple:
in fact, the UV theory for the imaginary time variable in the presence of a
fixed UV cutoff on the spatial variables is trivially
convergent, see \cite[Appendix C]{GM} or \cite{tesiporta} for more technical details on this issue.
The quantities $F_{0}$ and $S^{(\geq0)}(\Phi)$ (normalized so that $S^{(\geq0)}(0)=0$)
are the contributions to the specific free energy and to the generating function of response functions,
respectively, coming from the ultraviolet integration. $\VV^{(0)}$ and $\BBB^{(0)}$
(that are both normalized in such a way that $\VV^{(0)}(0,0)=\BBB^{(0)}(0,0,\Phi)=0$)
are the effective interaction and the effective source term, whose structure will be
explicitly spelled out below.

The integration of the IR effective theory is performed by using an iterative procedure, based on the following
decomposition:
\bea && \hat g_\o^{(\le 0)}(\kk')=\sum_{h\le 0}\hat g_\o^{(h)}(\kk')\;,\qquad \hat g_\o^{(h)}(\kk')=
f_h(\kk')\hat S_0(\kk'+\pp_F^\o)\;,\nn\\
&& \hat w^{(\le 0)}(\pp)=\sum_{h\le 0}\hat w^{(h)}(\pp)\;,\qquad
\hat w^{(h)}(\pp)=\frac{f_h(\pp)}{2|\pp|}\;,\label{deco}\eea
where $f_h(\pp)=\c(2^{-h}|\pp|)-\c(2^{-h+1}|\pp|)$.
At each step we integrate the propagators $\hat g^{(h)}$ and $\hat w^{(h)}$, $h=0,-1,-2,\ldots,$
corresponding to degrees of freedom on momentum scale of order $2^{0},2^{-1},2^{-2},\ldots$,
the result of the integration defining the new effective interaction and source terms at scale $h$.
At each step, we identify the marginal and relevant terms in these effective potentials and correspondingly
define the effective coupling constants at scale $h$. Finally, after having inserted the
quadratic fermionic relevant and marginal terms in the gaussian Grassmann integration (so defining a
flowing dressed fermionic propagator), we proceed to the next integration step. After the integration of
the scales $0,-1,\ldots,-h+1$, we get (see below for an inductive proof)
\be e^{\WW(\Phi)} = e^{-\b L^{2} F_{h} + S^{(\geq
h)}(\Phi)} \int P(d\Psi^{(\leq h)})P(dA^{(\leq
h)})e^{\VV^{(h)}(\sqrt{Z_h}\Psi^{(\leq h)},A^{(\leq h)})+
\BBB^{(h)}(\sqrt{Z_h}\Psi^{(\le h)},A^{(\le h)},\Phi)}\;,\label{3.6}
\ee
where $P(d\Psi_{\o}^{(\le h)})$ and $P(d A^{(\le h)})$ have propagators
\be \hat g^{(\leq h)}_{\o}(\kk') = -
\frac{\chi_h(\kk')}{Z_h(\kk')}\begin{pmatrix} ik_0 & v_h(\kk')\O^*(\vec k'+
\vec p_F^{\,\o})\\  v_h(\kk')\O(\vec k'+
\vec p_F^{\,\o}) & ik_0 \end{pmatrix}^{\!\!\!-1}
\;,\qquad \hat w^{(\leq h)}(\pp) =\frac{\chi_{h}(\pp)}{2|\pp|}\;,\label{3.2}
\ee
with $\chi_h(\pp) = \chi(2^{-h}|\pp|)$
and $Z_h(\kk'), v_h(\kk')$
the effective wave function renormalization and Fermi velocity at scale $h$, to be inductively defined below.
Moreover, if $Z_h=Z_h(\V0)$, the effective interaction and the effective source can be written as sums
over monomials of the fields (we recall that all the external fields $\Phi^{(a)}$ with $a\neq K$ are set to
zero, for notational simplicity):
\bea&&\VV^{(h)}(\sqrt{Z_h}\Psi,A) = \sum_{\substack{n,m\geq 0:\\ n+m\ge 1}}
(Z_h)^n\int \Big[
\prod_{i=1}^{2n}\hat \Psi^{\e_i}_{\kk'_{i},\s_i,\r_i,\o_{i}}\Big] \Big[
\prod_{i=1}^{m}\hat A_{\m_i,\pp_{i}}\Big]\hat
W^{(h)}_{2n,m,0}(\ul\kk',\ul\pp)
\d_{\ul\o}(\ul\kk',\ul\pp)\;,\label{3.7}\\
&&{\cal B}^{(h)}(\sqrt{Z_h}\Psi,A,\Phi) = \!\!\! \sum_{\substack{n,m\geq 0:\\ n+m\ge 1}}\sum_{p\ge 1}(Z_h)^n
\int \Big[\prod_{i=1}^{2n}\hat \Psi^{\e_i}_{\kk'_{i},\s_i,\r_i,\o_{i}}\Big] \Big[
\prod_{i=1}^{m}\hat A_{\m_i,\pp_{i}}\Big]\Big[ \prod_{i=1}^{p}\hat
\Phi^{K}_{j_i,\qq_{i}} \Big]\hat
W^{(h)}_{2n,m,p}(\ul\kk',\ul\pp,\ul\qq)
\d_{\ul\o}(\ul\kk',\ul\pp,\ul\qq)\;,\nn
 \eea
where the integral sign is a shorthand for the sum over momenta and over the field labels,
the underlined variables indicate a collection of variables (e.g., $\ul\kk'=(\kk_1',\ldots,\kk_{2n}')$)
and $\d_{\ul\o}$ enforces momentum conservation; note that, precisely because
of momentum conservation, $\hat W^{(h)}_{2n,m,p}(\ul\kk',\ul\pp,\ul\qq)$ explicitly depends
on $2n+m+p-1$ variables rather than on $2n+m+p$ (the ``missing" momentum, which can be
eliminated using the delta, can be chosen arbitrarily among the variables $(\ul\kk',\ul\pp,\ul\qq)$).
In Eq.(\ref{3.7}) the kernels
$\hat W^{(h)}_{2n,m,p}$ also depend on the choice of the field labels $\ul\e,\ul\s,\ul\r,\ul\o,\ul\m$,
but we dropped these indices to avoid an overwhelming notation.

\subsection{Emergent relativistic theory}

In order to make the emergent relativistic structure of the theory apparent, it is
convenient to rewrite the fermionic propagator in Eq.(\ref{3.2}) as
\be \hat g^{(\leq h)}_{\o}(\kk') = \frac1{\b L^2}\media{\Psi^-_{\kk',\s,\o}\Psi^+_{\kk',\s,\o}}_h=
\frac{\chi_h(\kk')}{Z_h(\kk')} \frac{1}{ik_0 \G^{0}_{\o} +iv_h(\kk') \vec k'\cdot \vec \G_{\o}}\big( 1 +
R_{h,\o}(\kk') \big) \ee
where $\kk'=(k_0,\vec k')$, $|\bar R_{h,\o}(\kk')| \leq (\const.)|\kk'|$ and
\be \G^{0}_{\o} = -\openone\;, \qquad \G^{1}_{\o} =i\s_2\;,\qquad \G^{2}_{\o} = i\o\s_1\;,\label{2.11s} \ee
with
\be \s_{1} = \begin{pmatrix} 0 & 1 \\ 1 & 0 \end{pmatrix}\;,\qquad
\s_{2} = \begin{pmatrix} 0 & -i \\ i & 0 \end{pmatrix}\;,\qquad
\s_{3} = \begin{pmatrix} 1 & 0 \\ 0 & -1 \end{pmatrix}\;. \ee
the standard Pauli matrices. We can introduce a ``Dirac'' 4-spinors, which makes the
relation between the quasi-particle fields $\Psi^\pm_\o$ and a theory of massless Dirac fermions more
transparent:
\be \lis\psi_{\kk',\s}^{(\le h)}=\big(-\Psi^{(\le h)+}_{\kk',\s,2,-}, -\Psi^{(\le h)+}_{\kk',\s,1,-},
\Psi^{(\le h)+}_{\kk',\s,1,+}, \Psi^{(\le h)+}_{\kk',\s,2,+}\big)\;,\qquad \psi_{\kk',\s}^{(\le h)}=\bp
\Psi^{(\le h)-}_{\kk',\s,1,+}\\
\Psi^{(\le h)-}_{\kk',\s,2,+}\\
\Psi^{(\le h)-}_{\kk',\s,2,-}\\
\Psi^{(\le h)-}_{\kk',\s,1,-}\ep\;.\label{psirel}\ee
The propagator of the $\psi,\lis\psi$ fields reads:
\be \frac1{\b L^2}\media{\psi^{(\le h)}_{\kk',\s}\,\lis\psi^{(\le h)}_{\kk',\s}}_h=
\frac{\c_h(\kk')}{Z_h(\kk')}\frac1{ik_0\g_0+iv_h(\kk')\,\vec k'\cdot\vec \g}(1+R_h(\kk'))\;,
\label{psipsibar}\ee
where $|R_h(\kk')|\le(\const.)|\kk'|$ and $\g_\m$, $\m=0,1,2$, are euclidean gamma matrices:
\be \g_0=\bp 0& \openone\\-\openone&0\ep\;,\qquad
\g_1=\bp 0& i\s_2\\i\s_2&0\ep\;,\qquad
\g_2=\bp 0& i\s_1\\i\s_1&0\ep\;,\label{gammamat}\ee
satisfying the anticommutation relations: $\{\g_\m,\g_\n\}=-2\d_{\m\n}$. For what follows,
it is also useful to define
$\g_3=\bp 0 & -i\s_3\\ -i\s_3 &0\ep$ and the corresponding fifth gamma matrix:
\be \g_5=\g_0\g_1\g_2\g_3=\bp\openone & 0 \\0&-\openone\ep\;,\label{gamma5}\ee
which anticommutes with all the other gamma matrices: $\{\g_\m,\g_5\}=0$, $\forall\m=0,1,2,3$.

Modulo the correction term $R_h$, the propagator in Eq.(\ref{psipsibar}) is the same as the one for
euclidean massless Dirac fermions in $2+1$ dimensions.
The analysis is therefore very similar to the one performed in
\cite{GMP1} for a system of interacting Dirac fermions coupled to
with a massless gauge field. In particular, the scaling dimension of the kernels $\hat W^{(h)}_{2n,m,p}$ of
the effective potential is the same, see \cite{GMP1}:
\be D = 3 - 2n - m - p\;,\label{dim} \ee
where we use the convention that positive scaling dimensions correspond to relevant operators and viceversa.

\subsection{Lattice symmetries}\label{sec4B}

An important difference between the present case and the one studied in \cite{GMP1} is that
here the propagator is not exactly equal to the Dirac one:
on the contrary, it differs from it by the correction term proportional to $R_h(\kk')$. Of course,
this correction term is dimensionally negligible: therefore, it does not change the power counting.
However, it violates some continuous relativistic symmetries used in \cite{GMP1} to exclude the
presence of several relevant and marginal terms in the RG flow. One may fear that the lack of
such symmetries might be responsible for the generation of new marginal or relevant terms,
which are absent in the relativistic Dirac model. These potentially dangerous terms can be controlled through a
careful analysis of the honeycomb lattice symmetries. In particular, it is proved in Appendix \ref{app1} that
the effective interaction, the effective source and the gaussian integrations at all scales $h\le 0$
are separately invariant under the following symmetry transformations. Again, we spell out the symmetries
only in the presence of the external field $\Phi^K$, the effect of the other external fields being discussed in
Appendix \ref{app1}. In the following formulas, we drop the scale label $h$ for notational simplicity;
moreover, we think of $\Psi^-_{\kk',\s,\o}$ ($\Psi^+_{\kk',\s,\o}$)
as being the the column (row) vector of components $\Psi^-_{\kk',\s,\r,\o}$ ($\Psi^+_{\kk',\s,\r,\o}$),
$\r=1,2$, and we think of $\hat A_{\pp}$ as being the column vector of components $\hat A_{\m,\pp}$, $\m=0,1,2$.
\begin{enumerate}
\item[(1)] \underline{Spin flip}:
$\hat\Psi^{\varepsilon}_{\kk',\s,\o}\to
\hat\Psi^{\varepsilon}_{\kk',-\s,\o}$, and $A_{\m,\pp},\hat\Phi^{K}_{j,\pp}$ are left invariant;
\item[(2)] \underline{Global $U(1)$}:
$\hat\Psi^{\varepsilon}_{\kk',\s,\o}\rightarrow e^{ i \varepsilon
\a_{\s}}\hat\Psi^{\varepsilon}_{\kk',\s,\o}$, with $\a_{\s}\in \RRR$
independent of $\kk'$, and $A_{\m,\pp},\hat\Phi^{K}_{j,\pp}$ are left invariant;
\item[(3)] \underline{Spin SO(2)}: $\begin{pmatrix}
\hat\Psi^{\e}_{\kk',\uparrow,\r,\o}
\\ \hat\Psi^{\e}_{\kk',\downarrow,\r,\o} \end{pmatrix} \rightarrow e^{i\th\s_{2}}\begin{pmatrix}
\hat\Psi^{\e}_{\kk',\uparrow,\r,\o} \\
\hat\Psi^{\e}_{\kk',\downarrow,\r,\o} \end{pmatrix}$, with $\th$
independent of $\kk'$, and $A_{\m,\pp},\hat\Phi^{K}_{j,\pp}$ are left invariant;
\item[(4)] \underline{Discrete spatial rotations}: if $T\kk = (k_0,e^{-i\frac{2\pi}{3}\s_2}\vec k)$
and $n_-=(1-\s_3)/2$,
\bea &&
\hat \Psi^{-}_{\kk',\s,\o}\to e^{i(\pp_F^\o+\kk')(\dd_{3} - \dd_{1})n_-}\hat\Psi^{-}_{T\kk',\s,\o}\;,\qquad
\hat\Psi^{+}_{\kk',\s,\o}\to
\hat\Psi^{+}_{T\kk',\s,\o} e^{-i(\pp_F^\o+\kk')(\dd_3 - \dd_1)n_-}\;,\nn\\
&&\hat A_{\pp}\to T^{-1}\,\hat A_{T\pp}\;,
\hskip4.truecm \hat\Phi^{K}_{j,\pp}\to \hat\Phi^{K}_{j+1,T\pp}\nn\eea
\item[(5)] \underline{Complex conjugation}: if $c$ is a generic constant appearing in the effective potentials
or in the gaussian integrations:
\be c\to c^*\;,\qquad \hat\Psi^{\varepsilon}_{\kk',\s,\o}\rightarrow
\hat\Psi^{\varepsilon}_{-\kk',\s,-\o}\;,\qquad \hat A_{\pp}\rightarrow -\hat
A_{-\pp}\;,\qquad \hat\Phi^{K}_{j,\pp}\to \hat\Phi^{K}_{j,-\pp}\;;\nn\ee
\item[(6.a)] \underline{Horizontal reflections}: if $R_{h}\kk = (k_0,-k_1,k_2)$ and
$r_h 1 = 1$, $r_h 2 = 3$, $r_h 3 = 2$,
\be\hat\Psi^{-}_{\kk',\s,\o}\to \s_{1}\hat\Psi^{-}_{R_h\kk',\s,\o}\;,\quad
\hat\Psi^{+}_{\kk',\s,\o}\to \hat \Psi^{+}_{R_h\kk',\s,\o}\s_1\;,\quad \hat A_{\pp}\to R_{h}\hat A_{R_{h}\pp}
e^{i\pp\dd_{1}}\;,\quad \hat\Phi^{K}_{j,\pp}\to \hat\Phi^{K}_{r_h j,R_h\pp}e^{-i\pp(\dd_j - \dd_1)}\;;\nn\ee
\item[(6.b)] \underline{Vertical reflections}: if $R_{v}\kk = (k_0,k_1,-k_2)$
and $r_{v} 1 = 1$, $r_v 2 = 3$, $r_v 3 =2$,
\be
\hat\Psi^{\e}_{\kk',\s,\o}\to\hat\Psi^{\e}_{R_v\kk',\s,-\o}\;,\qquad \hat
A_{\pp}\to R_{v}\hat A_{R_{v}\pp}\;,\qquad \hat\Phi^{K}_{j,\pp}\to \hat\Phi^{K}_{r_v j, R_v\pp}\;;\nn
\ee
\item[(7)] \underline{Particle-hole}: if $P\kk = (k_0, -\vec k)$,
\be\hat\Psi^{\varepsilon}_{\kk',\s,\o}\rightarrow i\hat\Psi^{-\varepsilon}_{P\kk',\s,-\o}\;,\qquad \hat A_{\pp}\to P\hat
A_{-P\pp}\;,\qquad \hat\Phi^{K}_{j,\pp}\to \hat\Phi^{K}_{j,-P\pp}\;;\nn\ee
\item[(8)] \underline{Time-reversal}: if $I\kk = (-k_0,\vec k)$,
\be\hat\Psi^{-}_{\kk',\s,\o}\to
-i\s_{3}\hat\Psi^{-}_{I\kk',\s,\o}\;,\qquad \hat\Psi^{+}_{\kk',\s,\o}\to
-i\hat\Psi^{+}_{I\kk',\s,\o}\s_3\;,\qquad  \hat A_{\pp}\to I\hat A_{I\pp}\;,\qquad
\hat \Phi^{K}_{j,\pp}\to \hat\Phi^{K}_{j,I\pp}\;.\nn\ee
\end{enumerate}
In the following subsection, the implications of these symmetries on the structure of the marginal
and relevant terms are discussed.

\subsection{Localization and the symmetry properties of the local terms}\label{sec4C}

In order to inductively prove Eq.(\ref{3.6}), we write
$\VV^{(h)} =\LL \VV^{(h)}+\RR \VV^{(h)}$ and ${\cal
B}^{(h)} =\LL {\cal B}^{(h)}+\RR {\cal B}^{(h)}$, where the $\LL$ operator
isolates the {\it local terms}, while $\RR$ isolates the {\it irrelevant terms}; according to Eq.\pref{dim}, we define
\be
\LL\hat W^{(h)}_{2n,m,p;\ul\o,\ul\m,\ul{j}}(\ul\kk',\ul\pp,\ul\qq) = \left\{\begin{array}{ll} \hat W^{(h)}_{2n,m,p;\ul\o,\ul\m,\ul{j}}
(\underline{\bf 0},\underline{\bf 0},\underline{\bf 0})\;,
& \quad\mbox{if $2n+m+p=3$}\;, \\
\big[1+(\underline{\bf k}',\ul\o)\cdot\partial_{(\underline{\bf k}',\ul\o)}+
\underline{\bf p}\cdot\partial_{\underline{\bf p}}+
\underline{\bf q}\cdot\partial_{\underline{\bf q}})
\hat W^{(h)}_{2n,m,p;\ul\o,\ul\m,\ul{j}}(
\underline{\bf 0},\underline{\bf 0},\underline{\bf 0})\;,
& \quad \mbox{if $2n+m+p=2$}\;, \\ 0\;, & \quad \mbox{otherwise.} \end{array}\right.\label{3.22b}
\ee
In the second line, $(\underline{\bf k}',\ul\o)\cdot\partial_{(\underline{\bf k}',\ul\o)}=
\sum_{i=1}^{2n}({\bf k}'_i,\o_i)\cdot\partial_{({\bf k}'_i,\o_i)}$, where
$({\bf k}',\o)\cdot\partial_{({\bf k}',\o)}$ is a shorthand for
\bea && ({\bf k}',\o)\cdot\partial_{({\bf k}',\o)}=k_0\dpr_{k_0}+
\O(\vec p_F^\o+\vec k')\,\dpr_{\vec k',\o}+\O^*(\vec p_F^\o+\vec k')\,\dpr^*_{\vec k',\o}\;,
\label{dpr4.18a}\\
&&{\rm with}\qquad \dpr_{\vec k',\o}=\frac12(-i\dpr_{k_1'}+\o\dpr_{k_2'})\;,\quad
\dpr^*_{\vec k',\o}=\frac12(i\dpr_{k_1'}+\o\dpr_{k_2'})\;.\nn\eea
In Appendix \ref{app3} it is proved that, thanks to the symmetry properties (1)--(8)
listed in the previous subsection, the only non-vanishing local terms with $2n+m+p=3$ are
either those with $(2n,m,p)=(0,2,1)$ (i.e., terms of the form $\Phi^KAA$)
or the {\it vertices}, with $2n=2$ and $m+p=1$ (i.e., terms of the form
$A\Psi^+\Psi^-$ or $\Phi^K\Psi^+\Psi^-$). The latter have the following explicit structure:
\bea && \LL\hat W^{(h)}_{2,1,0;(\o,\o),\m}(\kk',\pp) = i\l_{\m,h}\G^\m_\o\;,
\qquad \LL\hat W^{(h),K}_{2,0,1;(\o,\pm\o),j}(\kk',\pp) = \frac{Z^{\pm}_{K,h}}{Z_h}\G^\pm_{\o,j}\;,
\label{symbad}\\
&&{\rm where}\qquad \G^\pm_{\o,j}:=
\Biggl(\begin{matrix} 0 & e^{\pm i\o\frac{2\p}{3}(j-1)}\\
e^{-i\o\frac{2\p}{3}(j-1)}&0\end{matrix}\Biggr)\nn\eea
and the apex $K$ added to the kernels with $p\neq 0$ is meant to remind the reader that
the external field $\Phi$ is of type $K$. The constants $\l_{\m,h}$, $Z^\pm_{K,h}$, $Z_h$
are real and $\l_{1,h}=\l_{2,h}$. Note that the kernel with $2n=2$ and $m=1$ with different omegas is zero simply because
the photon field $A$, as well as the external field $J$, has an UV cutoff that makes
the modes corresponding to momenta $\pp$ close to
$\pm(\pp_F^+-\pp_F^-)$ and to their images over $\L^*$ vanishing, see
Remark after item 2 in Section \ref{sec2.2}.

Moreover,  the only non-vanishing local terms with $2n+m+p=2$ are either those with
$(2n,m,p)=(2,0,0)$ (i.e., terms of the form $\Psi^+\Psi^-$) or those with $(2n,m,p)=(0,2,0)$
(i.e., terms of the form $AA$). They have the following explicit structure (see Appendix
\ref{app3} for a proof):
\be  \LL\hat W^{(h)}_{2,0,0;(\o,\o)}(\kk') =\begin{pmatrix} iz_{0,h}&z_{1,h}\O^*(\vec p_F^\o+\vec k')\\
z_{1,h}\O(\vec p_F^\o+\vec k')&iz_{0,h}\end{pmatrix}\;,
\qquad \LL\hat W^{(h)}_{0,2,0;(\m,\n)}(\pp) =\n_{\m,h}\d_{\m\n}\;,\label{local2}\ee
where
$z_{\m,h}$ and $\n_{\m,h}$ are real and $\n_{1,h} = \n_{2,h}$.

\subsection{The single-scale RG step: the inductive integration procedure}

The splitting into local and irrelevant terms is used in the inductive integration of the
generating functional in the following way: we rewrite the integral in the r.h.s. of Eq.\pref{3.6} as
\bea &&\int P(d\Psi^{(\leq h)})P(dA^{(\leq h)})
e^{\VV^{(h)}(\sqrt{Z_h}\Psi^{(\leq h)},A^{(\leq h)}) +
{\cal B}^{(h)}(\sqrt{Z_h}\Psi^{(\leq h)},A^{(\leq h)},\Phi)}=\nn
\\
&&=e^{\b L^2 t_h}\int \widetilde P(d\Psi^{(\leq h)})P(dA^{(\leq
h)})e^{\widetilde \VV^{(h)}(\sqrt{Z_h}\Psi^{(\leq h)},A^{(\leq h)})+
{\cal B}^{(h)}(\sqrt{Z_h}\Psi^{(\leq h)},A^{(\leq h)},\Phi)}\;,\label{4.25}\eea
where $\widetilde \VV^{(h)}=\VV^{(h)}-\LL_\psi\VV^{(h)}$, with
$\LL_\psi\VV^{(h)}$ the contribution to $\LL\VV^{(h)}$ that is
quadratic in the fermionic fields (i.e., the one corresponding to the first term in Eq.(\ref{local2}))
and $t_h$ is a normalization constant. Moreover, $\widetilde
P(d\Psi^{(\leq h)})$ has a propagator given by the same expression as Eq.\pref{3.2} but for the fact
that $Z_h(\kk')$ and $v_h(\kk')$ are replaced by $Z_{h-1}(\kk')$ and
$v_{h-1}(\kk')$, respectively, where
\be Z_{h-1}(\kk')=Z_{h}(\kk')+Z_h z_{0,h}\c_h(\kk')\;,\quad Z_{h-1}(\kk') v_{h-1}(\kk') = Z_{h}(\kk')
v_{h}(\kk') + Z_h z_{1,h} \chi_{h}(\kk')\;,\label{3.16a}
\ee
and $Z_h= Z_h(\V0)$, $v_h= v_h(\V0)$. After this,
defining $Z_{h-1} =  Z_{h-1}(\V0)$, we {\it rescale} the
fermionic field by setting
\bea \widetilde \VV^{(h)}(\sqrt{Z_{h}}\Psi^{(\leq h)}, A^{(\leq
h)}) &=:& \hat \VV^{(h)}(\sqrt{Z_{h-1}}\Psi^{(\leq h)}, A^{(\leq
h)})\nn\\
\BBB^{(h)}(\sqrt{Z_{h}}\Psi^{(\leq h)}, A^{(\leq
h)},\Phi) &=:& \hat {\cal B}^{(h)}(\sqrt{Z_{h-1}}\Psi^{(\leq h)},
A^{(\leq h)},\Phi) \;.\label{3.17} \eea
By using Eqs.(\ref{symbad})-(\ref{local2}), the local part of the rescaled effective
potential can be written as
\be\LL \hat \VV^{(h)}(\sqrt{Z_{h-1}}\Psi^{(\leq h)}, A^{(\leq h)})
= \frac1{\b|\SS_L|}\sum_{\m,\pp}\,\big[Z_{h-1}e_{\m,h}\hat
\jmath^{(\leq h)}_{\m,\pp}\hat A^{(\leq h)}_{\m,\pp}
 - 2^{h}\n_{\m,h}\hat A^{(\leq h)}_{\m,-\pp}\hat A^{(\leq h)}_{\m,\pp}\big]  \label{3.14b}
\ee
where $e_{\m,h}:=
\frac{Z_{h}}{Z_{h-1}}\l_{\m,h}$ and, if $v_{h-1}:= v_{h-1}(\V0)$,
\be \hat\jmath^{(\leq h)}_{0,\pp} := \frac{i}{\b L^2}\sum_{\o,\s,\kk'}\,
\hat\Psi^{(\leq
h)+}_{\kk'+\pp,\s,\o}\G^{0}_{\o}\hat\Psi^{(\leq
h)-}_{\kk',\s,\o}\;,\quad \vec{\jmath}^{\,(\leq h)}_{\pp} :=
\frac{iv_{h-1}}{\b L^2}\sum_{\o,\s,\kk'}\, \hat\Psi^{(\leq
h)+}_{\kk'+\pp,\s,\o}\vec \G_{\o}\hat\Psi^{(\leq
h)-}_{\kk',\s,\o}\;.\label{3.15aa} \ee
Note that, by
using the notation defined in Eqs.(\ref{psirel}),(\ref{gammamat}),
the density and the current in Eq.(\ref{3.15aa}) can be rewritten in the familiar relativistic form as
\be \hat\jmath^{(\leq h)}_{0,\pp} := \frac{i}{\b L^2}\sum_{\s,\kk'}\,
\lis\psi^{(\leq
h)}_{\kk'+\pp,\s}\g_{0}\psi^{(\leq
h)}_{\kk',\s}\;,\quad \vec{\jmath}^{\,(\leq h)}_{\pp} :=
\frac{iv_{h-1}}{\b L^2}\sum_{\s,\kk'}\, \lis\psi^{(\leq
h)}_{\kk'+\pp,\s}\vec \g\psi^{(\leq
h)}_{\kk',\s}\;.\label{3.23} \ee
Finally, by using Eq.(\ref{symbad}) and the properties stated right before this equation,
we find that the local part of the effective source term is given by
\bea &&\LL \hat\BBB^{(h)}(\sqrt{Z_{h-1}}\Psi^{(\leq h)},
A^{(\leq h)},\Phi) = \frac1{\b^2|\SS_L|^2}\sum_{\substack{\pp,\qq\\ j,\m_1,\m_2}}
\l^{K}_{j,(\m_1,\m_2),h}\hat \Phi^{K}_{j,\qq}\hat A^{(\leq h)}_{\m_{1},\pp}\hat A^{(\leq
h)}_{\m_2,-\pp-\qq} + \nn\\&& + \frac1{\b^2L^2|\SS_L|}\sum_{\substack{\kk',\pp\\
\o,\s,j}}\Big[
Z^{+}_{K,h}\hat \Phi^K_{j,\pp}\hat \Psi^{(\leq h)+}_{\kk'+\pp,\s,\o}\G^+_{\o,j}
\hat \Psi^{(\leq h)-}_{\kk',\s,\o} + Z^{-}_{K,h}\hat \Phi^K_{j,\pp_F^{\o}-\pp_F^{-\o}+\pp}
\hat \Psi^{(\leq h)+}_{\kk'+\pp,\s,\o}\G^-_{\o,j}
\hat \Psi^{(\leq h)-}_{\kk',\s,-\o} \Big]\;.\label{effsou} \eea
Note that, in contrast to what happens in a relativistic QFT, the effective source term
Eq.(\ref{effsou}) contains marginal terms that
were not present in the original functional integral, i.e., the terms $\Phi^KAA$ in the first line
of Eq.(\ref{effsou}). These potentially dangerous
terms can be shown to be harmless by using the lattice symmetries (1) -- (8), see Section
\ref{secflow} for a discussion of this point (nevertheless, let us anticipate that the reason why these
terms do not create troubles is that they are ``almost zero", precisely because they vanish
in the relativistic approximation; therefore, their naive dimensional bound can be improved
and they can be shown to be effectively irrelevant).
%

After the rescaling Eq.(\ref{3.17}), we rewrite the r.h.s. of Eq.\pref{4.25} as
\be e^{-\b L^2 t_h}\int P(d\Psi^{(\leq h-1)})P(dA^{(\leq h-1)})\int
P(d\Psi^{(h)})P(dA^{(h)} )e^{\hat\VV^{(h)}(\sqrt{Z_{h-1}}\Psi^{(\leq h)},A^{(\leq h)})+
\hat\BBB^{(h)}(\sqrt{Z_{h-1}}\Psi^{(\leq h)},A^{(\leq h)},\Phi)}\;,\label{4.25a}
\ee
where $P(d\Psi^{(\leq h-1)})$, $P(dA^{(\leq h-1)})$ have propagators given
by (\ref{3.2}) with $h$ replaced by $h-1$, while $P(dA^{(h)})$, $P(d\Psi^{(h)})$ have propagators
\be
\hat w^{(h)}(\pp) := \frac{f_h(\pp)}{2|\pp|}\;,\qquad \frac{\hat g^{(h)}_{\o}(\kk')}{Z_{h-1}} := \frac{\tilde f_h(\kk')}{Z_{h-1}}\frac{1}{ik_0 \G^{0}_{\o} + iv_{h-1}(\kk')\vec k'\cdot \vec\G_{\o}}(1 + R'_{h,\o}(\kk'))
\;,\label{prop4.29}\ee
where $\tilde f_{h}(\kk') := Z_{h-1}f_{h}(\kk')/Z_{h-1}(\kk')$ and $|R'_{h,\o}(\kk')|\le (\const.)|\kk'|$. \\

{\bf Remark.} The single scale propagator can be decomposed as a sum of a Dirac-like
propagator $\hat g^{(h)}_{D,h}(\kk')$, which is the propagator obtained by setting $R'_{h,\o}=0$ in
the second definition in
Eq.(\ref{prop4.29}), plus a rest, which has a better infrared behavior. We shall correspondingly
write $\hat g^{(h)}_{\o}(\kk') = \hat g^{(h)}_{D,\o}(\kk') + r^{(h)}_{\o}(\kk')$. This decomposition
will be useful in the following, as already anticipated by the comment after Eq.(\ref{effsou}).\\

At this point, we can finally integrate the fields on scale $h$ and, defining
\bea && e^{-\b L^{2} F_{h-1} + S^{(\geq
h-1)}(\Phi)} e^{\VV^{(h-1)}(\sqrt{Z_{h-1}}\Psi^{(\leq h-1)},A^{(\leq h-1)})+
\BBB^{(h-1)}(\sqrt{Z_{h-1}}\Psi^{(\le h-1)},A^{(\le h-1)},\Phi)}
:=\label{RGstep}\\
&&\qquad =e^{-\b L^2(F_h+ t_h)+ S^{(\geq
h)}(\Phi)}\int
P(d\Psi^{(h)})P(dA^{(h)} )e^{\hat\VV^{(h)}(\sqrt{Z_{h-1}}\Psi^{(\leq h)},A^{(\leq h)})+
\hat\BBB^{(h)}(\sqrt{Z_{h-1}}\Psi^{(\leq h)},A^{(\leq h)},\Phi)}\;,\nn\eea
our inductive assumption Eq.(\ref{3.6}) is reproduced at scale $h-1$. Note that Eq.(\ref{RGstep})
can be thought as a recursive definition for the effective potential.
The integration in Eq.(\ref{RGstep}) is performed by expanding in series
the exponential in the r.h.s. and by
integrating term by term with respect to the gaussian integration
$P(d\psi^{(h)}) P(dA^{(h)})$. This procedure gives rise to an
expansion for the effective interaction and source terms in terms of the
renormalized parameters
$\{e_{\m,k},\n_{\m,k},Z_{k-1},v_{k-1},Z^\pm_{K,k},\l^K_{j,\ul\m,k}\}_{h< k\le 0}$, which can be
conveniently represented as a sum over Gallavotti-Nicol\`o (GN) trees \cite{G84}; the value of each GN tree can be thought of as a sum over connected labelled Feynman diagrams, i.e., every GN
tree represents a set of Feynman diagrams characterized by the same hierarchical structure
of the scale labels associated to the propagators, see \cite[Section 2.2]{GMP1} for a thorough
discussion of this expansion.

We will call $\{e_{\m,k},\n_{\m,k}\}_{k\le 0}$ the {\it effective couplings}
or {\it running coupling constants}. The constants $e_{\m,h}$ play the role
of {\it effective charges}, while $\n_{\m,h}$ play the role of {\it effective photon masses}.
It will be shown below that the effective charges stay constant under the RG flow (more precisely,
$e_{\m,h}$ are essentially independent of $\m$ and $h$) and that the effective photon mass
is vanishing (more precisely, $\n_{\m,h}$ is small uniformly in $h$, on the ``right scale"). \\

{\bf Remark.} Note the unusual dependence of the effective charges on the index $\m$:
the global symmetries (1)--(8) discussed above ensure that $e_{1,h}=e_{2,h}$ but they do not
a priori guarantee that $e_{0,h}=e_{1,h}$. This situation is in striking contrast with
what happens in QFT, where Lorentz invariance guarantees such a
property to be valid at all scales. However, in the next section we will show
that, thanks to lattice WIs, $e_{0,h}$ and $e_{1,h}$, even if not exactly equal to each other
at all scales, admit the same limit as $h\to-\io$; namely, $e_{0,-\io}=e_{1,-\io}=e+O(e^2)$.\\

The expansion in GN trees and labelled Feynman diagrams allows us to obtain the following
inductive estimate on the kernels $\hat W_{2n,m,p}$ in Eq.(\ref{3.7}).
Let $\bar\e_h =
\max_{h< k\leq 0} \{|e_{\m,k}|,|\n_{\m,k}|\}$ be small enough. If
$Z_{k}/Z_{k-1} \le e^{C\bar \e_h^2}$ and ${C}^{-1}\le v_{k-1}\le 1$,
for all $h< k\le 0$ and a suitable constant $C>0$, then the $N$-th order contribution
to $\hat W_{2n,m,p}$ in the effective couplings (to be denoted by $\hat W^{N;(h)}_{2n,m,p}$)
admits the following bound (the ``$N!$ bound"):
\be ||W^{N;(h)}_{2n,m,p}|| \le (\const.)^N\bar\e_h^N
\Big(\frac{N}2\Big)!\;,\label{4.11a}\ee
where $||W^{N;(h)}_{2n,m,p}||:=\; 2^{-h(3-2n-m-p)}\sup
|W^{N;(h)}_{2n,m,p}(\ul\kk',\ul\pp,\ul\qq)|$ and the sup is performed with respect to the
momenta and the field labels. The proof of Eq.(\ref{4.11a}) can be found in
\cite[Section 2.4]{GMP1}. The basic ingredient in the proof is a dimensional estimate
of the kernels, which follows from the bounds:
\bea && |\hat w^{(h)}(\pp)|\leq (\const.)2^{-h}\;,\qquad
\frac1{\b|\SS_L|}\sum_\pp |\hat w^{(h)}(\pp)|\le (\const.) 2^{2h}\;,\nn\\
&& \|\hat g^{(h)}_{\o}(\kk')\| \leq (\const.)2^{-h}\;,\qquad
\frac1{\b L^2}\sum_{\kk'}\|\hat g^{(h)}_{\o}(\kk')\| \leq (\const.)2^{2h}\;,\label{hgt}\eea
that is, every propagator on scale $h$ is associated to a factor $2^{-h}$ and every loop integral on
scale $h$ is associated to a factor $2^{3h}$. The estimate Eq.(\ref{4.11a}) follows from:
(i) counting the number of propagators and loop
integrals on each scale $h$ for every given labelled Feynman diagram; (ii) realizing that
the corresponding dimensional estimate is uniform in the Feynman diagram, for all the
diagrams associated to the same GN tree; (iii)
performing the sum over scale labels for every fixed GN tree; (iv) counting the number
of Feynman diagrams associated to each GN tree and the total number of GN trees contributing to
order $N$ in renormalized perturbation theory. See
\cite[Proof of Theorem 2.1]{GMP1} for more details.

The bound Eq.(\ref{4.11a})
tells us that the $N$-th order contribution to the effective potential is {\it finite in norm}, uniformly in
$h$. {\it If the effective couplings remain small in the infrared}, informations obtained from
our renormalized expansion by lowest order truncations are reliable at weak
coupling. The importance of having an expansion with finite
coefficients should not be underestimated. The naive perturbative
expansion in $e$ the fine structure constant is plagued by {\it
logarithmic infrared divergences} and higher orders are more and
more divergent. More precisely, one can find classes of diagrams
of order $N$ in $e$ contributing to the effective potential on scale $h$
whose size grows like $O(|h|^n)$. Therefore, finite order truncations of the naive
perturbation theory do not give a priori any reliable information on the IR behavior
of the theory, not even at weak coupling.

Regarding the combinatorial factor in the r.h.s. of Eq.(\ref{4.11a}), we note that the $(N/2)!$
dependence is compatible with Borel summability of the theory. However,
summability does not follow from our bounds. Constructive estimates on large fields
for the bosonic sector (in the spirit of, e.g., \cite{Bry}) combined with determinant estimates for the
fermionic sector (in the spirit of, e.g., \cite{GM}) may allow a full non-perturbative construction of
the theory. However, this goes beyond the scope of this paper.

Let us conclude this section by adding a comment, which will be useful for the study of the
flow of the effective parameters discussed in the next sections. Consider the splitting
$\hat g^{(h)}_{\o}(\kk') = \hat g^{(h)}_{D,\o}(\kk') + r^{(h)}_{\o}(\kk')$ mentioned in the remark
after Eq.(\ref{prop4.29}). As mentioned there, the rest $r^{(h)}_\o$ is better behaved in the infrared
than the relativistic propagator $\hat g^{(h)}_{D,\o}$. More precisely,
\be \| r^{(h)}_{\o}(\kk') \| \leq \const.\;,\qquad \frac1{\b L^2}\sum_{\kk'}\| r^{(h)}_{\o}(\kk') \|\le (\const.) 2^{3h}\;, \label{hg}
\ee
which should be compared with the second line of Eq.(\ref{hgt}). It is apparent that $r^{(h)}_\o$
is associated to a dimensional gain $2^h$ as compared to the leading term
$\hat g^{(h)}_{D,\o}(\kk')$. This implies that if we decompose $W^{N;(h)}_{2n,m,p}$ as
\be \hat W^{N;(h)}_{2n,m,p}=\hat W^{N;(h),D}_{2n,m,p}+
\widetilde W^{N;(h)}_{2n,m,p}\label{dec}\ee
where $W^{N;(h),D}_{2n,m,p}$ is obtained from
$W^{N;(h)}_{2n,m,p}$ by replacing all the propagators
$\hat g^{(h)}_{\o}(\kk')$ by $\hat g^{(h)}_{D,\o}(\kk')$ (and by neglecting the contributions
coming from the UV propagators on scale $h=1$), then $\widetilde W^{N;(h)}_{2n,m,p}$
is dimensionally negligible in the IR as compared to the ``relativistic" contribution
$\hat W^{N;(h),D}_{2n,m,p}$; i.e., $\widetilde W^{N;(h)}_{2n,m,p}$ admits a bound similar to
Eq.(\ref{4.11a}), with an extra factor (dimensional gain) proportional to $2^{\th h}$, with
$0<\th<1$. This follows from the improved dimensional estimate on the propagator $r^{(h)}_\o$,
Eq.(\ref{hg}), and from the fact that ``long GN trees are exponentially depressed", i.e.,
the property referred to as ``short memory property", see \cite[Section 2.4]{GMP1}.

\section{Ward Identities and the flow of the renormalized parameters}\label{secWI}
\setcounter{equation}{0}
\renewcommand{\theequation}{\ref{secWI}.\arabic{equation}}

We have seen that the effective potentials (and, similarly, the correlation functions)
can be written as series in the effective charges $e_{\m,h}$ and the
effective masses $\n_{\m,h}$ with bounded coefficients at all orders,
uniformly in the infrared cut-off, provided that the ratios $Z_h/Z_{h-1}$ remain close to 1
and that effective Fermi velocity remains bounded away from zero along the RG flow.
Of course, such expansions are useful only if the running coupling constants remain small for all
values of $h$, a fact that we are going to prove to be true, thanks to
exact lattice WIs. In this section we first study the flow of the running coupling constants
$\{e_{\m,h},\n_{\m,h}\}_{h\le 0}$ and next the one of the other renormalized parameters.
\subsection{The flow of the electric charge and of the photon mass}\label{secWIpart1}
The key idea is to get informations on the running coupling constants
$\{e_{\m,h},\n_{\m,h}\}_{h\le 0}$ by using the WIs Eq.(\ref{WI87}) for the sequence of
reference models $\WW^{0,h^*}(\Phi,J,\l)$, where the scale of the bosonic IR
cutoff $h^*$ is thought of as a parameter. For each choice of the IR cutoff $h^*$, the
generating functional $\WW^{0,h^*}(\Phi,J,\l)$ is computed by a multiscale integration
procedure similar to the one described in the previous section, with the important difference
that after the integration of the scale $h^*$ we are left with a
purely fermionic theory, which is {\em super-renormalizable}: in
fact, setting $m=0$ in the formula Eq.(\ref{dim}) for the scaling dimension
of the kernels of the effective potentials, we see that the
scaling dimension of the reference model {\it below} the cutoff $h^*$ is
$D=3-2n-p$. In particular, the kernels of the effective interaction (i.e.,
those with $p=0$) are always negative once the two-legged subdiagrams have been
renormalized; as shown in \cite{GM,GMP3,GMP3long},
the effect of the integration of the scales $\le h^*$ is
just  to renormalize by a small finite amount the effective parameters $Z_{h^*},v_{h^*},
Z^\pm_{K,h^*}, \l^K_{j,\ul\m,h^*}$. Let us denote by
$\{e^{[h^*]}_{\m,h},\n^{[h^*]}_{\m,h}
\}_{h\le 0}$
the running coupling constants of the reference model with infrared cut-off on scale $h^*$.
Of course, if $h\ge h^*$,
\be \{e^{[h^*]}_{\m,h},\n^{[h^*]}_{\m,h}
\}_{h^*\le h\le 0}=
\{e_{\m,h},\n_{\m,h}
\}_{h^*\le h\le 0}\;,\ee
where the constants in the r.h.s. are those of the
model without IR cutoff, i.e., $e_{\m,h}:=e^{[-\infty]}_{\m,h}$ and $\n_{\m,h}:=\n^{[-\infty]}_{\m,h}$.
On the other hand, as proved in \cite[Appendix B]{GMP1},
the two- and three-points correlation functions of the reference model are proportional to
the inverse wave function renormalization and to the effective charges, i.e.,
if $\kk'$ and $\pp$ are such that $|\kk'| = 2^{h^*}$ and $|\pp|\le
2^{h^*}$, then
\bea
&&\hat \SS^{0,h^*}_{2,0}(\pp_F^\o+\kk') = \frac{\hat g^{(h^*)}_{\o}(\kk')}{Z_{h^*-1}}
\big(1 +\bar B_{\o,h^*}(\kk')\big)\;,\label{WI4}\\
&&\hat\SS^{0,h^*}_{2,1;0}(\pp_F^\o+\kk',\pp) = i\frac{\hat g^{(h^*)}_{\o}(\kk'+\pp)}{Z_{h^*-1}}\Big(
e_{0,h^*}\G^{0}_{\o} + e{B}^0_{\o,h^*}(\kk',\pp)\Big)\hat g^{(h^*)}_{\o}(\kk')\;,\label{WI13bis} \\
&&\hat\SS^{0,h^*}_{2,1;l}(\pp_F^\o+\kk',\pp) = iv_{h^*-1}\frac{\hat g^{(h^*)}_{\o}(\kk'+\pp)}{Z_{h^*-1}}\Big(
e_{1,h^*}\, \G^l_{\o} +e {B}^l_{\o,h^*}(\kk',\pp)\Big)
\hat g^{(h^*)}_{\o}(\kk')\;,\qquad l\in\{1,2\}\;,\label{WI13}\eea
where the correction terms $\bar B_{\o,h^*}(\kk')$ and ${B}^\m_{\o,h^*}(\kk',\pp)$
are of order $\bar\e_{h^*}^{2}$ (recall that $\bar\e_h=\max\{|e_{\m,k}|,|\n_{\m,k}|\}_{h\le k\le 0}$),
uniformly in $h^*$, for all $|\kk'|=2^{h^*}$ and $|\pp|\le 2^{h^*}$. For later use, let us note that
$\bar B_{\o,h^*}(\kk')$ is differentiable in $\kk'$ and its derivatives computed at $|\kk'|=2^{h^*}$
are dimensionally bounded by $(\const.)2^{-h^*}\bar\e_{h^*}^2$.

Thanks to Eqs.(\ref{WI4})--(\ref{WI13}), we see that informations on the mutual relations between
the correlation functions of the reference model with cutoff $h^*$ (which are provided by the WIs)
imply relations between the effective charges on scale $h$. Regarding the
photon mass, an equation similar in spirit to Eqs.(\ref{WI4})-(\ref{WI13}) is valid, namely,
if $|\pp|\le 2^{h^*}$,
\be \hat\SS^{0,h^*}_{0,2;(\m,\n)}(\pp) = 2^{h^*}\Big(\n_{\m,h^*}\d_{\m\n} +
B^{\m\n}_{h^*}(\pp)\Big)\;,\label{WI3b}\ee
where the correction term $B^{\m\n}_{h^*}(\pp)$ is of order $\bar\e_{h^*}^{2}$, uniformly in $h^*$,
for all $|\pp|\le 2^{h^*}$. The proof of Eq.(\ref{WI3b}) can be worked out along the same lines of
\cite[Appendix B]{GMP1} and is based on the following remarks:
for all scales $h>h^*$, by construction, the external field $J$ appears in the
effective potential in the combination $A+J$, see Eq.(\ref{gen}). As discussed in the
previous section, the corresponding kernel, $\hat W^{(h)}_{0,2,0;(\m,\n)}(\pp)$
is equal to $\hat W^{(h)}_{0,2,0;(\m,\n)}(\pp)=2^h\Big(\n_{\m,h}\d_{\m\n} +\bar
B^{\m\n}_{h}(\pp)\Big)$, with $\bar B^{\m\n}_{h}(\pp)$ a bounded correction, of second
or higher order in the effective coupling constants. Therefore,
after the integration of all the scales $h\ge h^*$, we are left with a kernel
$JJ$ equal to the r.h.s. of Eq.(\ref{WI3b}), with a slightly different
correction term $\bar B^{\m\n}_{h^*}(\pp)$ replacing $B^{\m\n}_{h^*}(\pp)$. From that scale on,
we are left with the super-renormalizable theory studied in \cite{GM,GMP3long},
in which the only
marginal interactions are the $J\Psi^+\Psi^-$ terms, whose coefficient is renormalized
by a finite amount under the RG flow from $h^*$ to $-\io$, see \cite{GMP3long}. Therefore, the
dominant correction terms to $ \hat\SS^{0,h^*}_{0,2;(\m,\n)}(\pp) $ coming from the integration
of the scales below $h^*$ are obtained by contracting two effective vertices $J\Psi^+\Psi^-$
on scale $h<h^*$, and then summing over $h$. The resulting contribution is dimensionally
bounded as $\sum_{h< h^*} O(2^h\bar\e_{h^*}^2)=O(2^{h^*}\bar\e_{h^*}^2)$. Higher order
corrections are bounded in a similar way, using the hierarchical structure of the GN trees, see
also \cite{GMP3long}.

Now, combining Eq.(\ref{WI3b}) with the first WI in Eq.(\ref{WI87}) computed at $\pp=p{\bf u}_\m$
(with ${\bf u}_\m$ the unit vector in direction $\m$ and $p=|\pp|\le 2^{h^*}$), we find:
\be\n_{\m,h^*}=-\lim_{p\to 0}B^{\m\m}_{h^*}(p{\bf u}_\m)=O(\bar\e_{h^*}^2)\;,\label{s5.1}\ee
which means that there is no spontaneous generation of the photon mass (i.e., the
photon field remains unscreened). If read in naive (non-renormalized) perturbation theory,
the above identity is equivalent to an infinite sequence of cancellations taking place
at all orders among the graphs contributing to the photon mass. At lowest order, the
cancellation takes place between the two graphs in Fig.\ref{fig2}, as discussed in Appendix \ref{phm}.
\begin{figure}[htbp]
\includegraphics[width=0.4\textwidth]{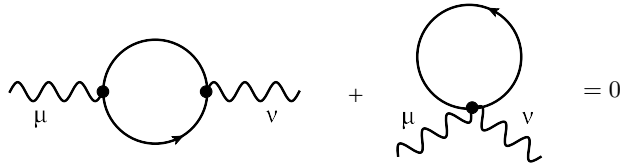}\makebox[1.5truecm]{\raisebox{.9truecm}{$=0$}}
\caption{The lowest order contributions to the
``mass'' of the field $A_{\m}$ with $\m=1,2$, which cancel out exactly when computed at
transferred momentum $\pp={\bf 0}$, as proved in Appendix \protect{\ref{phm}}.}
\label{fig2}\end{figure}

A similar argument can be applied to control the flow of the effective charge.
In fact, note that by computing the second WI in Eq.(\ref{WI87}) at $(\kk',\pp)=(2^{h^*}{\bf u}_\m,
p{\bf u}_\m)$ and by taking
the limit $p\to 0$, we find:
\be \hat S^{0,h^*}_{2,1;\m}(\pp_F^\o+2^{h^*}{\bf u}_\m,{\bf 0})=-e\dpr_\m\hat S^{0,h^*}_{2,0}(\pp_F^\o
+2^{h^*}{\bf u}_\m)\;.\label{s5.4}\ee
By plugging Eqs.(\ref{WI4})--(\ref{WI13}) into Eq.(\ref{s5.4}), we get
\bea && e_{0,h^*}=e\Big(1+B^0_{\o,h^*}(\pp_F^\o+2^{h^*}{\bf u}_0,{\bf 0})-2^{h^*}\dpr_0\bar B_{\o,h^*}
(\pp_F^\o+2^{h^*}{\bf u}_0)\Big)=:e(1+\bar A_{0,h^*})=e(1+O(\bar\e_{h^*}^2))\;,\label{s5.5}\\
&& e_{1,h^*}=e\Big(1+B^1_{\o,h^*}(\pp_F^\o+2^{h^*}{\bf u}_1,{\bf 0})\G^1_\o-2^{h^*}\dpr_1\bar B_{\o,h^*}
(\pp_F^\o+2^{h^*}{\bf u}_1)\Big)=:e(1+\bar A_{1,h^*})=e(1+O(\bar\e_{h^*}^2))\;,\label{s5.5bis}\eea
which tell us that {\it the effective charges $e_{\m,h^*}$ remain close to the unperturbed value
$e_{\m,0}=e$} at all scales $h^*\le 0$ and at all orders in renormalized perturbation theory.
If read in naive perturbation theory, Eqs.(\ref{s5.5})-(\ref{s5.5bis})
are equivalent to infinitely many cancellations taking place
at all orders among the logarithmically divergent graphs contributing to the dressing
of the electric charge. At lowest order, the
cancellation takes place between the graphs in Fig.\ref{fig1}, as discussed in
Appendix \ref{appch}.
\begin{figure}[htbp]
\centering
\includegraphics[width=.48\textwidth]{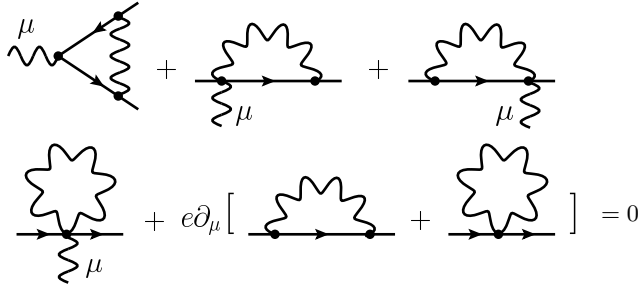}\makebox[-4ex]{\raisebox{7.ex}{$=0$}}
\caption{The lowest order contributions to the dressing of the electric charge,
which cancel out exactly when computed at the Fermi points $\kk=\pp_F^\o$ and at
transferred momentum $\pp={\bf 0}$, as proved in Appendix \ref{appch}.}\label{fig1}
\end{figure}

The corrections $\bar A_{\m,h^*}$ in Eqs.(\ref{s5.5})-(\ref{s5.5bis})
are given by sums over GN trees of order two or higher in the
effective couplings; their $N$-th order is bounded by $(\const.)^N(\bar\e_{h^*})^N(N/2)!$
Moreover, $|\bar A_{\m,h^*}-\bar A_{\m,-\io}|=O(\bar\e^2_{-\io}(v_{-\io}-v_h))+
O(\bar\e^2_{-\io}2^{h/2})$, see \cite[Section 4 and Appendix B]{GMP1} for a proof
of these facts. Note that a priori $\bar A_{0,-\io}\neq \bar A_{1,-\io}$; however, the approximate
Lorentz invariance of the theory combined with the flow equation for the Fermi velocity
will allow us to show that $\bar A_{0,-\io}=\bar A_{1,-\io}$, see the next subsection for a discussion of this
point. \\

{\bf Remark.} The discussion in this section is very similar to the corresponding one
for a model of Dirac fermions with electromagnetic interactions in the continuum in the
presence of a fixed rotationally invariant UV cutoff, see \cite[Section 3]{GMP1}. The key difference
is that the analogues of Eqs.(\ref{s5.5})-(\ref{s5.5bis}) in \cite[Section 3]{GMP1} have an extra term
in the l.h.s., i.e., they read $e_{\m,h^*}(1-\a_\m)=e(1+A_{\m,h^*})$,
with $| A_{\m,h^*}- A_{\m,-\io}|=O(\bar\e^2_{-\io}(v_{-\io}-v_h))+
O(\bar\e^2_{-\io}2^{h/2})$ and $|A_{0,-\io}-A_{1,-\io}|\le (\const.)e^4$; see
\cite[Eqs.(4.13)-(4.14)]{GMP1}. The extra constants $\a_\m$ in the l.h.s. are due to the corrections
to the WIs produced by the momentum cutoff used in \cite{GMP1}, which explicitly breaks
gauge invariance; as shown in \cite{GMP1}, $\a_0\neq\a_1$, the difference being of
second order in
the electric charge, which implies that Lorentz invariance is not recovered asymptotically in the IR.
This has to be contrasted with the case studied here, where exact lattice gauge invariance is
preserved by the RG flow, so that no correction terms appear in the l.h.s. of
Eqs.(\ref{s5.5})-(\ref{s5.5bis}) and Lorentz invariance spontaneously emerge in the deep IR, as
proved in the next subsection.

\subsection{The flow of the effective parameters}\label{secflow}

We are now left with studying the evolution of the effective parameters
$v_h,Z_h,Z^{\pm}_{K,h},\l^{K}_{j,\ul\m,h}$ under the RG flow. If we use the analogue
of the decomposition Eq.(\ref{dec}), we can write the flow equation for these parameters as:
\bea &&\frac{v_{h-1}}{v_h}=1+\b^v_{h}+r^v_{h}\;,\qquad
\frac{Z_{h-1}}{Z_h}=1+\b^{z}_h+r^z_{h}\;,\label{pp0}\\
&&\frac{Z^{\pm}_{K,h-1}}{Z^{\pm }_{K,h}}=1 + \b^{\pm }_{h}
+r^{\pm}_{h}\;,\qquad \l^{K}_{j,\ul\m,h-1}=\l^{K}_{j,\ul\m,h}+r^\l_{j,\ul\m,h}\;,\label{pp} \eea
where $\b^{\#}_h$ are the relativistic part of the {\it beta function} that, by definition,
are obtained by replacing all the propagators $\hat g^{(h)}_\o(\kk')$ contributing to
the r.h.s. of the flow equation by their relativistic part $\hat g^{(h)}_{D,\o}(\kk')$, while
$r^\#_h$ are the rests, which are smaller by a factor $2^{\th h}$, $\th\in(0,1)$,
as compared to the corresponding dominant terms (the reason is the same
as the one sketched after Eq.(\ref{dec})).
Note that both $\b^\#_h$ and $r^\#_h$ are functions of the
whole sequence of effective parameters
$\{e_{\m,k},\n_{\m,k},Z_{k-1},v_{k-1},Z^\pm_{K,k},\l^K_{j,\ul\m,h}\}_{h\le k\le 0}$ that
are bounded respectively by $O(\bar \e_h^2)$ and $O(2^{\th h}\bar \e_h^2)$, provided that
the ratios $Z_k/Z_{k-1}$ are close to 1 and that $v_k$ are bounded away from zero, for all
$h\le k\le 0$.

In the second equation in Eq.(\ref{pp}), we used the fact that the dominant part
$\b^\l_{j,\ul\m,h}$ is exactly zero for all choices of $j,\ul\m,h$, by a parity argument
(inspection of perturbation theory immediately shows that all the contributions  to $\b^\l_{j,\ul\m,h}$
are given by integrals of odd functions of $\kk'$ over a domain that is invariant under $\kk'\to-\kk'$).
The bound on $r^\l_{j,\ul\m,h}$ immediately shows that $\l^K_{j,\ul\m,h}$ is uniformly bounded by
$O(\bar\e_h^2)$ in the IR.

Regarding the flow of $v_h,Z_h$, we note that modulo the correction terms $r^v_h$ and $r^z_h$,
they are the same as those for Dirac fermions in the continuum, derived and written down in
\cite{GMP1}. In particular, using \cite[Eq.(3.14)]{GMP1}, we can write:
\be \frac{v_{h-1}}{v_h}=1 +\frac{ \log 2}{4\p^2} \Biggl[\frac85
e^2(1-v_h)(1+A'_h)+\frac43e(1+B'_h)
(e_{0,h}-e_{1,h})\Biggr]+r^{v}_h\;,\label{5.11b}\ee
where $A'_h$ is a sum of contributions that are finite at all orders
in the effective couplings, which are either of order two or more in the
effective charges, or vanishing at $v_k=1$; similarly, $B'_h$ is a sum of
contributions that are finite at all orders in the effective couplings, which
are of order two or more in the effective charges. The proof of Eq.(\ref{5.11b}) is based on
the remark that $\b_{h,v}$ would be vanishing if $v_h\=1$ and $e_{0,h}\=e_{1,h}$,
by Lorentz invariance (see \cite[Section 3.3]{GMP1} for more details); the numerical coefficients
follow from an explicit computation, see \cite[Appendix C]{GMP1}.
From (\ref{5.11b}) it is apparent that $v_h$ tends as
$h\to-\io$ to a limit value
\be v_{-\io}=1+\frac{5}{6e}(e_{0,-\io}-e_{1,-\io})(1+C'_{-\io})
\label{veff}\ee
with $C'_{-\io}$ a sum of contributions that are finite at all
orders in the effective couplings, which are of order two or more
in the effective charges. The fixed point (\ref{veff}) is found
simply by requiring that in the limit $h\rightarrow-\infty$ the
argument of the square brackets in (\ref{5.11b}) vanishes.

Consider now the identities Eqs.(\ref{s5.5})-(\ref{s5.5bis}). In a
Lorentz invariant theory (i.e., in a theory where all the propagators $\hat g^{(h)}_\o(\kk')$
are replaced by their Dirac approximations $\hat g^{(h)}_{D,\o}(\kk')$, the Fermi velocity is
equal to the speed of light, $v_h\=1$, and the charges $e_{\m,h}$ are $\m$-independent,
$e_{0,h}=e_{1,h}$), we would get that $A_{0,h} = A_{1,h}$. Therefore, using the
GN trees representation for $A_{0,h}, A_{1,h}$, their approximate Lorentz symmetry and the
short memory property (see \cite[Section 2.4]{GMP1}), we find:
\be e_{0,h-1} - e_{1,h-1} = E_{1,h}(e_{0,h} - e_{1,h}) + E_{2,h}(1
- v_{h}) + E_{3,h}\;,\label{LI1} \ee
where $E_{1,h}$ and $E_{2,h}$ are $O(\bar\e_h^2)$ and $E_{3,h}=O(\bar\e_h 2^{\th
h})$, for $\th\in(0,1)$. If we combine Eq.(\ref{LI1}) with Eq.(\ref{5.11b}) we get
\be e_{0,-\io} - e_{1,-\io} = E_{-\io}(e_{0,-\io} -
e_{1,-\io})\;,\label{LI2} \ee
with $E_{-\io} = O(\bar\e_{-\io}^{2})$, which implies the {\it spontaneous emergence of Lorentz
invariance} in the deep IR, i.e.,
\be e_{0,-\io}=e_{1,-\io}\quad{\rm and}\quad v_{-\io}=1\;. \ee
Using these informations into Eq.(\ref{5.11b}), we see that
the approach of $v_h$ to the speed of light is {\it anomalous}, i.e.,
\be  1-v_h\simeq A(v)\, 2^{\tilde\h h} \;,\qquad  \tilde\h=\frac{2e^2}{5\pi^2}+O(e^4)\;, \label{1-v}\ee
where $A(v)$ is a function of $e$ and $v$ that vanishes linearly at $v=1$, i.e.,
for $v$ close to 1, $A(v)=(1-v)\big(1+O(1-v)+O(e^2)\big)$. The ``$\simeq$" in the first equation means that the ratio of the two sides tends to 1 as $h\to-\infty$.

Regarding the flow of the other renormalization parameters, a second order computation
of the dominant contribution to the beta function (see \cite[Appendix C]{GMP1} and Appendix
\ref{secexp}), the fact that $e_{\m,h}$ are close to $e_{-\io}=e(1+O(e^2))$,  asymptotically
as $h\to-\io$, together with Eq.(\ref{1-v}), shows that
\bea && \b^z_{h}=\frac{e^2}{12\pi^2}\log 2+O(e^2(1-v)2^{ce^2h})+O(e^4)\;,\label{oth5cbis}\\
&&\b_h^+=
\frac{e^2}{12\pi^2}\log 2+O(e^2(1-v)2^{ce^2h})+O(e^4)\;,\qquad \b_h^-= \frac{3
e^{2}}{4\pi^2}\log 2+O(e^2(1-v)2^{ce^2h})+O(e^4)\;,\label{oth5c} \eea
for some $c>0$. Eqs.(\ref{oth5cbis})-(\ref{oth5c}) imply that
\be Z_h\simeq B^0(v)\, 2^{-h\eta}\;,\qquad
Z_{K,h}^{+}\simeq B^+(v)\,2^{-h\eta^+_K}\;,\qquad
Z_{K,h}^-\simeq B^-(v)\,2^{-h\eta^-_K}\;,\label{s5.10}\ee
with
\be \eta=\frac{e^2}{12\pi^2}+O(e^4)\;,\qquad \eta^+_K=
\frac{e^{2}}{12\pi^2}+O(e^4)\;,\qquad
\h^-_K= \frac{3
e^{2}}{4\pi^2}+O(e^4)\;,\label{s5.11}\ee
and $B^\#(v)=1+O(1-v)+O(e^2)$, which concludes the study of the flow of the renormalized parameters.

\section{The Kekul\'e response function}\label{sec3ac}
\setcounter{equation}{0}
\renewcommand{\theequation}{\ref{sec3ac}.\arabic{equation}}

\begin{figure}[hbtp]
\centering
\includegraphics[width=.3\textwidth]{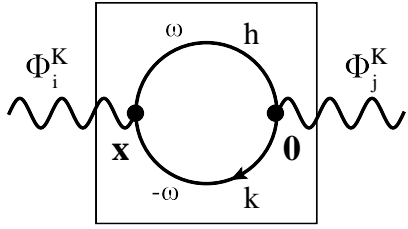}
\caption{Leading contribution to the Kekul\' e response function; a sum over $h\leq 0$ is understood.}\label{figkekbis}
\end{figure}

In this section we explicitly compute the Kekul\' e response function
(the other responses can be obtained in a similar way and will be discussed in the next section),
\be
R^{(K)}_{ij}(\xx) = \lim_{\b\to\infty}\lim_{L\to\infty}
\lim_{h^{*}\to-\io}\frac{\partial^{2}}{\partial \Phi^{K}_{i,\xx}\partial\Phi^{K}_{j,\V0}}\WW^{0,h^{*}}
(\Phi,0,0)\Big|_{\Phi=0}\;.\label{reskek1}\ee
The iterative construction of the generating functional described in Section \ref{sec3} induces
an explicit representation for the Kekul\'e response function in terms of GN trees, completely
analogous to the one described in \cite[Proof of Proposition 1]{GMP3long}. In particular,
using the analogue of the decomposition in Eq.(\ref{dec}), we can rewrite
\be
R^{(K)}_{ij}(\xx) = R^{(K),D}_{ij}(\xx) +\widetilde R^{(K)}_{ij}(\xx) \;,\label{3ac.1}\ee
where $R^{(K),D}_{ij}(\xx)$ is obtained by replacing the lattice by the Dirac propagators in the
expansion for $R^{(K)}_{ij}$ and $\widetilde R^{(K)}_{ij}(\xx)$ is the rest. Using the same strategy
leading to the bounds Eq.(\ref{4.11a}) and, more specifically, to
\cite[Eqs.(2.81)--(2.84)]{GMP3long}, we find that the $N$-th order
contribution in renormalized perturbation theory to $R^{(K),D}_{ij}(\xx)$ is bounded, for
all $M\ge 0$, by
\be\big| R^{N;(K),D}_{ij}(\xx)\big|\le |e|^N\Big(\frac{N}2\Big)!\sum_{h=-\infty}^0\sum_{\bar h=h}^0
\sum_{\o=\pm}\Big(\frac{Z_{K,h}^\o}{Z_h}\Big)^22^h 2^{3\bar h}
\frac{(C_M)^N}{1+(2^{\bar h}|\xx|)^M}\;,\label{3ac.2}\ee
for a suitable constant $C_M$. The factors $2^h 2^{3\bar h}\big[1+(2^{\bar h}|\xx|)^M\big]^{-1}$
represent the dimensional bound on all the labelled Feynman diagrams corresponding to the
same GN tree, where: (i) $h$ is the lowest among the scales of the propagators in the diagram; (ii)
$\bar h$ is the lowest among the scales of the propagators in a path connecting
the two special vertices of type $\Phi^K\Psi^+\Psi^-$; (iii) $2^h=2^{h(3-2n-m-p)}
\big|^{n=m=0}_{p=2}$ is the scaling dimension of the graph; (iv) $2^{3\bar h}$ is the dimensional
gain coming from the fact that the locations $\xx$ and $\V0$ of the external fields $\Phi^K$
are {\it fixed} rather than integrated over the whole space-time domain
(as it is the case for the Feynman diagrams contributing to the thermodynamic functions, where all
the space-time labels of the vertices are integrated over the whole space);
(v) $\big[1+(2^{\bar h}|\xx|)^M\big]^{-1}$ is the decay factor coming from the propagators on a
path connecting the two special vertices of type $\Phi^K\Psi^+\Psi^-$. Now,
picking $M=5$, exchanging
the order of summations over $h$ and $\bar h$, and summing over $h$ gives
(recall the asymptotic relations Eq.(\ref{s5.10}))
\bea  \big| R^{N;(K),D}_{ij}(\xx)\big|&\le&(\const.)^N |e|^N\Big(\frac{N}2\Big)!\sum_{\bar h=-\infty}^0
\frac{2^{\bar h(4+2\h-2\h^-_K)}}{1+(2^{\bar h}|\xx|)^5}\quad \Rightarrow\nn\\
\Rightarrow\quad \big| R^{N;(K),D}_{ij}(\xx)\big|&\le&(\const.)^N |e|^N\Big(\frac{N}2\Big)!
\frac1{1+|\xx|^{4+2\h-2\h^-_K}}\;,\label{3ac.3}\eea
where we used the fact that $\h^-_K>\h^+_K$. The correction term $\widetilde R^{(K)}_{ij}(\xx)$
admits a similar bound, with a dimensional gain factor that implies a faster decay in real space.
Using the symmetries in Appendix \ref{app1}
together with relativistic invariance (see symmetry \cite[(7)]{GMP1})
and proceeding as in Appendix \ref{app3}, we find that the symmetry structure of
$R^{N;(K)}_{ij}(\xx)$ is:
\be R^{N;(K)}_{ij}(\xx) = a_K^{(N)}e^N\frac{\cos\big(\vec p_F^{+}(\vec x-\vec\d_i+\vec\d_j)\big)}
{|\xx|^{4-\x^{(K)}}}+{\it faster}\ {\it decaying}\ {\it terms},\label{3ac.4}\ee
where the faster decaying correction terms come from: (i) the irrelevant terms that, scale by scale,
produce corrections smaller by a factor $O(2^h)$ as compared to the dominant terms;
(ii) the Lorentz-symmetry breaking terms, i.e., the terms proportional to $1-v_h$
that come from a rewriting of the effective Fermi velocity $v_h$
in the definition of $\hat g_{D,\o}$ as $v_h=1-(1-v_h)$; these produce corrections that,
scale by scale, are smaller by a factor $O(e^22^{(\const.)e^2h})$ as compared to the dominant
terms. In Eq.(\ref{3ac.4}), $\x^{(K)}=2\h^-_K-2\h=\frac{4e^2}{3\p^2}+O(e^4)$ and $a_K^{(N)}$ is
a suitable constant, bounded in absolute value by $(\const.)^N\Big(\frac{N}2\Big)!$
The $0$-th order constant, which gives the dominant contribution to the Kekul\'e response
function, is given by the value of the graph in Fig.\ref{figkekbis}.

More explicitly,
\bea && R^{0;(K),D}_{ij}(\xx)=-2 \sum_{h,k=-\io}^0\sum_{\o}e^{i(\pp_F^\o-\pp_F^{-\o})\xx}
\frac{(Z^-_{K,h\vee k})^2}{Z_hZ_k}\int \frac{d\kk}{2\p|\BBB|}\int\frac{d\pp}{2\p|\BBB|}
f_h(\kk)f_k(\pp)e^{i(\kk-\pp)\xx}\cdot\nn\\
&&\qquad \cdot
\Tr\Big\{\frac{-ik_0\G^0_\o+i\vec k\cdot\vec\G_\o}{\kk^2}\G^-_{\o,i}
\frac{-ip_0\G^0_{-\o}+i\vec p\cdot\vec\G_{-\o}}{\pp^2}\G^-_{-\o,j}\Big\}+O(e^2|\xx|^{-4+\x^{(K)}})+
{\it faster}\ {\it decaying}\ {\it terms}\;,\label{3ac.5}\eea
where $h\vee k=\max\{h,k\}$. Recalling that $\G^-_{\o,j}=e^{-i\o\th_j}\s_1=
e^{i\pp_F^\o(\dd_j-\dd_1)}\s_1$ and using the fact that
$\Tr\{\G^\n_\o\s_1\G^\n_{-\o}\s_1\}=2$ for all $\n\in\{0,1,2\}$ and that
$\Tr\{\G^\m_\o\s_1\G^\n_{-\o}\s_1\}=0$ for $\m\neq \n$, we can rewrite
\be R^{0;(K),D}_{ij}(\xx)=4 \sum_{h,k\le 0}\sum_\o e^{-i\pp_F^\o(\xx-\dd_i+\dd_j)}
\frac{(Z^-_{K,h\vee k})^2}{Z_hZ_k}\int \frac{d\kk}{2\p|\BBB|}\int\frac{d\pp}{2\p|\BBB|}
f_h(\kk)f_k(\pp)e^{i(\kk-\pp)\xx}\frac{\kk\cdot\pp}{|\kk|^2|\pp|^2}\;,\label{3ac.6}\ee
modulo corrections $O(e^2|\xx|^{-4+\x^{(K)}})$ or decaying faster than $|\xx|^{-4+\x^{(K)}}$
at infinity. Using
Eq.(\ref{s5.10}), we can replace $\frac{(Z^-_{K,h\vee k})^2}{Z_hZ_k}$ by
$\big(\frac{B^-(v)}{B^0(v)}\big)^22^{-2\h^-_K(h\vee k)}2^{\h h+\h k}$,
modulo faster decaying corrections.
Finally, rewriting, e.g., $2^{\h h}=|\xx|^{-\h}\big[1+\big((2^h|\xx|)^{\h}-1\big)\big]$, it is easy
to realize that the contributions to the response function coming from the terms like
$|\xx|^{-\h}\big[(2^h|\xx|)^{\h}-1\big]$ are $O(e^2|\xx|^{-4+\x^{(K)}})$, see
\cite[Section 4.3.2]{tesiporta}. Therefore, using
the identity $\sum_{h\le 0}f_h(\kk)=\c(\kk)$,
\be R^{0;(K),D}_{ij}(\xx)=4 \Big(\frac{B^-(v)}{B^0(v)}\Big)^2|\xx|^{2(\h^-_K-\h)}
\sum_\o e^{-i\pp_F^\o(\xx-\dd_i+\dd_j)}
\int \frac{d\kk}{2\p|\BBB|}\c(\kk)e^{i\kk\xx}\frac{\kk}{|\kk|^2}\cdot
\int\frac{d\pp}{2\p|\BBB|}\c(\pp)e^{-i\pp\xx}\frac{\pp}{|\pp|^2}\;,\label{3ac.7}\ee
modulo the aforementioned corrections. Finally, using the fact that $\int \frac{d\kk}{2\p|\BBB|}\c(\kk)e^{i\kk\xx}\frac{\kk}{|\kk|^2}=\frac{i3\sqrt3}{8\p}\frac{\xx}{|\xx|^3}+$ {\it faster decaying terms},
we can rewrite:
\be R^{0;(K),D}_{ij}(\xx)= \frac{27}{8\p^2} \Big(\frac{B^-(v)}{B^0(v)}\Big)^2\frac{\cos\Big(\pp_F^+(\xx-\dd_i+\dd_j)\Big)}{|\xx|^{4-\x^{(K)}}}+O(e^2|\xx|^{-4+\x^{(K)}})+
{\it faster}\ {\it decaying}\ {\it terms}\;.\label{3ac.8}\ee
Using the fact that $B^\#(v)=1+O(1-v)+O(e^2)$ we finally
get Eq.(\ref{RK}).

As we already remarked in the introduction,
the effect of the interaction is that of {\it enhancing} the response function of the Kekul\'e distortion,
but still preserving its integrability, so that $\hat R^{(K)}_{ij}(\pp)$ is finite. However, $\partial_{\pp}
\hat R^{(K)}_{ij}(\pp)$ is {\it singular} in $\pp=\pp_F^{\o}$. In fact, the Fourier
transform of Eq.(\ref{RK}) gives:
\be
\hat R^{(K)}_{ij}(\pp) = (\const.)\Big[ e^{-i\pp_F^{+}(\dd_i - \dd_j)}|\pp - \pp_F^{+}|^{1-\xi^{(K)}}
+ e^{-i\pp_F^{-}(\dd_i - \dd_j)}|\pp - \pp_F^{-}|^{1 - \xi^{(K)}} \Big]
+{\it more}\ {\it regular}\ {\it terms}\;,\label{3ac.9}\ee
which implies that $\partial_{\pp}\hat R^{(K)}_{ij}(\pp)$ diverges like $|\pp-\pp_F^{\o}|^{-\xi^{(K)}}$ at the Fermi points, as claimed in Section \ref{sec2.1.1c}.

\section{Other response functions}\label{secoth}
\setcounter{equation}{0}
\renewcommand{\theequation}{\ref{secoth}.\arabic{equation}}

An analysis analogous to the one for the Kekul\'e response function can be worked out for
the other responses and is sketched below. Regarding the CDW and AF responses, the
symmetries of the model imply that the relevant terms of the form $\Phi^{CDW}A$ or $\Phi^{AF}A$
are vanishing, while the marginal terms of the form $\Phi^{CDW}\Psi^+\Psi^-$
and $\Phi^{AF}\Psi^+\Psi^-$ have the following structure:
\bea
&&\LL \hat\BBB^{(h)}_{CDW}(\sqrt{Z_{h-1}}\Psi^{(\leq h)},
A^{(\leq h)},\Phi) =\label{oth1}\\
&&=\frac1{\b^2L^2|\SS_L|}\sum_{\substack{\kk',\pp\\
\o,\s,j}}\Big[
Z^{+}_{CDW,h}\hat \Phi^{CDW}_{j,\pp}\hat \Psi^{(\leq h)+}_{\kk'+\pp,\s,\o}\s_3
\hat \Psi^{(\leq h)-}_{\kk',\s,\o} + Z^{-}_{K,h}\hat \Phi^{CDW}_{j,\pp_F^{\o}-\pp_F^{-\o}+\pp}
\hat \Psi^{(\leq h)+}_{\kk'+\pp,\s,\o}\big[e^{i\o\th_j n_-}\s_3\big]
\hat \Psi^{(\leq h)-}_{\kk',\s,-\o} \Big]\;,\nn\eea
\bea
&&\LL \hat\BBB^{(h)}_{AF}(\sqrt{Z_{h-1}}\Psi^{(\leq h)},
A^{(\leq h)},\Phi) =\label{oth2}\\
&&=\frac1{\b^2L^2|\SS_L|}\sum_{\substack{\kk',\pp\\
\o,\s,j}}\s\Big[
Z^{+}_{AF,h}\hat \Phi^{AF}_{j,\pp}\hat \Psi^{(\leq h)+}_{\kk'+\pp,\s,\o}\s_3
\hat \Psi^{(\leq h)-}_{\kk',\s,\o} + Z^{-}_{K,h}\hat \Phi^{AF}_{j,\pp_F^{\o}-\pp_F^{-\o}+\pp}
\hat \Psi^{(\leq h)+}_{\kk'+\pp,\s,\o}\big[e^{i\o\th_j n_-}\s_3\big]
\hat \Psi^{(\leq h)-}_{\kk',\s,-\o} \Big]\;.\nn\eea
Similar expressions are valid for the other external fields $\Phi^a$. Note that besides the marginal terms spelled out in the previous equation, there may also be marginal terms of the form
$\Phi A A$, $\Phi \Phi A$ or $p_\m \Phi A$; the flow of the corresponding coupling constants
can be easily controlled along the same lines followed to prove the boundedness of the flow of
$\l^K_{j,\ul\m,h}$ in Section \ref{secflow}.

As seen in the previous section, the flow of the renormalization constants
$Z^\pm_{CDW,h}$ and $Z^\pm_{AF,h}$ control the large distance decay of the CDW and AF
response functions. In Appendix \ref{secothexp} it is shown that
\be \frac{Z^{+}_{CDW,h-1}}{Z^{+}_{CDW,h}}=1 +
\frac{3e^2}{4\pi^2}\log 2+O(e^4) +\cdots\;,\qquad
 \frac{Z^{-}_{CDW,h-1}}{Z^{- }_{CDW,h}}=1 +
\frac{e^2}{12\pi^2}\log 2+O(e^4) +\cdots\;, \label{zetasf}\ee
which implies that
\be Z_{CDW,h}^\pm\simeq B^\pm_{CDW}(v)\, 2^{-h\eta^\pm_{CDW}}\;,\label{s7.1}\ee
with
\be \eta^+_{CDW}=
\frac{3e^{2}}{4\pi^2}+O(e^4)\;,\qquad
\h^-_{CDW}= \frac{
e^{2}}{12\pi^2}+O(e^4)\;,
\label{s7.2}\ee
and $B^\pm_{CDW}(v)=1+O(1-v)+O(e^2)$. The exponents for the $AF$ are exactly the same,
simply because the kernels of the marginal source terms $\Phi^{AF}\Psi^+\Psi^-$ are the same
as the corresponding terms of the form $\Phi^{CDW}\Psi^+\Psi^-$, see Eqs.(\ref{oth1})-(\ref{oth2}).
Therefore, proceeding as in the previous section, we get, for $a=CDW,AF$,
\bea
&&R^{(a)}_{ij}(\xx)=\frac{27}{8\pi^2}\Big(\frac{B^+_{a}(v)}{B^0(v)}\Big)^2
\frac1{|\xx|^{4 - \xi^{(a)}}}+O(e^2|\xx|^{-4+\x^{(a)}})+{\it faster}\ {\it decaying}\ {\it terms}\;,
\label{s7.3a}\\
&& \xi^{(CDW)} =\x^{(AF)}= 2(\eta^{+}_{CDW} - \eta) = \frac{4e^{2}}{3\pi^2} + O(e^4)\;.
\label{s7.3}\eea
As for the Kekul\'e response function, the anomalous decay in Eq.(\ref{s7.3a}) implies
the presence of a singularity in the first derivative of the corresponding Fourier transforms,
in analogy with Eq.(\ref{3ac.9}).

It is now clear how to extend such analysis to all other responses. The large distance
asymptotic behavior depends on the specific values of the critical exponents, which
are determined at leading order by a second order computation along the lines of
Appendix \ref{app2d}: one first identifies the structure of the local part of the effective source terms,
in analogy with Eqs.(\ref{effsou}),(\ref{oth1}),(\ref{oth2}); for each such term
two renormalization constants
appear, corresponding to processes with transferred momentum equal to $\V0$ or to $\pp_F^\o-
\pp_F^{-\o}$; the flow of each of them is given by a diagram like the one in Fig.\ref{figexp} and can
be computed as in Appendix \ref{app2d}. The resulting  values of the critical exponents
$\x^{(channel)}=2(\h_{channel}-\h)$ at second order in the electric charge
are summarized in Table \ref{tabexp}.
\renewcommand{\arraystretch}{1.3}
\begin{table}[htbp]
\centering
\begin{tabular}{|l|l|}
\hline
$channel$    & $critical\ exponent$\\
\hline
 $a^+_{\o,\s} b_{\o,\s} +b^+_{\o,\s} a_{\o,\s}$ &  $0$\\
 \hline
 $a^+_{\o,\s} b_{-\o,\s} +b^+_{\o,\s} a_{-\o,\s}$ & $4 e^{2}/(3\pi^2)$ \\
\hline
 $ia^+_{\o,\s} b_{\o,\s} -ib^+_{\o,\s} a_{\o,\s}$ & $0$ \\
 \hline
 $ia^+_{\o,\s} b_{-\o,\s} -ib^+_{\o,\s} a_{-\o,\s}$ & $0$ \\
 \hline
$a^+_{\o,\s} a_{\o,\s}+b^+_{\o,\s} b_{\o,\s}$ & $0$ \\
 \hline
$a^+_{\o,\s} a_{-\o,\s}+b^+_{\o,\s} b_{-\o,\s}$ & $0$  \\
  \hline
$a^+_{\o,\s} a_{\o,\s}-b^+_{\o,\s} b_{\o,\s}$ & $4e^2/(3\p^2)$  \\
 \hline
$a^+_{\o,\s} a_{-\o,\s}-b^+_{\o,\s} b_{-\o,\s}$ & $0$  \\
\hline
\end{tabular}
\qquad\qquad
\renewcommand{\arraystretch}{1.3}
\begin{tabular}{|l|l|}
\hline
$channel$    & $critical\ exponent$\\
 \hline
$\s(a^+_{\o,\s} a_{\o,\s}+b^+_{\o,\s} b_{\o,\s})$ & $0$  \\
 \hline
$\s(a^+_{\o,\s} a_{-\o,\s}+b^+_{\o,\s} b_{-\o,\s})$ & $0$  \\
  \hline
$\s(a^+_{\o,\s} a_{\o,\s}-b^+_{\o,\s} b_{\o,\s})$ & $4 e^{2}/(3\pi^2)$  \\
 \hline
$\s(a^+_{\o,\s} a_{-\o,\s}-b^+_{\o,\s} b_{-\o,\s})$ & $0$  \\
 \hline
$\o(a^+_{\o,\s} a_{\o,\s}-b^+_{\o,\s} b_{\o,\s})$ & $4 e^{2}/(3\pi^2)$  \\
\hline
 $i\s (a^+_{\o,\s} a^+_{\o,-\s} +b^+_{\o,\s} b^+_{\o,-\s})+c.c.$ & $-e^2/(3\p^2)$ \\
 \hline
 $i\s \o(a^+_{\o,\s} a^+_{\o,-\s} +b^+_{\o,\s} b^+_{\o,-\s})+c.c.$ &  $-e^2/(3\p^2)$\\
\hline
 $a^+_{\o,\s} b^+_{-\o,\s}-b^+_{\o,\s} a^+_{\o,\s}+ c.c.$ & $-e^2/(3\p^2)$\\
 \hline
 $i\s (a^+_{\o,\s} b^+_{-\o,-\s}-b^+_{\o,\s} a^+_{-\o,-\s})+ c.c.$ & $-e^2/(3\p^2)$ \\
\hline
\end{tabular}
\caption{Lowest order contributions to the anomalous exponents. Summation over the
valley and spin indices is understood.}\label{tabexp}
 \end{table}

The interaction removes the degeneracy of the critical exponents and, therefore, we can identify
the excitations whose response functions decay slowest at infinity, which correspond to the
order parameters in putative strong coupling broken phases. In particular:
\begin{enumerate}
\item the dominant exponents correspond to the second and seventh channel in the left table
and to the third and fifth channel in the right table. The first three are the K, CDW and AF
local order parameters discussed in the introduction. The fifth channel in the right table
was introduced in \cite{Ha}. Microscopically, this order parameter
may be understood as a specific pattern of circulating currents
and its enhancement in the presence of interactions is in agreement with the expectation that a
time-reversal broken fixed point should emerge in the strong coupling regime \cite{HJR}.
\item The last four channels in the right table are Majorana-type masses that correspond to
inter-node and intra-node (uniform and non-uniform) Cooper pairings \cite{K2}. Their critical exponents are all negative, which means that superconducting order is unfavored at
intermediate to strong coupling. There are other possible Cooper pairings besides those
explicitly reported in the table and it turns out that all their exponents are either equal to or
smaller than $-e^2/(3\p^2)$ at second order.
\item The exponents of the density-density and current-current response functions
at zero transferred momentum (third and fifth channel in the left table) are vanishing.
Actually, using the second WI in Eq.(\ref{WI87}) it can be easily proved that these two exponents
are zero at all orders in renormalized perturbation theory. This is indeed a necessary  prerequisite
for having finite conductivity and a semi-metallic behavior also in the presence of interactions,
as expected \cite{Sac}.
\end{enumerate}

\section{Renormalization Group analysis in presence of a mass term}\label{seckek}
\setcounter{equation}{0}
\renewcommand{\theequation}{\ref{seckek}.\arabic{equation}}

As discussed in the introduction, it is natural to investigate the effects of a mass term coupled
to the local order parameters associated to the largest critical exponents. In this section we
discuss how the RG construction is modified by the presence of a mass.
Consider, e.g.,  the Hamiltonian in Eq.(\ref{hd0}) (similar considerations are
valid for CDW, AF masses or for the ``Haldane mass", i.e., the mass associated to the local
order parameter in the fifth line of the right part of Table \ref{tabexp}).
The model is still invariant under the same symmetries (1)--(8) described in Section
\ref{sec4B}, the only novelty being that now the value of $j_0$ appearing in the very definition
of the Kekul\'e mass, see Eq.(\ref{hd0}), should also be changed under the symmetry
transformations; i.e.,
\begin{itemize}
\item under (4), $j_0\rightarrow j_0 - 1$; \item
under (6.b), $j_0 \rightarrow r_v j_0$;
\item $j_0$ is left invariant in all other cases.
\end{itemize}
In the presence of the mass term, the multiscale integration of the generating functional
is performed in a way very similar to the one described in Section \ref{sec3} for the
massless case.
After the integration of the ultraviolet degrees of freedom and the definition
of the quasi-particle fields, one immediately realizes that the presence of the
mass produces new local relevant terms of the form
$\sum_{\o,\s}\D^{(j_0)}_{\o,h}a^{(\le h)+}_{\kk',\s,\o} b^{(\le h)-}_{\kk',\s,-\o}+c.c.$,
which can be step by step inserted into the definition of the fermionic
propagator.
The symmetries of the model imply the following conditions on the complex
constants $\D^{(j_0)}_{\o,h}$:
\begin{enumerate}
\item using (4), we find that $\D^{(j_0)}_{\o,h}=e^{-i\o \frac{2\p}3}
\D^{(j_0-1)}_{\o,h}$;
\item using (5), we find that
$\D^{(j_0)}_{\o,h}=\big[\D^{(j_0)}_{-\o,h}\big]^*$;
\item using (6.b), we find that $\D^{(j_0)}_{\o,h}=
\D^{(r_vj_0)}_{-\o,h}$.
\end{enumerate}
This fixes the structure of the effective mass term, in the form:
\be \LL_{mass}^{(K)}\VV^{(h)}(\Psi,A)=\D^K_h\frac{1}{\b L^2}\sum_{\kk',\o,\s}
\hat\Psi^{(\le h)+}_{\kk',\o,\s}\G^-_{\o,j_0}\hat\Psi^{(\le h)-}_{\kk',-\o,\s}\;,\label{s8.1} \ee
with $\D^K_h$ a real constant such that $\D^K_0=\D_0$ is the same as the one in Eq.(\ref{hd0}).
The term in Eq.(\ref{s8.1}) is the only extra contribution to
the local part of the effective action. In fact, the localization procedure can be now
modified by requiring that the
localization operator extracts from the kernels with $2n+m+p=2$ (or $=3$) the first two (or the first)
terms of a Taylor series in $(\ul\kk,\ul\pp,\ul\qq)$ {\it and} in $\{\D^K_k\}_{k\le 0}$: this is because
all the terms in perturbations theory proportional to the mass itself have a dimensional gain of
$O(2^{(1-\const.e^2) h})$ with respect to their ``massless" bound, see
\cite[Sections 3.2 and 3.3]{GM05} or \cite[Section 5.3.2]{tesiporta}. Therefore, with this new
definition of localization, the only possible extra local term in the effective action is a quadratic
bilinear in the fermionic fields as the one in Eq.(\ref{s8.1}). Note also that such term
can be rewritten in relativistic form as:
\be  \LL_{mass}^{(K)}\VV^{(h)}(\Psi,A)=
- \D^K_h\frac{1}{\b L^2}\sum_{\kk',\s}\lis\psi^{(\le h)+}_{\kk',\s}\big[
e^{i\th_{j_0}\g_5}\g_5\big]\psi^{(\le h)-}_{\kk',\s}\;,\label{s8.2} \ee
with $\th_{j_0}=\frac{2\p}{3}(j_0-1)$. \\

{\bf Remark.} The masses corresponding to the CDW, AF and Haldane mass terms can
be worked out along the same lines. It turns out that these effective masses have the form:
\bea && \LL_{mass}^{(CDW)}\VV^{(h)}(\Psi,A)=\D_h^{CDW}\frac{1}{\b L^2}\sum_{\kk',\o,\s}
\hat\Psi^{(\le h)+}_{\kk',\o,\s}\s_3\hat\Psi^{(\le h)-}_{\kk',\o,\s}\;,\label{s8.3} \\
&& \LL_{mass}^{(AF)}\VV^{(h)}(\Psi,A)=\D_h^{AF}\frac{1}{\b L^2}\sum_{\kk',\o,\s}\s
\hat\Psi^{(\le h)+}_{\kk',\o,\s}\s_3\hat\Psi^{(\le h)-}_{\kk',\o,\s}\;,\label{s8.3bis} \\
&& \LL_{mass}^{(H)}\VV^{(h)}(\Psi,A)=\D_h^{H}\frac{1}{\b L^2}\sum_{\kk',\o,\s}\o
\hat\Psi^{(\le h)+}_{\kk',\o,\s}\s_3\hat\Psi^{(\le h)-}_{\kk',\o,\s}\;,\label{s8.4} \eea
which can be rewritten in relativistic form as
\bea &&
\LL_{mass}^{(CDW)}\VV^{(h)}(\Psi,A)=
i\D^{CDW}_h\frac{1}{\b L^2}\sum_{\kk',\s}\lis\psi^{(\le h)+}_{\kk',\s}\g_3\psi^{(\le h)-}_{\kk',\s}\;,\label{s8.5} \\
&&\LL_{mass}^{(AF)}\VV^{(h)}(\Psi,A)=i\D^{AF}_h\frac{1}{\b L^2}\sum_{\kk',\s}\s\lis\psi^{(\le h)+}_{\kk',\s}\g_3\psi^{(\le h)-}_{\kk',\s}\;,\label{s8.6}\\
&&\LL_{mass}^{(H)}\VV^{(h)}(\Psi,A)=i\D^H_h\frac{1}{\b L^2}\sum_{\kk',\s}\lis\psi^{(\le h)+}_{\kk',\s}\g_3\g_5\psi^{(\le h)-}_{\kk',\s}\;,\label{s8.7}\eea
which are the same as those considered in \cite{HJR,GSC}.\\

As mentioned above, the fermionic bilinear in Eq.(\ref{s8.1}) is inserted step by step into the
definition of the fermionic propagator. As a consequence, the effective propagator at scale $h$
acquires a mass gap of size $\D^K_h$. Of  course, this gap is not visible as long as
$|\D^K_h|\ll 2^h$. Therefore, if the bare mass $\D_0$ is small, the multiscale integration
and the dimensional bounds remain unchanged up to a scale $h_0$ such that
$\D^K_{h_0}\simeq 2^{h_0}$. At that point the propagator has
a mass of size comparable with $2^{h_0}$ itself, and we can integrate the
fermionic degrees of freedom associated to scales $\le h_0$ in a single step. From that on,
we are left with a purely bosonic theory. The symmetries of the theory can be still
used to prove that the only local terms in the effective action for such a bosonic theory
are photon mass terms of the form $\sum_\m 2^h\tilde\n_{\m,h}A^{(\le h)}_\m A^{(\le h)}_\m$,
$h\le 0$, with $\tilde\n_{\m,h}$ independent of $j_0$, see \cite[Section 5.3.2]{tesiporta}.
Since the Kekul\'e mass term in the Hamiltonian does not break gauge invariance,
the same argument used in Section \ref{secWIpart1} to control the flow of $\tilde \n_{\m,h}$ can
be repeated here to show that $\tilde \n_{\m,h}=O(e^2)$, which allows one to safely integrate
all scales up to $-\io$.

We are left with the problem of computing the scale $h_0$ that separates the massless and
massive regimes. In order to do this, we need to control the flow of $\D^K_h$,
which is driven, as usual, by a beta function equation:
\be \frac{\D^K_{h-1}}{\D^K_h}=\frac{Z_h}{Z_{h-1}}\big(1+\b^\D_{K,h}\big)\;.
\label{s8.8}\ee
Proceeding once again as in Appendix \ref{app2d}, we find that
\be \b^{\D}_{K,h}=\frac{3e^2}{4\p^2}\log 2+O(e^4)+O(e^22^h)+O(e^2(1-v)2^{(\const.)e^2h})\;.
\label{s8.9}\ee
The resulting flow is:
\be \D^K_h\simeq \D_0(v)2^{-\h_\D h}\;,\quad {\rm with}\quad
\h_\D=\h^-_K-\h=\frac{2e^2}{3\p^2}+O(e^4)\label{s8.10}\ee
and $\D_0(v)=\D_0(1+O(e^2)+O(1-v))$. Therefore, the equation for the {\it dressed
electron mass} $\D=2^{h_0}$ becomes:
\be \D=2^{h_0}=\D_0(v)2^{-\h_\D h_0}\quad \Rightarrow\quad \D=\big[\D_0(v)\big]^{1/(1+\h_\D)}
\;,\label{s8.11}\ee
which proves Eq.(\ref{aa}).

\section{Gap equation}\label{appC}
\setcounter{equation}{0}
\renewcommand{\theequation}{\ref{appC}.\arabic{equation}}

The variational equation corresponding to the minimization problem Eq.(\ref{var_eq}) is:
\be \phi_{\vec x,j}=\frac{g^2}{\k}\media{\z^K_{\vec x,j}}^{\phi}\;,\label{s9.1}\ee
where $\media{\cdot}^{\phi}$ is the statistical average in the presence of the phonon
field $\phi_{\vec x,j}$. We now want to check that the distortion $\phi^{(j_0)}_{\vec x,
j}$ defined in Eq.(\ref{phi0*}) is a stationary point of the total
energy, provided $\phi_0$ and $\D_0$ are chosen properly.
The stationarity condition Eq.(\ref{s9.1}) with $\phi_{\vec x,j}=\phi_{\vec x,j}^{(j_0)}$
in the limit $\b,L\to\infty$ is equivalent to
the two following coupled self-consistent equations for $\phi_0$ and $\D_0$:
\bea && \phi_0 =\frac{g^2}{\k}\sum_{n\ge 0}\frac{(ie)^n}{n!}
\int\frac{d\pp}{(2\p)^3} \frac{d\kk}{2\p|\BBB|}\langle
\big[\h^j_{\cdot}(\vec\d_j\cdot\vec A_{\cdot})\big]^{*n}_\pp\,
\hat a^+_{\kk+\pp,\s}
\hat b^-_{\kk,\s}\rangle^{(j_0)}\,e^{-i\kk(\dd_j-\dd_1)}+c.c.\;,\label{s9.2}\\
&& \frac{\D_0}3=\frac{g^2}{\k} 
\sum_{n\ge 0}\frac{(ie)^n}{n!}
\int\frac{d\pp'}{(2\p)^3}\frac{d\kk'}{2\p|\BBB|}\langle
\big[\h^j_{\cdot}(\vec\d_j\cdot\vec A_{\cdot})\big]^{*n}_{\pp'}\,
\hat a^+_{\kk'+\pp'+\pp_F^{-\o},\s}
\hat b^-_{\kk'+\pp_F^\o,\s}\rangle^{(j_0)}\, e^{-i\kk'(\dd_j-\dd_1)} e^{-i\pp_F^\o(\dd_{j_0}-\dd_1)}
+c.c.\;,\nn\eea
where $\big[\h^j_{\cdot}(\vec\d_j\cdot\vec A_{\cdot})\big]^{*n}_\pp$ was defined in
Eq.(\ref{adeltaj}), $\media{\cdot}^{(j_0)}$ is the statistical average in the presence of the phonon
field $\phi^{(j_0)}_{\vec x,j}$ and we denoted by $\hat a^\pm_{\kk,\s},\hat b^\pm_{\kk,\s}$
the first and second components of the spinor $\hat\Psi^\pm_{\kk,\s}$, respectively.
These equations are well defined provided that the right hand sides of the two equations are independent of $j$, $j_0$ and $\o$. Using the symmetries of the model, we find that (see Appendix \ref{appAgap} for
a proof)
\bea && \langle\!\langle\hat a^+_{\kk,\s}\hat b^-_{\kk,\s}\rangle\!\rangle^{(j_0)}_j:=
\sum_{n\ge 0}\frac{(ie)^n}{n!}
\int\frac{d\pp}{(2\p)^3} \langle
\big[\h^j_{\cdot}(\vec\d_j\cdot\vec A_{\cdot})\big]^{*n}_\pp\,
\hat a^+_{\kk+\pp,\s}\hat b^-_{\kk,\s}\rangle^{(j_0)}=
\O(\vec k) A(\kk)\;,\label{s9.3}\\
&& \langle\! \langle\hat a^+_{\kk'+\pp_F^{-\o},\s}\hat b^-_{\kk'+\pp_F^\o,\s}\rangle\!
\rangle^{(j_0)}_j:=\sum_{n\ge 0}\frac{(ie)^n}{n!}
\int\frac{d\pp'}{(2\p)^3}\langle
\big[\h^j_{\cdot}(\vec\d_j\cdot\vec A_{\cdot})\big]^{*n}_{\pp'}\,
\hat a^+_{\kk'+\pp'+\pp_F^{-\o},\s}\hat b^-_{\kk'+\pp_F^\o,\s}\rangle^{(j_0)}
=\O(\vec k') B(\kk')e^{i\pp_F^\o(\dd_{j_0}-\dd_1)}\;,
\nn\eea
with $A(\kk)$ and $B(\kk')$ two functions, independent of $\o$ and
$j_0$, transforming as follows under the discrete symmetries of the model:
\bea &&A(\kk)=A(T\kk)=A(I\kk)=
A^*(R_h\kk)=A(R_v\kk)\;,\label{s9.5}\\
&&B(\kk')=B(T\kk')=B(I\kk')=
B^*(R_h\kk')=B(R_v\kk')\;,\label{s9.6}\eea
The dimensional bounds following from the multiscale integration,
combined with an explicit lowest order computation, show that, if
$|\kk'|\simeq 2^h$ with $h\ge h_0$ (here $h_0$ is the scale defined in Eq.(\ref{s8.11})),
\bea && \langle\!\langle\hat a^+_{\kk'+\pp_F^\o,\s}\hat b^-_{\kk'+\pp_F^\o,\s}
\rangle\!\rangle^{(j_0)}_j=
\frac1{Z_h} \frac{v_h\O(\vec k'+\vec p_F^\o)}{k_0^2+v_h^2|\O(\vec
k'+\vec p_F^\o)|^2}
(1+A'(\kk'))\;,\label{s9.7}\\
&&  \langle\! \langle\hat a^+_{\kk'+\pp_F^{-\o},\s}\hat b^-_{\kk'+\pp_F^\o,\s}\rangle\!
\rangle^{(j_0)}_j=
\frac{\O(\vec k')}3\frac{\D^K_h}{Z_h} \frac{e^{i\pp_F^\o(\dd_{j_0}-\dd_1)}} {k_0^2+v_h^2|\O(\vec k'+\vec
p_F^\o)|^2}(1+ B'(\kk'))\;,\label{s9.8}\eea
where the correction terms $A'(\kk')$ and $B'(\kk')$ satisfying the same symmetry
properties as $A(\kk)$ and $B(\kk')$ in Eqs.(\ref{s9.5})-(\ref{s9.6})
and of order $O(e^2)+O(2^h)+O(\D_h 2^{-h})$.
If we plug Eqs.(\ref{s9.7})-(\ref{s9.8}) into Eq.(\ref{s9.2}),
we are led to the self-consistent equations
\bea && \phi_0 \simeq 2 \frac{g^2}{\k}\sum_\o
\int_{\D\lesssim|\kk'|\lesssim 1} \frac{d\kk'}{2\p|\BBB|}
\frac1{Z(\kk')} \frac{v(\kk')|\O(\vec k'+\vec
p_F^\o)|^2}{k_0^2+v(\kk')^2|\O(\vec k'+
\vec p_F^\o)|^2}\;,\label{7.9}\\
&& \D_0\simeq 6\frac{g^2}{\k}\int_{\D\lesssim|\kk'|\lesssim 1}
\frac{d\kk'}{2\p|\BBB|}\frac1{Z(\kk')}\frac{\D(\kk')}{k_0^2+v(\kk')^2|\O(\vec
k'+ \vec p_F^\o)|^2}\;,\label{7.10}\eea
where the ``$\simeq$" indicates that we are neglecting higher order corrections and,
for $\kk'$ small, $Z(\kk')\sim|\kk'|^{-\h}$,
$\D(\kk')\sim \D_0 |\kk'|^{-\h_\D}$ and $1-v(\kk')\sim (1-v) |\kk'|^{\tilde\h}$. It is apparent that
Eq.(\ref{7.9}) can be solved by fixing $\phi_0$ to be a suitable (positive)
constant of order $g^2$.

On the other hand, Eq.(\ref{7.10}) is equivalent to
Eq.(\ref{nonsimpl}) and leads to Eq.(\ref{simpl3}) and to the conclusions spelled out after
Eq.(\ref{simpl3}). In particular, at small coupling, the critical phonon coupling $g_c$ above
which the gap equation admits a non trivial solution scales like $g_c\sim\sqrt{v}$. This can
be proved by observing that the range of $\r$ such that the denominator of the integrand
in the r.h.s. of Eq.(\ref{simpl3}) is larger than (say) $\frac32v$ is contained in the interval $[0,\r^*]$,
where $1-(1-v)(\r^*)^{\tilde\h}=\frac32v$, which gives, for small $v$, an exponentially small
$\r^*$, i.e., $\r^*\sim e^{-v/(2\tilde\h)}$; in other words, at small coupling, the range of momenta
such that the effective Fermi velocity is substantially different from the bare one is exponentially small, and gives no relevant contribution to the r.h.s. of Eq.(\ref{simpl3}). However,
the gap equation can be naturally extrapolated at intermediate to strong coupling. As
noted in the introduction, the larger the charge, the smaller $g_c$; possibly,
$g_c$ goes to zero for large enough values of the electric charge. \\
\\

{\bf Acknowledgements.} A.G. and V.M. acknowledge financial support from the
ERC Starting Grant CoMBoS-239694.

\appendix
%

\section{Functional integral representation and gauge invariance}\label{app1b}
\setcounter{equation}{0}
\renewcommand{\theequation}{\ref{app1b}.\arabic{equation}}

In this Appendix we sketch the derivation of the functional integral representation presented in Section
\ref{sec2.2} and prove its gauge invariance properties.

\subsection{Functional integral representation}

Using standard methods of many body theory \cite{NeOr} and the explicit form
of the electron and photon propagators discussed in Section \ref{sec2_model}, we can
write the partition function $\Tr{e^{-\b H_\L}}$ of our model in the Coulomb gauge as
\be \frac{\Tr\{e^{-\b H_\L}\}}{\Tr\{e^{-\b(H^0_\L+H^f_\L)}\}}=
\int P(d\Psi)P^{(C)}(d\vec A)e^{\lis\VV(\Psi,\vec A)}\;, \label{trace}\ee
which should be understood as an identity between power series in the electric charge $e$. In Eq.(\ref{trace}),
$P(d\Psi)$ is the same Grassmann gaussian integration as in Eq.(\ref{Pdpsi}), $P^{(C)}(d\vec A)$ is the gaussian
measure
\be P^{(C)}(d\vec A) =\frac1{\NN_{A}}\prod_{\pp\in\DD^+_{\b,L}}\prod_{ i=1,2}
d{\rm Re}A_{i,\pp}d{\rm Im}A_{i,\pp} \exp\Big\{-\frac1{2\b|\SS_L|}
\sum_{\substack{\pp\in\DD_{\b,L}\\ i,j=1,2}}\hat A_{i,\pp}\big[\hat w^{(C)}(\pp)]_{ij}^{-1}\hat
A_{j,-\pp}\Big\}\;,\label{PdACou}\ee
with $\hat w^{(C)}_{ij}(\pp)$ the covariance matrix in Eq.(\ref{1.15}) and, if
and $ \f_{\r\r'}(\xx)=\d(x_0)\f(\vec x+(\r-\r')\vec\d_1)$,
\bea
\lis\VV(\Psi,\vec A)&=&t\sum_{\substack{\s=\uparrow,\downarrow\\ j=1,2,3}}
\int d\xx\Big[\big(e^{ie\int_0^1ds\, \vec\d_j\vec A_{\xx+s\dd_j}}-1\big)
\Psi^{+}_{\xx,\s,1}\Psi^{-}_{\xx + \dd_{j} - \dd_1,\s,2} +
\big(e^{ie\int_0^1ds\, \vec\d_j\vec A_{\xx+s\dd_j}}-1\big)\Psi^{+}_{\xx+\dd_j-\dd_1,\s,2}
\Psi^{-}_{\xx,\s,1}\Big] -\nn\\
&-&\frac{e^2}2\sum_{\substack{\r,\r'=1,2\\ \s,\s'=\uparrow\downarrow}}\int d\xx\int d\yy\,
\Psi^+_{\xx,\s,\r}\Psi^-_{\xx,\s,\r}\f_{\r\r'}(\xx-\yy)\Psi^+_{\yy,\s',\r'}\Psi^-_{\yy,\s',\r'}\;,
\label{ham_c}\eea
The quartic fermionic term in $\lis\VV$ can be eliminated by a Hubbard-Stratonovich
transformation, at the cost of introducing an extra component $A_{0,\xx}$ of the
photon field. This yields a representation of the partition function in terms of a functional
integral analogous to Eq.(\ref{gen}):
\be \int P(d\Psi)P^{(C)}(d\vec A)e^{\lis\VV(\Psi,A)}=\int P(d\psi)P^\x(dA)e^{\VV(\Psi,A)}\Big|_{\x=1}
\;,\label{hubstrat}\ee
where $\VV(\Psi,A)$ is given by Eq.(\ref{1.15uy}) and $P^\x(dA)$ by Eq.(\ref{PdA}), with $\hat w^{\x,h^*}$ replaced
by $\hat w^{\x}:=\hat w^{\x,-\infty}$.
The functional integral representation for the observables can be derived along the same lines.

\subsection{Derivation of The Ward Identities}
The derivation of the Ward Identities involves manipulation of the
functional integral. Therefore, it is crucial to define with care the cutoffs used to regularize the
functional integral and how they are eliminated.
First of all, keeping the IR cutoff at scale $2^{h^*}$ fixed,
we introduce an ultraviolet cut-off, by replacing the
fermionic and bosonic propagator by $\chi_M(k_0)\hat S_0(\kk)$
and $\chi_M(p_0) w_{\m\n}^{\x,h^*}(\pp)$, with $\chi_M(k_0)=\chi(2^{-M}|k_0|)$. Next, we replace the compact
support cut-off functions by new functions, exponentially close to them but with
full support in $\mathbb{R}$; i.e., we replace $\chi(t)$ by a smooth function $\chi^\e(t)$ such that
$\chi^{\e}(t)=\chi(t)=1$ for $0\leq t\leq \frac12$ and $\c(t)<\chi^{\e}(t)\le \c(t)+\e e^{-t}$ for $t>\frac12$.
Finally, we let the Grassmann field live on a finite lattice, both in the space and time variables:
pick $N\in \mathbb{N}$ and define:
\be\L_{\b}^{(N)} =\{x_0 = \frac{\b n}{N}:\ 0\leq n < N\}\times \L\;,\qquad
\BBB^{(N)}_{\b,L} := \{k_0 =
\frac{2\pi}{\b}(n+\frac{1}{2}):\  0\leq n < N\}\times\BBB_{L} \;,\nn\ee
Let $\tilde\Psi$ be a Grassmann  field on $\L_{\b}^{(N)}$
with antiperiodic boundary conditions in the $x_0$ variable, with propagator
$\tilde S_0(k_0,\vec k):=\hat S_0(d_N(k_0),\vec k)$, where
$d_N(k_0)=i\frac{e^{-ik_0\b/N}-1}{\b/N}$. We shall think of Eq.(\ref{gen}) as the $M,N\to\infty$ and
$\e\to0$ limit of a regularized functional, with the limits taken in the following order:
\be e^{\WW^{\x,h^*}(\Phi,J,\l)}=\lim_{M\to\infty}\lim_{\e\to0}
\lim_{N\to\infty}e^{\WW^{\x,h^*}_{M,N}(\Phi,J,\l)}\;,\quad
e^{\WW^{\x,h^*}_{M,N}(\Phi,J,\l)}=\int P_{M,N}(d\tilde\Psi) P_M^{\x,h^*}(d A)
e^{\tilde\VV(\tilde\Psi, A+J)+\BBB(\tilde \Psi, A+J,\Phi)+(\l,\tilde \Psi)}\;,\label{tilde_gen}\ee
where the interaction and source terms are the same as in Section \ref{sec2.2}, with the only differences that
$\int d\xx$ should be understood as a shorthand for $\frac{\b}{N}\sum_{\xx\in\L^{(N)}_\b}$ and the term in the
second line of Eq.(\ref{1.15uy}) should be replaced by
\be- \frac{N}{\b}\sum_{\s=\uparrow\downarrow}\int d\xx\Big[
\tilde\Psi^{+}_{\xx,\s,1}\tilde\Psi^{-}_{(x_0+\frac{\b}{N},\vec x),\s,1}\big(e^{ie\int_0^{\b/N}
A_{0,(x_0+s,\vec x)}}-1\big) +
\tilde\Psi^{+}_{\xx,\s,2}\tilde\Psi^{-}_{(x_0+\frac{\b}N,\vec x),\s,2}
\big(e^{ie\int_0^{\b/N} A_{0,(x_0+s,\vec x+\vec\d_1)}}-1\big)\Big]\;,\nn\ee
Now, the regularized functional $e^{\WW^{\x,h^*}_{M,N}(\Phi,J,\l)}$ is a non-singular finite dimensional integral.
Therefore, we can freely perform unitary change of variables of the fields $\tilde\Psi$, without affecting
the value of the integral. In particular, by performing the unitary {\it phase
transformation} $\tilde \Psi^{\pm}_{\xx,\s,\r}\rightarrow\tilde\Psi^{\pm}_{\xx,\s,\r} e^{\pm i e \a(\xx+(\r-1)\dd_1)}$,
we find:
\be 0 = \frac{\partial}{\partial\hat \a_{\pp}}\WW^{\xi,h^{*}}_{M,N}(\Phi,J + \partial\a,\l \eu^{ i e \a})
\Big|_{\a=0}+\D^\e_{M,N}(\pp;\Phi,J,\l)\;,\label{1.23aa} \ee
where
\be \D^\e_{M,N}(\pp;\Phi,J,\l) = \frac{\partial}{\partial\phi_\pp}\log \int P_{M,N}(d\tilde\Psi) P_M^{\x,h^*}(d A)
e^{\tilde\VV(\tilde\Psi, A+J)+\BBB(\tilde \Psi, A+J,\Phi)+(\l,\tilde \Psi)+\CC(\tilde\Psi,\phi)}
\Big|_{\phi=0}\;,\label{corr1} \ee
and
\bea &&\CC(\Psi,\phi) := \frac{-ie}{\b^2 L^2|\SS_L|}\sum_{\substack{\kk\in\BBB^{(N)}_{\b,L}\\\pp\in\DD_{\b,L}}}
\phi_{\pp}
\hat{\Psi}^{+}_{\kk+\pp,\s}
C_{M,N}^\e(\kk,\pp)\hat{\Psi}^{-}_{\kk,\s}\;,\label{corr2}\nn\\
&&C^\e_{M,N}(\kk,\pp) = \Big[\big(\chi^\e_{M}(k_0 + p_0)\big)^{-1} - 1\Big] B_N(\kk+\pp)
\G_0(\vec p) - \Big[\big(\chi^\e_{M}(k_0)\big)^{-1} - 1\Big]
\G_0(\vec p) B_N(\kk)\;,\label{1.22} \eea
where $B_N(\kk)=[\hat S_0(d_N(k_0),\vec k)]^{-1}$. The correction
term $\D^\e_{M,N}$ is due to the presence of the imaginary time UV
cut-off in $P_{M,N}(d\tilde\Psi)$, which slightly breaks gauge
invariance. It can be studied by a multiscale analyisis of the
functional integral in Eq.(\ref{corr1}), using ideas similar to
(but much simpler than) those of \cite[Appendix D]{GMP1} (using
methods first developed in \cite{BM0,BM}), where the correction to
the WIs due to a fixed fermionic UV cutoff at momenta of order 1
were studied. Here the main difference is that the UV cutoff is
only on the imaginary time coordinates and it cuts off momenta of
scale $2^M$. By the support properties of the kernel
$C^\e_{M,N}(\kk,\pp)$ all the contributions to
$\D^\e_{M,N}(\kk,\pp)$ have at least one loop momentum flowing at
scale $2^M$. Moreover, since the scaling dimensions in the UV
theory for the time variables are all negative, see \cite[Appendix
C]{GM} and \cite{tesiporta}, it is easy to show that all such
contributions go exponentially to zero as $M,N\to\infty$:
\be\lim_{M\to\infty}\lim_{\e\to0}\lim_{N\to\infty}\D^\e_{M,N}(\pp;\Phi,J,\l)=0\;,\label{goto0}\ee
for all fixed $\pp$, which proves Eq.(\ref{1.23}). More details on the proof of Eq.(\ref{goto0})
can be found in \cite{tesiporta}.

\subsection{Independence of the functional integral on the choice of $\x$ }

In this subsection we show that the averages of gauge
covariant operators do not depend on the choice of the gauge
fixing parameter $\xi$ appearing in the bosonic integration
measure. This allows us to work in the technically convenient {\it
Feynman gauge}, corresponding to the choice $\xi=0$.

Let $F(\Psi,A)$ be a gauge invariant function, i.e., $F(\Psi,A)=F(\Psi e^{ie \a},A+\dpr\a)$.
We want to show that
\be \dpr_\x\frac{\int P(d\Psi)P^{\x,h^*}(dA)e^{\VV(\Psi,A)}F(\Psi,A)}{\int P(d\Psi)P^{\x,h^*}(dA)
e^{\VV(\Psi,A)}}=0\;.\label{invxi}\ee
As in the derivation of the WIs, the proof requires a number of manipulation of the
functional integral. In principle, we should proceed as in the previous subsection, by first
introducing proper UV cutoffs in the time variable, then keep track of the possible correction
terms produced by the manipulations and finally discuss the vanishing of the corrections in
the limit where the UV cutoffs are removed. However, the result is that, once again, these
correction terms vanish in the limit, so here we will neglect them from the very first moment.

With obvious notation, let us denote the gaussian measure in Eq.(\ref{PdA}) by
\be P^{\x,h^*}(d A) =\frac{\DD\! A\, e^{-\frac12(A,(w^{\x,h^*})^{-1}A)}}{\int
\DD\! A\, e^{-\frac12(A,(w^{\x,h^*})^{-1}A)}}\;,\label{PdAshort}\ee
and, given a function of $\Psi,A$, let us define
\be \mathbb{E}_A^\x(\OO(\Psi,A))=\frac{\int \DD\! A\, e^{-\frac12(A,(w^{\x,h^*})^{-1}A)}\OO(\Psi,A)}{\int
\DD\! A\, e^{-\frac12(A,(w^{\x,h^*})^{-1}A)}}\;.\label{Oave}\ee
Note that, letting $\OO_1:=e^{\VV}F$ and $\OO_2:=e^{\VV}$, Eq.(\ref{invxi}) can be rewritten as
\be \dpr_\x \frac{\int P(d\Psi)\mathbb{E}_A^\x(\OO_1(\Psi,A))}{\int P(d\Psi)\mathbb{E}_A^\x(\OO_2
(\Psi,A))}=0\;,\label{rew}
\ee
where $\OO_1$ and $\OO_2$ are both gauge invariant functions. So in order to prove
Eq.(\ref{invxi}) it is enough to prove that
$\int P(d\Psi)\dpr_\x\mathbb{E}_A^\x(\OO(\Psi,A))=0$, with $\OO$ a gauge invariant function.
To this purpose, note that
\be \dpr_\x\mathbb{E}_A^\x(\OO(\Psi,A))=-\frac1{2\b|\SS_L|}
\sum_{\substack{\pp\in\DD_{\b,L}\\ \m,\n=0,1,2}}
\dpr_\x[\hat w^{\x,h^*}(\pp)]^{-1}_{\m\n}\Big[
\mathbb{E}_A^\x(\hat A_{\m,\pp}\hat A_{\n,-\pp}\OO(\Psi,A))
-\hat w^{\x,h^*}_{\m\n}(\pp)\mathbb{E}_A^\x(\OO(\Psi,A))\Big]\;.\label{marf}\ee
Integrating by parts the first term in square brackets, we find:
\bea \dpr_\x\mathbb{E}_A^\x(\OO(\Psi,A))&=&-\frac1{2\b|\SS_L|}
\sum_{\substack{\pp\in\DD_{\b,L}\\ \m,\n,\n_1,\n_2=0,1,2}}
\hat w^{\x,h^*}_{\n_1\m}(\pp)\dpr_\x[\hat w^{\x,h^*}(\pp)]^{-1}_{\m\n}
\hat w^{\x,h^*}_{\n\n_2}(\pp)
\mathbb{E}_A^\x\big(\frac{\dpr^2\OO(\Psi,A)}{\dpr\hat A_{\n_1,\pp}\dpr\hat A_{\n_2,-\pp}}\big)
\nn\\
&=& \frac1{2\b|\SS_L|}
\sum_{\substack{\pp\in\DD_{\b,L}\\ \m,\n=0,1,2}}
\dpr_\x\hat w^{\x,h^*}_{\m\n}(\pp)\mathbb{E}_A^\x\big(\frac{\dpr^2\OO(\Psi,A)}
{\dpr\hat A_{\m,\pp}\dpr\hat A_{\n,-\pp}}\big)\;.\label{merf}\eea
Using the explicit form of $\hat w^{\x,h^*}_{\m\n}(\pp)$, Eq.(\ref{pho_fey7}),
the last expression can be further rewritten as
\bea \dpr_\x\mathbb{E}_A^\x(\OO(\Psi,A))&=& -\frac1{2\b|\SS_L|}
\sum_{\substack{\pp\in\DD_{\b,L}\\ \m,\n=0,1,2}}
\int_{\mathbb{R}}\frac{dp_3}{2\p}\frac{\c(|p|)-\c(2^{-h^*}|\pp|)}{(\pp^2+p_3^2)(|\vec p|^2+p_3^2)}\cdot\label{almth}\\
&&\cdot \big[p_\m p_\n-p_0(p_\m n_\n+p_\n n_\m)\big]
\frac{\dpr^2}{\dpr\hat J_{\m,\pp}\dpr\hat J_{\n,-\pp}}\mathbb{E}_A^\x\big(\OO(\Psi,A+J)\big)
\Big|_{J=0}\;.\nn\eea
On the other hand, by the gauge invariance of $\OO$ and performing the unitary phase
transformation $\Psi^{\pm}_{\xx,\s,\r}\rightarrow\Psi^{\pm}_{\xx,\s,\r}
e^{\pm i e \a(\xx+(\r-1)\dd_1)}$, as in the previous subsection, we find
$$\int P(d\Psi)\mathbb{E}_A^\x\big(\OO(\Psi,A)\big)=
\int P(d\Psi)\mathbb{E}_A^\x\big(\OO(\Psi,A+\dpr\a)\big)\;,$$
which implies
\be \frac{\partial}{\partial\hat\a_{\pp}}\int P(d\Psi)\mathbb{E}_A^\x\big(\OO(\Psi,A+\dpr\a)\big)\Big|_{\a=0}=-i\sum_{\m=0,1,2}\int P(d\Psi)\,p_\m
\frac{\partial}{\partial\hat J_{\m,\pp}}
\mathbb{E}_A^\x\big(\OO(\Psi,A+J)\big)\Big|_{J=0}=0\;.\label{1.28} \ee
Integrating Eq.(\ref{almth}) with respect to $\int P(d\Psi)$ and using Eq.(\ref{1.28}) finally
gives the desired cancellation, $$\int P(d\Psi) \dpr_\x\mathbb{E}_A^\x(\OO(\Psi,A))=0\;.$$
%

\section{Symmetry transformations}\label{app1}
\setcounter{equation}{0}
\renewcommand{\theequation}{\ref{app1}.\arabic{equation}}

In this Appendix we prove that both the gaussian integrations and
the effective potentials are invariant under the symmetries (1)--(8) listed in
Section \ref{sec4B}. As already done in Section \ref{sec3}, we restrict for simplicity
to the case that $J=\l=0$ and that all the external fields $\Phi^a$ but the one with $a=K$ are set
to 0. The invariance of the effective potentials on scale $h$ and of the single-scale integrations
under the stated transformations follows from the fact that the bare interaction and source terms,
Eqs.(\ref{1.15uy})-(\ref{1.16}), and all the single scale integrations $P(d\Psi^{(h)})P(dA^{(h)})$
are separately invariant under the same transformations. Moreover, since the single scale
integrations are obtained from the bare integration $P(d\Psi)P^{0,h^*}(dA)$ by recursively
including the local terms $\LL_\psi\VV^{(h)}$ (which  is invariant under
the symmetries (1)--(8), see Eq.(\ref{local2}) and \cite[Proof of Lemma 1]{GMP1}), one can immediately convince oneself that the
desired invariance properties of the effective potentials at all scales follow from the invariance of
the bare interaction, bare source term and bare integration under the analogue of the symmetries
(1)--(8) written in terms of the fields $\hat\Psi^\pm_{\kk,\s}$ (rather then in terms of the
quasi-particle fields $\hat\Psi^\pm_{\kk',\s,\o}$, as done in Section \ref{sec4B}).
The symmetries (1)--(8), if rewritten in terms of  $\hat\Psi^\pm_{\kk,\s}$, read as follows.

\begin{enumerate}
\item[(1)] \underline{Spin flip}:
$\hat\Psi^{\varepsilon}_{\kk,\s}\to
\hat\Psi^{\varepsilon}_{\kk,-\s}$, and $A_{\m,\pp},\hat\Phi^{K}_{j,\pp}$ are left invariant;
\item[(2)] \underline{Global $U(1)$}:
$\hat\Psi^{\varepsilon}_{\kk,\s}\rightarrow e^{ i \varepsilon
\a_{\s}}\hat\Psi^{\varepsilon}_{\kk,\s}$, with $\a_{\s}\in \RRR$
independent of $\kk$, and $A_{\m,\pp},\hat\Phi^{K}_{j,\pp}$ are left invariant;
\item[(3)] \underline{Spin SO(2)}: $\begin{pmatrix}
\hat\Psi^{\e}_{\kk,\uparrow,\cdot}
\\ \hat\Psi^{\e}_{\kk,\downarrow,\cdot} \end{pmatrix} \rightarrow e^{i\th\s_{2}}\begin{pmatrix}
\hat\Psi^{\e}_{\kk,\uparrow,\cdot} \\
\hat\Psi^{\e}_{\kk,\downarrow,\cdot} \end{pmatrix}$, with $\th$ independent of $\kk$,
and $A_{\m,\pp},\hat\Phi^{K}_{j,\pp}$ are left invariant;
\item[(4)] \underline{Discrete spatial rotations}: if $T\kk = (k_0,e^{-i\frac{2\pi}{3}\s_2}\vec k)$
and $n_-=(1-\s_3)/2$,
\be
\hat \Psi^{-}_{\kk,\s}\to e^{i\kk(\dd_{3} - \dd_{1})n_-}\hat\Psi^{-}_{T\kk,\s}\;,\quad \hat\Psi^{+}_{\kk,\s}\to
\hat\Psi^{+}_{T\kk,\s} e^{-i\kk(\dd_3 - \dd_1)n_-}\;,\quad \hat A_{\pp}\to T^{-1}\hat A_{T\pp}\;,\quad \hat\Phi^{K}_{j,\pp}\to \hat\Phi^{K}_{j+1,T\pp}\;;\nn
\ee
\item[(5)] \underline{Complex conjugation}: if $c$ is a generic constant appearing in the effective
potentials or in the gaussian integrations:
\be c\to c^{*}\;,\qquad \hat\Psi^{\varepsilon}_{\kk,\s}\rightarrow
\hat\Psi^{\varepsilon}_{-\kk,\s}\;,\qquad \hat A_{\pp}\rightarrow -\hat
A_{-\pp}\;,\qquad \hat\Phi^{K}_{j,\pp}\to \hat\Phi^{K}_{j,-\pp}\;;\nn\ee
\item[(6.a)] \underline{Horizontal reflections}: if $R_{h}\kk = (k_0,-k_1,k_2)$ and
$r_h 1 = 1$, $r_h 2 = 3$, $r_h 3 = 2$,
\be
\hat\Psi^{-}_{\kk,\s}\to \s_{1}\hat\Psi^{-}_{R_h\kk,\s}\;,\qquad \hat\Psi^{+}_{\kk,\s}\to \hat \Psi^{+}_{R_h\kk,\s}\s_1\;,\qquad \hat A_{\pp}\to R_{h}\hat A_{R_{h}\pp} e^{i\pp\dd_{1}}\;,\qquad
\hat\Phi^{K}_{j,\pp}\to \hat\Phi^{K}_{r_h j,R_h\pp}e^{-i\pp(\dd_j - \dd_1)}\;,\nn\ee
\item[(6.b)] \underline{Vertical reflections}: if $R_{v}\kk = (k_0,k_1,-k_2)$ and
$r_v 1 = 1$, $r_v 2 = 3$, $r_v 3 = 2$,
\be \hat\Psi^{\e}_{\kk,\s}\to\hat\Psi^{\e}_{R_v\kk,\s}\;,\qquad \hat
A_{\pp}\to R_{v}\hat A_{R_{v}\pp}\;,\qquad \hat\Phi^{K}_{j,\pp}\to \hat\Phi^{K}_{r_v j, R_v\pp}\;;\nn
\ee
\item[(7)] \underline{Particle-hole}: if $P\kk=(k_0,-\vec k)$,
\be\hat\Psi^{\varepsilon}_{\kk,\s}\rightarrow i\hat\Psi^{-\varepsilon}_{P\kk,\s}\;,\qquad
\hat A_{\pp}\to P\hat A_{-P\pp}\;,\qquad \hat\Phi^{K}_{j,\pp}\to \hat\Phi^{K}_{j,-P\pp}\;,\nn\ee
\item[(8)] \underline{Time-reversal}: if $I\kk=(-k_0,\vec k)$,
\be\hat\Psi^{-}_{\kk,\s}\to
-i\s_{3}\hat\Psi^{-}_{I\kk,\s}\;,\qquad \hat\Psi^{+}_{\kk,\s}\to
-i\hat\Psi^{+}_{I\kk,\s}\s_3\;,\qquad  \hat A_{\pp}\to I\hat A_{I\pp}\;,\qquad \hat \Phi^{K}_{j,\pp}\to
\hat\Phi^{K}_{j,I\pp}\;.\nn\ee
\end{enumerate}
Now, the proof of the fact that $P(d\Psi)$ is invariant under (1)--(8) has already been discussed
in \cite[Section 3.1]{GM}. The invariance of $P^{0,h^*}(dA)$ under (1)--(8) is obvious, and so is the
invariance of $\VV(\Psi,A)$ and $\BBB(\Psi,A,\Phi)$ under (1)--(3). Therefore, we are left with
proving the invariance of  $\VV(\Psi,A)$ and $\BBB(\Psi,A,\Phi)$ under (4)--(8). As a preliminary
step, let us rewrite the interaction and source term in momentum space:
\bea \VV(\Psi, A)&=&\frac{1}{\b^2 L^2|\SS_L|}\sum_{\kk,\pp,\s}
\Big\{t\sum_{j=1}^3\sum_{n\ge 1}\frac{(ie)^n}{n!}
\big[\h^j_{\cdot}(\vec\d_j\cdot\vec A_{\cdot})\big]^{*n}_\pp
\hat \Psi^+_{\kk+\pp,\s}\Biggl(\begin{matrix}0& e^{-i\kk(\dd_j-\dd_1)}\\
\,(-1)^ne^{i(\kk+\pp)(\dd_j-\dd_1)}&0\end{matrix}\Biggr)
\hat\Psi^-_{\kk,\s}\nn\\
&&\hskip2.4truecm-ie\hat A_{0,\pp}\hat \Psi^+_{\kk+\pp,\s}
\G_0(\pp)\hat \Psi^-_{\kk,\s}\Big\}\;,\label{intmom}\\
\BBB(\Psi,A,\Phi)&=&\frac{1}{\b^3 L^2|\SS_L|^2}\sum_{\kk,\pp,\qq,\s}
\sum_{j=1}^3\hat \Phi^K_{j,\qq}
\sum_{n\ge 0}\frac{(ie)^n}{n!}
\big[\h^j_{\cdot}(\vec\d_j\cdot\vec A_{\cdot})\big]^{*n}_\pp
\hat \Psi^+_{\kk+\pp+\qq,\s}\Biggl(\begin{matrix}0& e^{-i\kk(\dd_j-\dd_1)}\\
\,(-1)^ne^{i(\kk+\pp+\qq)(\dd_j-\dd_1)}&0\end{matrix}\Biggr)
\hat\Psi^-_{\kk,\s}\;,\nn\eea
where, if $\h^j_\pp=(1-e^{-i\pp\dd_j})/(i\pp\dd_j)$,
\be \big[\h^j_{\cdot}(\vec\d_j\cdot\vec A_{\cdot})\big]^{*n}_\pp:=\frac1{(\b|\SS_L|)^{n-1}}\sum_{\pp_1+\cdots+\pp_n=\pp}
\h^j_{\pp_1}(\vec\d_j\cdot\vec A_{\pp_1})\cdots \h^j_{\pp_n}
(\vec\d_j\cdot\vec A_{\pp_n})\;.\label{adeltaj}\ee
and we recall that $\G_0(\vec p)=\Biggl(\begin{matrix} 1&0\\0&e^{-i\pp\dd_1}\end{matrix}\Biggr)$.
In the second of Eq.(\ref{intmom}), $\big[\h^j_{\cdot}(\vec\d_j\cdot\vec A_{\cdot})\big]^{*0}_\pp$
should be interpreted as equal to $\b|\SS_L|\d_{\pp,\V0}$.
Let us neglect the spin index, which plays no role in the following, and let us denote by
\be (*):=t\sum_{\kk,\pp}
\sum_{j=1}^3\sum_{n\ge 1}\frac{(ie)^n}{n!}
\big[\h^j_{\cdot}(\vec\d_j\cdot\vec A_{\cdot})\big]^{*n}_\pp
\hat \Psi^+_{\kk+\pp}\Biggl(\begin{matrix}0& e^{-i\kk(\dd_j-\dd_1)}\\
\,(-1)^ne^{i(\kk+\pp)(\dd_j-\dd_1)}&0\end{matrix}\Biggr)
\hat\Psi^-_{\kk}\;,\label{first}\ee
the term in the first line of Eq.(\ref{intmom}) and, similarly,
\bea &&(**):=-ie\sum_{\kk,\pp}\hat A_{0,\pp}\hat \Psi^+_{\kk+\pp}
\G_0(\pp)\hat \Psi^-_{\kk}\;,\label{second}\\
&&(K):=\sum_{\kk,\pp,\qq}\sum_{j=1}^3\hat \Phi^K_{j,\qq}
\sum_{n\ge 0}\frac{(ie)^n}{n!}
\big[\h^j_{\cdot}(\vec\d_j\cdot\vec A_{\cdot})\big]^{*n}_\pp
\hat \Psi^+_{\kk+\pp+\qq}\Biggl(\begin{matrix}0& e^{-i\kk(\dd_j-\dd_1)}\\
\,(-1)^ne^{i(\kk+\pp)(\dd_j-\dd_1)}&0\end{matrix}\Biggr)
\hat\Psi^-_{\kk}\;.\label{third}\eea
Let us prove that these terms are separately invariant under the symmetries (4)--(8).
\paragraph*{Symmetry (4).} The term $(*)$ is changed under (4) as
\be (*)\to t\sum_{\kk,\pp}
\sum_{j=1}^3\sum_{n\ge 1}\frac{(ie)^n}{n!}
\big[\h^{j+1}_{\cdot}(\vec\d_j\cdot(T^{-1}\vec A_{\cdot}))\big]^{*n}_{T\pp}
\hat \Psi^+_{T(\kk+\pp)}\Biggl(\begin{matrix}0& e^{-i\kk(\dd_j-\dd_3)}\\
\,(-1)^ne^{i(\kk+\pp)(\dd_j-\dd_3)}&0\end{matrix}\Biggr)
\hat\Psi^-_{T\kk}\;,\label{bb4.1}\ee
where we used that $\h^j_{\pp}=\h^{j+1}_{T\pp}$. Using the fact that
$\big[\h^{j+1}_{\cdot}(\vec\d_j\cdot(T^{-1}\vec A_{\cdot}))\big]^{*n}_{T\pp}=
\big[\h^{j+1}_{\cdot}(\vec\d_{j+1}\cdot\vec A_{\cdot})\big]^{*n}_{T\pp}$ and
$\pp\cdot\dd_j=(T\pp)\cdot\dd_{j+1}$, we see that the r.h.s. of Eq.(\ref{bb4.1}) can be rewritten as
\be t \sum_{\kk,\pp}
\sum_{j=1}^3\sum_{n\ge 1}\frac{(ie)^n}{n!}
\big[\h^{j+1}_{\cdot}(\vec\d_{j+1}\cdot\vec A_{\cdot})\big]^{*n}_{T\pp}
\hat \Psi^+_{T(\kk+\pp)}\Biggl(\begin{matrix}0& e^{-iT\kk(\dd_{j+1}-\dd_1)}\\
\,(-1)^ne^{iT(\kk+\pp)(\dd_{j+1}-\dd_1)}&0\end{matrix}\Biggr)
\hat\Psi^-_{T\kk}\;,\label{bb4.2}\ee
which is the same as $(*)$, as apparent by the change of variables $(T\kk,T\pp)\to(\kk,\pp)$
in the sum. The proof of the invariance of the term $(K)$ under (4) is
exactly the same. Regarding the term $(**)$, it is changed as
\be (**)\to -ie\sum_{\kk,\pp}\hat A_{0,T\pp}\hat \Psi^+_{T(\kk+\pp)}
\Biggl(\begin{matrix}1&0\\0&e^{-i\pp\dd_3}\end{matrix}\Biggr)\hat \Psi^-_{T\kk,\s}\;,\label{bb4.3}\ee
which is the same as $(**)$, simply because $\Biggl(\begin{matrix}1&0\\0&e^{-i\pp\dd_3}\end{matrix}\Biggr)
=\G_0(T\pp)$ and we can perform the change of variables $(T\kk,T\pp)\to(\kk,\pp)$ in the sum.
\paragraph*{Symmetry (5).} The term $(*)$ is changed as
\be (*)\to t\sum_{\kk,\pp}
\sum_{j=1}^3\sum_{n\ge 1}\frac{(-ie)^n}{n!}
\big[\h^j_{\cdot}(\vec\d_j\cdot(-\vec A_{\cdot}))\big]^{*n}_{-\pp}
\hat \Psi^+_{-\kk-\pp}\Biggl(\begin{matrix}0& e^{+i\kk(\dd_j-\dd_1)}\\
\,(-1)^ne^{-i(\kk+\pp)(\dd_j-\dd_1)}&0\end{matrix}\Biggr)
\hat\Psi^-_{-\kk}\;,\label{bb4.4}\ee
where we used that $(\h^j_\pp)^*=\h^j_{-\pp}$. The r.h.s. of Eq.(\ref{bb4.4}) is the same
as $(*)$, as apparent by the change of variables $(\kk,\pp)\to(-\kk,-\pp)$. The proof of the
invariance of the term $(K)$ under (5) is exactly the same. Moreover, the term $(**)$ is changed as
\be (**)\to+ie\sum_{\kk,\pp}(-\hat A_{0,-\pp})\hat \Psi^+_{-\kk-\pp}
\G_0^*(\pp)\hat \Psi^-_{-\kk}\;,\label{1.36} \ee
which is the same as $(**)$, because $\G^*_0(\pp)=\G_0(-\pp)$
and we can perform the change of variables $(\kk,\pp)\to(-\kk,-\pp)$ in the sum.
\paragraph*{Symmetry (6.a).} Using the fact that $R_h\dd_j=-\dd_{r_hj}$ and $\h^j_\pp=
e^{-i\pp\dd_j}\h^{r_hj}_{R_h\pp}$,
the term $(*)$ is changed as
\be (*)\to t\sum_{\kk,\pp}
\sum_{j=1}^3\sum_{n\ge 1}\frac{(ie)^n}{n!}
\big[\h^{r_hj}_{\cdot}(-\vec\d_{r_hj}\cdot\vec A_{\cdot})\big]^{*n}_{R_h\pp}
e^{-i\pp(\dd_j-\dd_1)}
\hat \Psi^+_{R_h(\kk+\pp)}\s_1\Biggl(\begin{matrix}0& e^{-i\kk(\dd_j-\dd_1)}\\
\,(-1)^ne^{i(\kk+\pp)(\dd_j-\dd_1)}&0\end{matrix}\Biggr)\s_1
\hat\Psi^-_{R_h\kk}\;,\label{bb4.5}\ee
which can be rewritten as
\be t\sum_{\kk,\pp}
\sum_{j=1}^3\sum_{n\ge 1}\frac{(ie)^n}{n!}
\big[\h^{r_hj}_{\cdot}(\vec\d_{r_hj}\cdot\vec A_{\cdot})\big]^{*n}_{R_h\pp}
\hat \Psi^+_{R_h(\kk+\pp)}\Biggl(\begin{matrix}0& e^{i\kk(\dd_j-\dd_1)}\\
\,(-1)^ne^{-i(\kk+\pp)(\dd_j-\dd_1)}&0\end{matrix}\Biggr)
\hat\Psi^-_{R_h\kk}\;,\label{bb4.6}\ee
which is the same as $(*)$, simply because $e^{i\kk(\dd_j-\dd_1)}=e^{-iR_h\kk(\dd_{r_hj}-\dd_1)}$
and we can perform the change of variables $(R_h\kk,R_h\pp)\to(\kk,\pp)$ in the sum.
The proof of the
invariance of the term $(K)$ under (5) is exactly the same. Moreover, the term $(**)$ is changed as
\be (**)\to -ie\sum_{\kk,\pp}\hat A_{0,R_h\pp}e^{i\pp\dd_1}\hat \Psi^+_{R_h(\kk+\pp)}\s_1
\G_0(\pp)\s_1\hat \Psi^-_{R_h\kk}\;,\label{bb4.7}\ee
which is the same as $(**)$, because $\s_1\G_0(\pp)\s_1=e^{-i\pp\dd_1}\G_0(R_h\pp)$
and we can perform the change of variables $(R_h\kk,R_h\pp)\to(\kk,\pp)$ in the sum.
\paragraph*{Symmetry (6.b).} Using the fact that $R_v\dd_j=\dd_{r_vj}$
and $\h^j_\pp=\h^{r_vj}_{R_v\pp}$, the term $(*)$ is changed as
\be (*)\to t\sum_{\kk,\pp}
\sum_{j=1}^3\sum_{n\ge 1}\frac{(ie)^n}{n!}
\big[\h^{r_vj}_{\cdot}(\vec\d_{r_vj}\cdot\vec A_{\cdot})\big]^{*n}_{R_v\pp}
\hat \Psi^+_{R_v(\kk+\pp)}\Biggl(\begin{matrix}0& e^{-i\kk(\dd_j-\dd_1)}\\
\,(-1)^ne^{i(\kk+\pp)(\dd_j-\dd_1)}&0\end{matrix}\Biggr)
\hat\Psi^-_{R_v\kk}\;,\label{bb4.8}\ee
which is the same as $(*)$, simply because $e^{-i\kk(\dd_j-\dd_1)}=e^{-iR_v\kk(\dd_{r_vj}-\dd_1)}$
and we can perform the change of variables $(R_v\kk,R_v\pp)\to(\kk,\pp)$ in the sum.
The proof of the
invariance of the term $(K)$ under (5) is exactly the same. Moreover, the term $(**)$ is changed as
\be -ie\sum_{\kk,\pp}\hat A_{0,R_v\pp}\hat \Psi^+_{R_v(\kk+\pp)}
\G_0(\pp)\hat \Psi^-_{R_v\kk}\;,\label{bb4.9}\ee
which is the same as $(**)$, because $\G_0(\pp)=\G_0(R_v\pp)$
and we can perform the change of variables $(R_v\kk,R_v\pp)\to(\kk,\pp)$ in the sum.
\paragraph*{Symmetry (7).} The term $(*)$ changes as
\bea  (*)&\to& t\sum_{\kk,\pp}
\sum_{j=1}^3\sum_{n\ge 1}\frac{(ie)^n}{n!}
\big[\h^j_{\cdot}(-\vec\d_j\cdot\vec A_{\cdot})\big]^{*n}_{-P\pp}
\hat \Psi^+_{P\kk}\Biggl(\begin{matrix}0& e^{-i\kk(\dd_j-\dd_1)}\\
\,(-1)^ne^{i(\kk+\pp)(\dd_j-\dd_1)}&0\end{matrix}\Biggr)^T
\hat\Psi^-_{P(\kk+\pp)}=\nn\\
&=&t\sum_{\kk,\pp}
\sum_{j=1}^3\sum_{n\ge 1}\frac{(ie)^n}{n!}
\big[\h^j_{\cdot}(\vec\d_j\cdot\vec A_{\cdot})\big]^{*n}_{-P\pp}
\hat \Psi^+_{P\kk}\Biggl(\begin{matrix}0& e^{-iP(\kk+\pp)(\dd_j-\dd_1)}\\
\,(-1)^ne^{iP\kk(\dd_j-\dd_1)}&0\end{matrix}\Biggr)
\hat\Psi^-_{P(\kk+\pp)}\;,\label{bb4.10}\eea
which is equal to $(*)$, as apparent after the change of variables $\big(P(\kk+\pp),-P\pp)\to
(\kk,\pp)$ in the sum.
The proof of the
invariance of the term $(K)$ under (5) is exactly the same. Moreover, the term $(**)$ is changed as
\be (**)\to -ie\sum_{\kk,\pp}\hat A_{0,-P\pp}\hat \Psi^+_{P\kk}
\G^T_0(\pp)\hat \Psi^-_{P(\kk+\pp)}\;,\label{bb4.11}\ee
which is the same as $(**)$, because $\G_0^T(\pp)=\G_0(\pp)=\G_0(-P\pp)$
and we can perform the change of variables $\big(P(\kk+\pp),-P\pp)\to
(\kk,\pp)$ in the sum.
\paragraph*{Symmetry (8).} The term $(*)$ changes as
\bea (*) &\to& -t\sum_{\kk,\pp}
\sum_{j=1}^3\sum_{n\ge 1}\frac{(ie)^n}{n!}
\big[\h^j_{\cdot}(\vec\d_j\cdot\vec A_{\cdot})\big]^{*n}_{I\pp}
\hat \Psi^+_{I(\kk+\pp)}\s_3\Biggl(\begin{matrix}0& e^{-i\kk(\dd_j-\dd_1)}\\
\,(-1)^ne^{i(\kk+\pp)(\dd_j-\dd_1)}&0\end{matrix}\Biggr)\s_3
\hat\Psi^-_{I\kk}=\nn\\
&=&t\sum_{\kk,\pp}
\sum_{j=1}^3\sum_{n\ge 1}\frac{(ie)^n}{n!}
\big[\h^j_{\cdot}(\vec\d_j\cdot\vec A_{\cdot})\big]^{*n}_{I\pp}
\hat \Psi^+_{I(\kk+\pp)}\Biggl(\begin{matrix}0& e^{-i\kk(\dd_j-\dd_1)}\\
\,(-1)^ne^{i(\kk+\pp)(\dd_j-\dd_1)}&0\end{matrix}\Biggr)\hat\Psi^-_{I\kk}
\;,\label{1.45} \eea
which is equal to $(*)$, because $e^{-i\kk(\dd_j-\dd_1)}=e^{-iI\kk(\dd_j-\dd_1)}$
and we can perform the change of variables $(I\kk,I\pp)\to(\kk,\pp)$ in the sum.
The proof of the
invariance of the term $(K)$ under (5) is exactly the same. Moreover, the term $(**)$ is changed as
\be (**)\to -ie\sum_{\kk,\pp}\hat A_{0,I\pp}\hat \Psi^+_{I(\kk+\pp)}\s_3
\G_0(\pp)\s_3\hat \Psi^-_{I\kk}\;,\label{bb4.12}\ee
which is the same as $(**)$, because $\s_3\G_0(\pp)\s_3=\G_0(\pp)=\G_0(I\pp)$
and we can perform the change of variables $(I\kk,I\pp)\to(\kk,\pp)$
in the sum. This concludes the proof of the invariance properties stated
in Section \ref{sec4B}.

\section{Symmetry properties of the kernels}\label{app3}
\setcounter{equation}{0}
\renewcommand{\theequation}{\ref{app3}.\arabic{equation}}

In this Appendix we exploit the lattice symmetries (1)--(8) listed in Section \ref{sec4B}
to prove the invariance properties Eqs.(\ref{symbad})-(\ref{local2}) of the local terms in the effective action
stated in Section \ref{sec4C}. We will start with the ``relevant'' terms with $2n+m+p=2$ (i.e.,
the terms of the form $AA$, $\Phi^KA$ or $\Psi^+\Psi^-$)
and we will then proceed with the
``marginal'' terms with $2n+m+p=3$ (i.e., the
terms of the form $A\Psi^+\Psi^-$, $\Phi^K\Psi^+\Psi^-$, $AAA$, $\Phi^KAA$ or $\Phi^K\Phi^KA$).
In the following, we shall drop all the unnecessary labels (including the scale and spin labels),
to avoid an overwhelming notation. We will think the operators $T,R_v,R_h,P,I$ appearing in the
symmetry transformations as $3\times3$ matrices acting on the $\m$-indices, with
\be
T=\begin{pmatrix}
1&0&0\\
0& \cos\frac{2\p}{3}&-\sin\frac{2\p}{3}\\
0& \sin\frac{2\p}{3}&\cos\frac{2\p}{3}\end{pmatrix}\;,\qquad
R_h=\begin{pmatrix}
1&0&0\\
0& -1&0\\
0&0&1\end{pmatrix}\;,\qquad
R_v=\begin{pmatrix}
1&0&0\\
0& 1&0\\
0&0&-1\end{pmatrix}\;,\qquad
P=-I=\begin{pmatrix}
1&0&0\\
0& -1&0\\
0&0&-1\end{pmatrix}\;.\label{TRPI}\ee
\subsection{The ``relevant'' terms}
The structure of the local terms of the form $\Psi^+\Psi^-$ has already been
studied in \cite[Lemma 3]{GM}, where the first equation of Eq.(\ref{local2})
was proved. Let us then look at the terms of the form $AA$ or $\Phi^KA$.
\\
\paragraph*{The $AA$ kernels.} The contribution to the effective potential
quadratic in the $A$ field is proportional to
\be \sum_\pp \hat A_{\pp} \hat
W(\pp)\hat A_{-\pp}\;,\label{A3.0} \ee
for a suitable $3\times 3$ matrix-valued kernel $\hat W(\pp)$. Imposing the invariance of Eq.(\ref{A3.0})
under symmetries (4)--(8), we find:
\be \hat W(\pp) \os{(4)}{=} T\hat W(T^{-1}\pp)T^{-1} \os{(5)}{=}
[\hat W(-\pp)]^{*} \os{(6.a)}{=} R_{h}\hat W(R_{h}\pp)R_{h}
\os{(6.b)}{=} R_{v}\hat W(R_{v}\pp)R_{v} \os{(8)}{=} I\hat
W(I\pp)I\;.\label{A3.1} \ee
Let $\hat W_{\m\n}(\V0) =: \n_{\m\n}$ and $\partial_{\a}\hat
W_{\m\n}(\V0) =: \n'_{\m\n\a}$; the properties (6.a), (6.b) and
(8) in Eq.(\ref{A3.1}) imply that $\n_{\m\n} = \d_{\m\n}\n_{\m}$, $\n'_{\m\n\a}=0$,
while (5) in Eq.(\ref{A3.1}) gives $\n_{\m} = \n_{\m}^{*}$; finally, the property (4) in Eq.(\ref{A3.1})
implies that $\n_{1} = \n_{2}$. This proves the second equation in Eq.(\ref{local2}).\\

\paragraph*{The $\Phi^KA$ kernels.}
The contribution to the effective potential of the form $\Phi^KA$ is proportional to
\be \sum_j\sum_\pp \hat\Phi^K_{j,\pp}\hat W^K_{j}(\pp)\hat A_{-\pp}\;,\label{A3.2a} \ee
for a suitable vector-valued kernel $\hat W^{K}_{j}(\pp) = \big(\hat W^{K}_{j,0}(\pp),\,
\hat W^{K}_{j,1}(\pp),\,\hat W^{K}_{j,2}(\pp)\big)$. Using the symmetries (7) and (8) we get:
\be
\hat W^{K}_{j}(\pp) \os{(7)}{=} \hat W^{K}_{j}(-P\pp)P \os{(8)}{=} \hat W^{K}_{j}(I\pp)I \;,\label{bb5.1}
\ee
which implies that $\hat W^K_j(\pp)=0$, because $P=-I$.
%
%
\subsection{The ``marginal'' terms}
In this subsection we study the structure of the terms of the form
$A\Psi^+\Psi^-$, $\Phi^K\Psi^+\Psi^-$, $AAA$, $\Phi^KAA$ or $\Phi^K\Phi^KA$.\\
\paragraph*{The $A\Psi^{+}\Psi^{-}$ kernels.} Since $A$ has an UV cutoff that suppresses modes
$\hat A_{\pp}$ with $\pp$ close to $\pm(\pp_F^+-\pp_F^-)$ (and to its images over $\L^*$),
the only non zero terms of the form $A\Psi^{+}\Psi^{-}$ are those with the two
fermi fields associated to the same $\o$ index. [Of course, if we were interested in studying
the theory without the UV cutoff on the photon field, then terms of the form $A\Psi^{+}_{\o}
\Psi^{-}_{-\o}$ would be allowed, and their structure could be investigated by the same methods
used here.]
The contribution to the effective potential of the form $A\Psi^{+}_{\o}\Psi^{-}_{\o}$ is
proportional to
\be \sum_{\kk',\pp,\o}\hat\Psi^{+}_{\kk'+\pp,\o}\big[\hat W_{\o}(\kk',\pp)\cdot\hat A_{\pp}\big]
\hat\Psi^{-}_{\kk',\o}\;,\label{A3.6}
\ee
for a suitable tensor-valued kernel $\hat W_{\o}(\kk',\pp)=(\hat W_{0,\o}(\kk',\pp),
\hat W_{1,\o}(\kk',\pp),\hat W_{2,\o}(\kk',\pp)$ (each component $\hat W_{\m,\o}(\kk',\pp)$
is a $2\times2$ matrix, acting on the $\r$ indices of the grassmann fields).
The invariance under the symmetries (4)--(8) implies:
\bea &&\hat W_{\o}(\kk',\pp) \os{(4)}{=} e^{-i(\pp_F^\o+\kk'+\pp)(\ddd_1 - \ddd_2)n_-}[\hat W_{\o}(T^{-1}\kk',T^{-1}\pp)T^{-1} ]e^{i(\pp_F^\o+\kk')(\ddd_1 -
\ddd_2)n_-} \os{(5)}{=} -\hat W_{-\o}^{*}(-\kk',-\pp) \os{(6.a)}{=}
\label{A3.7}\\&&\os{(6.a)}{=} e^{-i\pp\ddd_1}\s_{1}[\hat W_{\o}(R_h \kk', R_h
\pp)R_{h}]\s_{1} \os{(6.b)}{=} \hat W_{-\o}(R_v \kk', R_v \pp)R_{v}
\os{(7)}{=} \hat W_{-\o}^{T}(P(\kk'+\pp), - P\pp)P \os{(8)}{=}
-\s_{3}[\hat W_{\o}(I\kk', I\pp)I]\s_{3}\;.\nn\eea
Let us define
\be
\hat W_{\m,\o}(\V0,\V0) = \sum_{\n=0}^3a^{\n}_{\m,\o}\s_{\n}\;,\qquad {\mathbf a}^{\n}_{\o} =
(a_{0,\o}^{\n},\,a_{1,\o}^{\n},\,a_{2,\o}^{\n})\;,\ee
with $\s_0=\openone$.
The properties (\ref{A3.7}) imply that (summation over repeated indices is implied):
\bea
{\mathbf a}^{\n}_{\o}\s_{\n} &\os{(4)}{=}& [{\mathbf a}^{\n}_{\o}T^{-1}]e^{-i \pp^{\o}_{F}(\ddd_1 - \ddd_2)n_-}\s_{\n}e^{i\pp_{F}^{\o}(\ddd_1 - \ddd_2)n_-}\;,\label{A3.8}\\
{\mathbf a}^{\n}_{\o}\s_\n &\os{(5)}{=}& -[\mathbf{a}^{0}_{-\o}]^*\s_0 -[\mathbf{a}^{1}_{-\o}]^*\s_1
+ [\mathbf{a}^{2}_{-\o}]^*\s_2 - [\mathbf{a}^{3}_{-\o}]^*\s_3\;,\label{A3.8-5}\\
{\mathbf a}^{\n}_{\o}\s_\n &\os{(6.a)}{=}& [{\mathbf a}^{0}_{\o}R_h]\s_0 + [{\mathbf a}^{1}_{\o}R_h]\s_1 - [{\mathbf a}^{2}_{\o}R_h]\s_2 - [{\mathbf a}^{3}_{\o}R_h]\s_3\;,\label{A3.8-6.a}\\
{\mathbf a}^{\n}_{\o}\s_\n &\os{(6.b)}{=}& [{\mathbf a}^{\n}_{-\o}R_v]\s_\n\;,\label{A3.8-6.b}\\
{\mathbf a}^{\n}_{\o}\s_\n &\os{(7)}{=}& [{\mathbf a}^{0}_{-\o}P]\s_0 + [{\mathbf a}^{1}_{-\o}P]\s_1 - [{\mathbf a}^{2}_{-\o}P]\s_2 + [{\mathbf a}^{3}_{-\o}P]\s_3\;.\label{A3.7c}\\
{\mathbf a}^{\n}_{\o}\s_\n &\os{(8)}{=}& -[{\mathbf a}^{0}_{\o}I]\s_0 + [{\mathbf a}^{1}_{\o}I]\s_1 + [{\mathbf a}^{2}_{\o}I]\s_2 - [{\mathbf a}^{3}_{\o}I]\s_3\;.\label{A3.7b} \eea
We recall that $T,R_h,R_v,P$ and $I$ are the matrices in Eq.(\ref{TRPI}); the notation, e.g.,
$[{\bf a}T^{-1}]$ indicates the row 3-vector obtained by matrix multiplication of the
row 3-vector ${\bf a}$ times the $3\times3$ matrix $T^{-1}$.
Properties Eqs.(\ref{A3.8-6.a}) and (\ref{A3.7b}) imply that
\be {\mathbf a}^{0}_{\o} = (a^{0}_{0,\o},\,0,\,0)\;,\;\;
{\mathbf a}^{1}_{\o} = (0,\,0,\,a^{1}_{2,\o})\;,\;\;
{\mathbf a}^{2}_{\o} = (0,\,a^{2}_{1,\o},\,0)\;,\;\;
{\mathbf a}^{3}_{\o} = (0,\,0,\,0)\;,\label{A3.7d} \ee
while from Eq.(\ref{A3.8-6.b}) we get that
\be a^{0}_{0,\o} = a^{0}_{0,-\o} =: -i\l_{0}\;,\quad a^{1}_{2,\o} = -a^{1}_{2,-\o} =: -\o \l_{2}\;,\quad a^{2}_{1,\o} = a^{2}_{1,-\o} =: -\l_{1}\;,\label{A3.7e} \ee
with $\l_{\m}\in \RRR$, thanks to Eq.(\ref{A3.8-5}). Therefore,
\be
\hat W_{0,\o}(\V0,\V0) = -i\l_0\s_0 =i\l_0\G^0_\o\;,\qquad \hat W_{1,\o}(\V0,\V0) = -\l_1\s_2=
i\l_1\G^1_\o\;,\qquad \hat W_{2,\o}(\V0,\V0) = -\l_2\o\s_1=i\l_2\G^2_\o\;.\label{A3.7f}
\ee
Moreover, thanks to Eq.(\ref{A3.8}):
\be {\mathbf a}^{1}_{\o}\s_{1} + {\mathbf a}^{2}_{\o}\s_2 =
[{\mathbf a}^{1}_{\o}T^{-1}]\begin{pmatrix} 0 & e^{+i\frac{2\pi}{3}\o} \\
e^{-i\frac{2\pi}{3}\o} & 0 \end{pmatrix} +
[{\mathbf a}^{2}_{\o}T^{-1}]\begin{pmatrix} 0 & -ie^{+i\frac{2\pi}{3}\o} \\ i
e^{-i\frac{2\pi}{3}\o} & 0 \end{pmatrix}\;,\label{A3.15} \ee
which gives $a^{1}_{2,\o} = \o a^{2}_{1,\o}$, i.e., $\l_{1} = \l_2$. This proves the
equation in the first line of Eq.(\ref{symbad}).\\

\paragraph*{The $\Phi^K\Psi^{+}\Psi^{-}$ kernels.} The contribution to the effective potential of the form $\Phi^K\Psi^{+}_{\o}\Psi^{-}_{\o'}$ is proportional to
\be \sum_{\substack{\kk',\pp\\
j,\o,\o'}}\hat\Phi^{K}_{j,\pp' + \pp_{F}^{\o} - \pp_{F}^{\o'}}\Psi^{+}_{\kk'+\pp',\o}\hat
W^{K}_{j,(\o,\o')}(\kk',\pp')\Psi^{-}_{\kk',\o'}\;,\label{A3.17}
\ee
for a suitable $2\times2$ matrix-valued potential $\hat W^{K}_{j,\ul\o}(\kk',\pp')$.
The symmetry properties (4)--(8) imply that:
\bea \hat W^{K}_{j,\ul\o}(\kk',\pp') &\os{(4)}{=}& e^{-i(\pp_F^\o+\kk'+\pp')(\dd_1 - \dd_2)n_-}\hat
W^{K}_{j-1,\ul\o}(T^{-1}\kk',T^{-1}\pp')e^{i(\pp_F^{\o'}+\kk')(\dd_1 - \dd_2)n_-} \os{(5)}{=}
[\hat W^{K}_{j,-\ul\o}(-\kk',-\pp')]^* \os{(6.a)}{=} \nn\\ &\os{(6.a)}{=}& e^{i(\pp_F^\o-\pp_F^{\o'}+\pp')
(\ddd_j - \ddd_1)}\s_1\hat
W^{K}_{r_h j,\ul\o}(R_h \kk', R_h \pp')\s_1 \os{(6.b)}{=} \hat W^{K}_{r_v j,-\ul\o}(R_v \kk', R_v \pp') \os{(7)}{=} \nn\\ &\os{(7)}{=}&
\big[\hat W^{K}_{j,(-\o',-\o)}(P(\kk' + \pp'),-P\pp')\big]^T \os{(8)}{=} -\s_{3}\hat W^{K}_{j,\ul\o}(I\kk',I\pp')\s_{3}\;,\label{A3.19} \eea
Let us define $\hat W^{K}_{j,\ul\o}(\V0,\V0) = \sum_{\n=0}^3b^{\n}_{j,\ul\o}\s_{\n}$. Then, property
(8) in the third line of Eq.(\ref{A3.19})
implies that $b^{0}_{j,\ul\o} = b^{3}_{j,\ul\o}=0$. Moreover,
using properties (6.b) and (7) in the second and third lines of Eq.(\ref{A3.19}),
\bea &&b^{1}_{j,\ul\o} \os{(6.b)}{=}b^{1}_{r_v j,-\ul\o}\;,\qquad\hskip.3truecm
 b^{2}_{j,\ul\o} \os{(6.b)}{=}
b^{2}_{r_v j,-\ul\o}\;,\nn\\
 &&b^{1}_{j,\ul\o} \os{(7)}{=} b^{1}_{j,(-\o',-\o)}\;,\qquad b^{2}_{j,\ul\o} \os{(7)}{=} - b^{2}_{j, (-\o',-\o)}\;,\nn\eea
which implies, in particular, that $b^2_{1,\ul\o}=0$ and $b^1_{1,(\o,\pm\o)}=b_{\pm}$,
for two suitable ($\o$-independent) constants $b_{\pm}$. Using property (5)
in Eq.(\ref{A3.19}),
$b^{1}_{j,\ul\o} \os{(5)}{=} (b^{1}_{j,-\ul\o})^*$, we see that both these constants are real,
$b_{\pm}\in\mathbb{R}$. Property (4) in Eq.(\ref{A3.19}) implies that, if $\ul\o=(\o,\o')$,
\be b^{1}_{j,\ul\o}\s_{1} + b^{2}_{j,\ul\o}\s_2 \os{(4)}{=} \Biggl(\begin{matrix}
0& e^{i\o'\frac{2\p}{3}}(b^1_{j-1,\ul\o}-ib^2_{j-1,\ul\o})\\
e^{-i\o\frac{2\p}{3}}(b^1_{j-1,\ul\o}+ib^2_{j-1,\ul\o})&0
\end{matrix}\Biggr)\;,\nn\ee
so that, finally,
\bea &&\hat W^K_{j,(\o,\pm\o)}({\bf 0},{\bf 0})=b_\pm\Biggl(\begin{matrix}
0& e^{\pm i\o\frac{2\p}{3}(j-1)}\\
e^{- i\o\frac{2\p}{3}(j-1)}&0
\end{matrix}\Biggr)\;,\label{bb5.22}\eea
which proves the second of Eq.(\ref{symbad}).\\

\paragraph*{The $AAA$, $\Phi^KAA$ and $\Phi^K\Phi^KA$ kernels.}
The contribution to the effective potential of the form $AAA$ or $\Phi^K\Phi^KA$ are
proportional to, respectively,
\be \sum_{\substack{\pp_1,\pp_2\\
\m_1,\m_2,\m_3}}\hat
A_{\m_1,\pp_1}\hat A_{\m_2,\pp_2} \hat A_{\m_3,-\pp_1-\pp_2}\hat
W_{(\m_1,\m_2,\m_3)}(\pp_1,\pp_2)\qquad {\rm or}\qquad
\sum_{\substack{\pp_1,\pp_2\\
\m,j_1,j_2}}\hat\Phi^{K}_{j_1,\pp_1}\hat\Phi^{K}_{j_2,\pp_2}\hat A_{\m,-\pp_1-\pp_2}\hat
W^{KK}_{(j_1,j_2),\m}(\pp_1,\pp_2)\;,\nn \ee
for suitable kernels $\hat W_{(\m_1,\m_2,\m_3)}(\pp_1,\pp_2)$, $\hat
W^{KK}_{(j_1,j_2),\m}(\pp_1,\pp_2)$. Using the invariance under the symmetry
(7)+(8) we find that $\hat W_{(\m_1,\m_2,\m_3)}(\pp_1,\pp_2)=-\hat W_{(\m_1,\m_2,\m_3)}(\pp_1,\pp_2)$ and $\hat
W^{KK}_{(j_1,j_2),\m}(\pp_1,\pp_2)=-\hat
W^{KK}_{(j_1,j_2),\m}(\pp_1,\pp_2)$, that is, they are both identically zero.
On the contrary, the contribution to the effective potential of the form
$\Phi^KAA$ is proportional to
\be \sum_{\substack{\pp_1,\pp_2\\
j,\m_1,\m_2}}\hat\Phi^{K}_{j,\pp_1}\hat A_{\m_1,\pp_2}\hat A_{\m_2,-\pp_1-\pp_2}\hat
W^{K}_{j,(\m_1,\m_2)}(\pp_1,\pp_2)\;,\nn\ee
for a suitable kernel $\hat W^{K}_{j,(\m_1,\m_2)}(\pp_1,\pp_2)$. Its local part
$\hat W_{j,(\m_1,\m_2)}^K({\bf 0},{\bf 0})=:\l^K_{j,(\m_1,\m_2)}$ satisfies:
\be \l^K_j\os{(4)}{=}T\l^K_{j-1}T^{-1}\os{(5)}{=}
\big[\l^K_j\big]^*\os{(6.a)}{=}
R_h\l^K_{r_hj}R_h\os{(6.b)}{=}
R_v\l^K_{r_vj}R_v\;,\nn\ee
which imply
\be
\l^K_1=\begin{pmatrix} a&0&0\\0&b&0\\0&0&c\end{pmatrix}\,\qquad
\l^K_2=\begin{pmatrix} a&0&0\\0&\frac14(b+3c)&\frac{\sqrt3}{4}(c-b)
\\0&\frac{\sqrt3}{4}(c-b)&\frac14(c+3b)\end{pmatrix}\;,\qquad
\l^K_j=\begin{pmatrix} a&0&0\\0&\frac14(b+3c)&\frac{\sqrt3}{4}(b-c)
\\0&\frac{\sqrt3}{4}(b-c)&\frac14(c+3b)\end{pmatrix}\;,\nn\ee
for some real constants $a,b,c$. This the general symmetry structure of the terms in the first line
of Eq.(\ref{effsou}).

\subsection{Symmetry structure of the kernels in the presence of the phonon field}\label{appAgap}

In this Appendix we prove Eqs.(\ref{s9.3}),(\ref{s9.5}),(\ref{s9.6}).
Let us define, for all $\kk,\kk'\neq \pp_F^\pm$,
\be A_j^{(j_0)}(\kk):=\frac{\langle\!\langle \hat a^+_{\kk,\s}\hat b^-_{\kk,\s}\rangle\!\rangle^{(j_0)}_j}{\O(\vec k)}
\;,\qquad B_{\o,j}^{(j_0)}(\kk'):=e^{-i\pp_F^\o(\dd_{j_0}-\dd_1)}
\frac{\langle\!\langle \hat a^+_{\kk'+\pp_F^{-\o}}\hat b^-_{\kk'+\pp_F^\o}\rangle\!\rangle^{(j_0)}_j}
{\O(\vec k')}\;.\label{A.1}\ee
Using the symmetries listed in Section \ref{sec4B} and in Appendix \ref{app1}
and proceeding as in Appendix \ref{app3}, we find:
\bea &&
A^{(j_0)}_j(\kk)\os{(4)}{=} A^{(j_0+1)}_{j+1}(T\kk)
\os{(5)}{=}\big[A^{(j_0)}_j(-\kk)\big]^* \os{(6.a)}{=} \big[A^{(j_0)}_{r_hj}(R_h\kk)\big]^*\os{(6.b)}{=}
A^{(r_vj_0)}_{r_vj}(R_v\kk)
\os{(7)}{=} \big[A^{(j_0)}_j(P\kk)\big]^*\os{(8)}{=} A^{(j_0)}_j(I\kk)\;,\nn\\
&&
B^{(j_0)}_{\o,j}(\kk')\os{(4)}{=} B^{(j_0+1)}_{\o,j+1}(T\kk')
\os{(5)}{=}\big[B^{(j_0)}_{-\o,j}(-\kk')\big]^* \os{(6.a)}{=} \big[B^{(j_0)}_{-\o,r_hj}(R_h\kk')\big]^*\os
{(6.b)}{=} B^{(r_vj_0)}_{-\o,r_vj}(R_v\kk)
\os{(7)}{=} \big[B^{(j_0)}_{-\o,j}(P\kk)\big]^*\os{(8)}{=} B^{(j_0)}_{\o,j}(I\kk)\nn\eea
Using the properties (6.a)+(6.b)+(7) in the latter equation, we find: $A^{(j_0)}_j(\kk)=A^{(r_vj_0)}_j(\kk)$ and $B^{(j_0)}_{\o,j}(\kk)=
B^{(r_vj_0)}_{-\o,j}(\kk)$. Combining this with property (4) we get:
\bea && A^{(1)}_1(\kk)=A^{(2)}_2(T\kk)=A^{(3)}_2(T\kk)=A^{(3)}_3(T^2\kk)=A^{(2)}_3(T^2\kk)\;,
\nn\\
&& A^{(2)}_1(\kk)=A^{(3)}_1(\kk)=A^{(3)}_2(T\kk)=A^{(2)}_2(T\kk)=A^{(1)}_3(T^2\kk)\;,
\qquad  A^{(3)}_1(\kk)=A^{(1)}_2(T\kk)=A^{(2)}_3(T^2\kk)\;,
\nn\eea
which implies that $A^{(j_0)}_j(\kk)=A_j(\kk)$. The analogous equations for
$B$ imply that $B^{(j_0)}_{\o,j}(\kk)=B_{j}(\kk)$. Now, using properties (4)+(6.b) gives
$A_1(\kk)=A_2(T\kk)=A_3(TR_v\kk)=A_1(T^2R_v\kk)=A_1(T^2\kk)$, which implies that $A_1(\kk)=A_1(T\kk)$; therefore, $A_j(\kk)$ is independent of $j$ and transforms as in Eq.(\ref{s9.5}).
The same argument holds for $B$.
Finally, it is straightforward to check that the
previous symmetries also imply that,
if $\kk=\pp_F^{\pm}$, then $\langle\!\langle \hat a^+_{\kk,\s}\hat b^-_{\kk,\s}\rangle\!\rangle^{(j_0)}_j
=\langle\!\langle \hat a^+_{\kk'+\pp_F^{-\o}}\hat b^-_{\kk'+\pp_F^\o}\rangle\!\rangle^{(j_0)}_j=0$,
which concludes the proof of Eq.(\ref{s9.3}) for all $\kk,\kk'$.

\section{Lowest order computations}\label{app2d}
\setcounter{equation}{0}
\renewcommand{\theequation}{\ref{app2d}.\arabic{equation}}

\subsection{The beta function for \protect{$Z_{K,h}^\pm$}}\label{secexp}

\begin{figure}[hbtp]
\centering
\includegraphics[width=.27\textwidth]{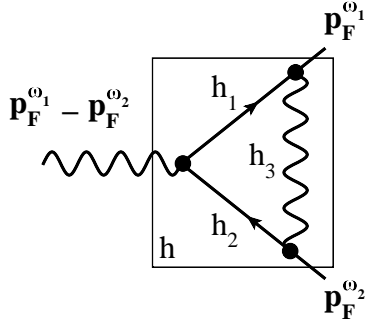}
\caption{The lowest order contribution to the beta function for $Z^{\o_1\o_2}_{K,h}$.
}\label{figexp}\end{figure}

Here we prove the two identities in Eq.(\ref{oth5c}) (the identity in Eq.(\ref{oth5cbis})
has been proved in \cite[Appendix C]{GMP1}).
The relativistic part of the second order beta function is given by the
sum over $h_1,h_2,h_3$ of the graph in Fig.\ref{figexp}, with the constraint that
$\min\{h_1,h_2,h_3\}=h$ and that $h_i-h\in\{0,1\}$. In the graph, the external wavy line
with transferred momentum $\pp_F^{\o_1}-\pp_F^{\o_2}$ represents $\hat \Phi^K_{j,
\pp_F^{\o_1}-\pp_F^{\o_2}}$ and the straight internal lines correspond to Dirac propagators on
scales $h_1,h_2$; if $\o_1=\o$ and $\o_2=\pm\o$, then
the graph represents a contribution to $\b^+_h\G_{\o,j}^+$ or $\b^-_h\G^-_{\o,j}$,
where $\G^\pm_{\o,j}$ was defined in Eq.(\ref{symbad}).

If we neglect corrections of order $O(e^4)$ and $O(e^2(1-v)2^{ce^2h})$ for some $c>0$,
in the computation of the graph we can replace $e_{\m,h}$ by $e$, $\tilde f_h(\kk')$ by
$f_h(\kk')$ (see Eq.(\ref{prop4.29}) for a definition of $\tilde f_h$) and $v_h$ by $1$. As a result,
we find:
\be \b^\pm_h\G^\pm_{\o,j}=-e^2\sum_{\n=0}^2\sum_{\substack{h_i\ge h\\ \min\{h_i\}=h}}
\int \frac{d\pp}{(2\p)^3} f_{h_1}(\pp)f_{h_2}(\pp)f_{h_3}(\pp) \frac1{2|\pp|}
 \G^\n_\o\frac1{ip_0\G^0_\o+i\vec p\cdot\vec \G_\o}\G^\pm_{\o,j}
 \frac1{ip_0\G^0_{\pm\o}+i\vec p\cdot\vec \G_{\pm\o}}\G^\n_{\pm\o}\;,\label{app2d.1}\ee
 modulo corrections $O(e^4)$ and $O(e^2(1-v)2^{(\const.)e^2h})$.
 It is convenient to pass to spherical coordinates. Note that, if
 $F_{h}(|\pp|) := f_{h}(\pp)^{3} + 3
f_{h}(\pp)^{2}f_{h+1}(\pp) + 3f_{h}(\pp)f_{h+1}(\pp)^{2}$, then
$\int_{0}^{+\infty} d\r\,\r^{-1} F_{h}(\r) = \log 2$. Therefore, after having integrated the
radial coordinates we are left with:
\bea && \b^\pm_h\G^\pm_{\o,j}=-\frac{e^2\log 2}{2(2\p)^3}\sum_{\n=0}^2
\int_{-1}^1d\cos\th\int_0^{2\p}d\phi \cdot\label{app2d.2}\\
&&\cdot \G^\n_\o\big(-i\G^0_\o\cos\th+i\G^1_\o\sin\th\cos\phi+i\G^2_\o\sin\th\sin\phi\big)
 \G^\pm_{\o,j}\big(-i\G^0_{\pm\o}\cos\th+i\G^1_{\pm\o}\sin\th\cos\phi+i\G^2_{\pm\o}\sin\th
 \sin\phi\big)\G^\n_{\pm\o}\;,\nn\eea
which implies
\be   \b^\pm_h=\frac{e^2\log 2}{12\p^2}\big[\G^{\pm}_{\o,j}\big]^{-1}\sum_{\n=0}^2
\G^\n_\o\Big(\G^0_\o \G^\pm_{\o,j}\G^0_{\pm\o}+
\G^1_\o \G^\pm_{\o,j}\G^1_{\pm\o}+\G^2_\o \G^\pm_{\o,j}\G^2_{\pm\o}\Big) \G^\n_\o\;.
\label{app2d.3}\ee
Now, if we pick the plus sign, $\G^+_{\o,j}=\s_1\cos\th_j-\o\s_2\sin\th_j$, where
$\th_j=\frac{2\p}{3}(j-1)$. Therefore, using the definitions of $\G^\m_\o$ in Eq.(\ref{2.11s}),
\be \G^0_\o\G^+_{\o,j}\G^0_\o=\G^+_{\o,j}\;,\qquad
\G^1_\o\G^+_{\o,j}\G^1_\o=\G^+_{-\o,j}\;,\qquad
\G^2_\o\G^+_{\o,j}\G^2_\o=-\G^+_{-\o,j}\;.\label{app2d.4}\ee
Plugging Eq.(\ref{app2d.4}) into Eq.(\ref{app2d.3}) we find $\b^+_h=\frac{e^2}{12\p^2}\log 2$, as
desired.

Similarly, picking the minus sign in Eq.(\ref{app2d.3}), $\G^-_{\o,j}=e^{-i\o\th_j}\s_1$, so that
\be \G^0_\o\G^-_{\o,j}\G^0_{-\o}=\G^-_{\o,j}\;,\qquad
\G^1_\o\G^-_{\o,j}\G^1_{-\o}= \G^-_{\o,j}\;,\qquad
\G^2_\o\G^-_{\o,j}\G^2_{-\o}=\G^-_{\o,j}\;.\label{app2d.5}\ee
Plugging Eq.(\ref{app2d.5}) into Eq.(\ref{app2d.3}) we find $\b^-_h=\frac{3e^2}{4\p^2}\log 2$.
This completes the proof of Eq.(\ref{oth5c}).

\subsection{The beta function for the other renormalization constants}\label{secothexp}

A similar computation can be performed for the beta function controlling the flow
of the other renormalization constants. In particular, the beta functions for $Z^+_{CDW,h}$,
$Z^-_{CDW,h}$ and $Z^+_{AF,h}$ and $Z^-_{AF,h}$ are given by expressions
analogous to Eq.(\ref{app2d.3}) with $\G^\m_{\o,j}$ replaced by $\s_3$ and
$e^{i\o\th_j n_-}\s_3$, respectively. Using the fact that
\be
\G^0_\o\s_3\G^0_\o=\s_3 \;,\qquad
\G^1_\o\s_3\G^1_\o=\s_3\;,\qquad
\G^2_\o\s_3\G^2_\o=\s_3\;,\nn\ee
we find that $\b^+_{CDW,h}=\frac{3 e^2}{4\p^2}\log2$, modulo subdominant corrections. Similarly,
\be \G^0_\o e^{i\o\th_j n_-}\s_3\G^0_{-\o}= e^{i\o\th_j n_-}\s_3\;,\qquad
\G^1_\o e^{i\o\th_j n_-} e^{i\o\th_j n_-}\s_3\G^1_{-\o}=e^{i\o\th_j n_+}\s_3 \;,\qquad
\G^2_\o e^{i\o\th_j n_-}\s_3\G^2_{-\o}=-e^{i\o\th_j n_+}\s_3 \;,\nn\ee
implies that $\b^-_{CDW,h}=\frac{e^2}{12\p^2}\log2$, modulo subdominant corrections. This
proves Eqs.(\ref{zetasf})--(\ref{s7.2}). The exponents in Table \ref{tabexp} are computed analogously and we will not belabor the details here.

\section{Lowest order check of the Ward Identities}\label{WIcheck}
\setcounter{equation}{0}
\renewcommand{\theequation}{\ref{WIcheck}.\arabic{equation}}

In this section we check at lowest order in non-renormalized
perturbation theory the validity of the WIs that we used in Section
\ref{secWI} to prove the infrared stability of the flows of the
effective charge and of the photon mass.

\subsection{Ward identity for the photon mass}\label{phm}

We start by checking the WI Eq.(\ref{s5.1}) for the photon mass.
At lowest order, the graphs contributing to the dressed photon
mass are depicted in Fig.\ref{fig2}.
As we are going to show here, the sum of the two graphs computed at zero transferred momentum
is {\it exactly vanishing}, for all choices of $\m,\n$.

The first and second order interactions (of the form $A\Psi^+\Psi^-$ and $AA\Psi^+\Psi^-$)
involved in the computation of the
two graphs are obtained by expanding the interaction $\VV(\Psi,A)$
in Eqs.(\ref{1.15uy})-(\ref{intmom}) up to second order in the electric charge:
\bea  \VV(\Psi, A)&=&\frac{ie}{\b^2 L^2|\SS_L|}\sum_{\kk,\pp,\s}
\Big\{-\hat A_{0,\pp}\hat \Psi^+_{\kk+\pp,\s}
\G_0(\pp)\hat \Psi^-_{\kk,\s}+t\sum_{j=1}^3\h^j_{\pp}(\vec\d_j\vec A_{\pp})
\hat \Psi^+_{\kk+\pp,\s}\Biggl(\begin{matrix}0& e^{-i\kk(\dd_j-\dd_1)}\\
-e^{i(\kk+\pp)(\dd_j-\dd_1)}&0\end{matrix}\Biggr)
\hat\Psi^-_{\kk,\s}\Big\}\nn\\
&-&\frac12\frac{te^2}{\b^3 L^2|\SS_L|^2}\sum_{\substack{\kk,\pp_1,\pp_2\\\s}}
\sum_{j=1}^3
\h^j_{\pp_1}\h^j_{\pp_2}(\vec\d_j\vec A_{\pp_1})(\vec\d_j\vec A_{\pp_2})
\hat \Psi^+_{\kk+\pp_1+\pp_2,\s}\Biggl(\begin{matrix}0& e^{-i\kk(\dd_j-\dd_1)}\\
e^{i(\kk+\pp_1+\pp_2)(\dd_j-\dd_1)}&0\end{matrix}\Biggr)
\hat\Psi^-_{\kk,\s}+O(e^3)\;.\nn\eea
Defining
\bea && \vec \G(\kk,\pp):=\frac23\sum_{j=1}^3\h^j_{\pp}\vec\d_j
\Biggl(\begin{matrix}0& e^{-i\kk(\dd_j-\dd_1)}\\
-e^{i(\kk+\pp)(\dd_j-\dd_1)}&0\end{matrix}\Biggr)\;,\nn\\
&& \G_{lm}(\kk,\pp_1,\pp_2)=\frac23
\sum_{j=1}^3\h^j_{\pp_1}\h^j_{\pp_2}(\vec\d_j)_l(\vec\d_j)_m
\Biggl(\begin{matrix}0& e^{-i\kk(\dd_j-\dd_1)}\\
e^{i(\kk+\pp_1+\pp_2)(\dd_j-\dd_1)}&0\end{matrix}\Biggr)\;,\nn\eea
and recalling that $v=\frac32t$, we can rewrite the interaction as
$ \VV(\Psi, A)= \VV_2(\Psi, A)+O(e^3)$, with
\bea  \VV_2(\Psi, A)&=&\frac{ie}{\b^2 L^2|\SS_L|}\sum_{\kk,\pp,\s}
\Big\{-\hat A_{0,\pp}\hat \Psi^+_{\kk+\pp,\s}
\G_0(\pp)\hat \Psi^-_{\kk,\s}+v\hat \Psi^+_{\kk+\pp,\s}\Big(\vec\G(\kk,\pp)\cdot\vec A_\pp\Big)
\hat\Psi^-_{\kk,\s}\Big\}\nn\\
&-&\frac12\frac{ve^2}{\b^3 L^2|\SS_L|^2}\sum_{\kk,\pp_1,\pp_2}\,
\sum_{\substack{l,m=1,2\\ \s=\uparrow\downarrow}} A_{l,\pp_1}A_{m,\pp_2}
\hat \Psi^+_{\kk+\pp_1+\pp_2,\s}\G_{lm}(\kk,\pp_1,\pp_2)
\hat\Psi^-_{\kk,\s}\;.\label{appe.4a}\eea
Since the term in the last line involves only the spatial components of the photon field,
the second graph in Fig.\ref{fig2} gives a non zero contribution only if both $\m$ and $\n$ are
different from zero.

If $\m=\n=0$, only the first graph survives, which in the $\b,L\to\infty$ limit,
if computed at $\pp={\bf 0}$, gives (using the fact that $\G_0({\bf 0})=\openone$)
\be +\frac{e^2}{2}\int \frac{d\kk}{2\p|\BBB|}\,\Tr\Big\{ \hat S_0(\kk)\hat{S}_0(\kk) \Big\} \;,
\label{appe.1}\ee
where $\hat S_0(\kk)$ was defined in Eq.(\ref{free1.1}) and $|\BBB|=8\p^2/(3\sqrt{3})$ is the
area of the first Brillouin zone. Using the fact that
\be \hat S_0(\kk)\hat{S}_0(\kk)=-i\dpr_0\hat S_0(\kk)\;,\label{appe.2}\ee
we see that Eq.(\ref{appe.1}) is the integral of a total derivative, which is zero.

Let us now consider the case where $\m=\n=2$. The sum of the two graphs
in the $\b,L\to\infty$ limit, if computed at $\pp={\bf 0}$, gives:
\be\frac{v^2e^2}{2}\int \frac{d\kk}{2\p|\BBB|}\,\Tr\Big\{ \hat S_0(\kk)\G_2(\kk,\V0)\hat{S}_0(\kk)
\G_2(\kk,\V0) \Big\}+\frac{v e^2}{2}\int \frac{d\kk}{2\p|\BBB|}\,
\Tr\Big\{ \hat S_0(\kk)\G_{22}(\kk,\V0,\V0)\Big\}\;.\label{appe.3}\ee
Using the fact that
\be \hat S_0(\kk)v\G_2(\kk,\V0)\hat{S}_0(\kk)=i\dpr_2\hat S_0(\kk)\;,\label{appe.4}\ee
and integrating by parts, we can rewrite Eq.(\ref{appe.3}) as
\be\frac{ve^2}{2}\int \frac{d\kk}{2\p|\BBB|}\,\Tr\Big\{ \hat S_0(\kk)
\big[-i\dpr_2\G_2(\kk,\V0)+\G_{22}(\kk,\V0,\V0)\big]\Big\}\;,\label{appe.5}\ee
which is zero, simply because the matrix in square brackets is identically zero. This proves that
the graphs in Fig.\ref{fig2} with $\m=\n=2$ cancel out exactly. Using the symmetry
(4), we find that the diagonal terms with $\m=\n=1$ cancel out, too. The non-diagonal
terms can be treated analogously.

\subsection{Ward identity for the effective charge}\label{appch}

Let us check at lowest order in non-renormalized
perturbation theory the WI for the effective charge,
Eqs.(\ref{s5.5})-(\ref{s5.5bis}). This amounts to check the cancellation of
the graphs depicted in Fig.\ref{fig1}. In order to compute these graphs we use the
bare fermionic propagator $\hat S_0(\kk)$ and the photon propagator
$\hat w^{0,h^*}_{\m\n}(\pp)=:\hat w^{(\ge h^*)}(\pp)\d_{\m\n}$ with IR
cutoff on scale $h^*$, see Eq.(\ref{pho_fey7}). As we are going to show, the
sum of these six graphs is exactly vanishing; therefore, the
dressed charge is equal to the bare one at lowest order.
Remarkably, this cancellation does not depend on the presence of
the infrared cutoff on the photons; this fact has been exploited
in Section \ref{secWI} to derive a WI for the effective
charge on all IR scales. We shall only consider the cases
$\m=0$ and $\m=2$; the renormalization of the charge corresponding
to $\m=1$ is equal to the case $\m=2$, thanks to the discrete rotational symmetry (4).

In order to compute the graphs in Fig.\ref{fig1} we need the interaction $\VV(\Psi,A)$
up to third order in $e$:
\be \VV(\Psi,A) = \VV_2(\Psi,A) +\frac16\frac{v(ie)^3}{\b^4 L^2|\SS_L|^3}
\sum_{\substack{\kk,\pp_1\\
\pp_2,\pp_3}}\,
\sum_{\substack{l,m,k=1,2\\ \s=\uparrow\downarrow}} A_{l,\pp_1}A_{m,\pp_2}A_{k,\pp_3}
\hat \Psi^+_{\kk+\pp_1+\pp_2+\pp_3,\s}\G_{lmk}(\kk,\pp_1,\pp_2,\pp_3)
\hat\Psi^-_{\kk,\s}+O(e^4)\;,\label{appe.6}\ee
where $\VV_2(\Psi,A)$ was defined in Eq.(\ref{appe.4a}) and
\be \G_{lmk}(\kk,\pp_1,\pp_2,\pp_3)=\frac23
\sum_{j=1}^3\h^j_{\pp_1}\h^j_{\pp_2}\h^j_{\pp_3}(\vec\d_j)_l(\vec\d_j)_m(\vec\d_j)_k
\Biggl(\begin{matrix}0& e^{-i\kk(\dd_j-\dd_1)}\\
-e^{i(\kk+\pp_1+\pp_2+\pp_3)(\dd_j-\dd_1)}&0\end{matrix}\Biggr)\;.\label{appe.7}\ee
\paragraph{Case $\m=0$.} In this case only the first and fifth graphs in Fig.\ref{fig1} are
non-vanishing. Their value in the $\b,L\to\infty$ limit computed at the Fermi points
$\kk=\pp_F^\o$ and at transferred momentum $\pp=\V0$ is
\bea && ie\sum_{\n=0}^2(i\bar e_\n)^2\int \frac{d\pp}{(2\p)^3}
\hat w^{(\ge h^*)}(\pp)\Big\{\G_\n(\pp_F^\o+\pp,-\pp)\hat S_0(\pp_F^\o+\pp)\G_0(\pp_F^\o+\pp,\V0)
\hat S_0(\pp_F^\o+\pp)\G_\n(\pp_F^\o,\pp)\Big\}+\nn\\
&&+e\sum_{\n=0}^2(i\bar e_\n)^2\int \frac{d\pp}{(2\p)^3}
\hat w^{(\ge h^*)}(\pp)\dpr_{k_0}\Big\{\G_\n(\kk+\pp,-\pp)\hat S_0(\kk+\pp)\G_\n(\kk,\pp)
\Big\}_{\kk=\pp_F^\o}\;,\label{appe.8}\eea
where we defined $\bar e_0=e$, $\bar e_{l}=v e$ for $l\in\{1,2\}$ and $\G_0(\kk,\pp):=-\G_0(\pp)$.
Now, in the first line we can rewrite:
\be \hat S_0(\pp_F^\o+\pp)\G_0(\pp_F^\o+\pp,\V0)
\hat S_0(\pp_F^\o+\pp)=i\dpr_0\hat S_0(\pp_F^\o+\pp)\;,\nn\ee
which partially cancel with the second line. We are only left with the terms in the second line
where the derivative $\dpr_{k_0}$ acts on the kernels $\G_\n$; however, these terms are
identically zero, simply because the kernels $\G_\n$ are by definition independent
of $k_0$.

\paragraph{Case $\m=2$.} Here the situation is more complicated, because of the presence of the
second, third, fourth and sixth graph in Fig.\ref{fig1}, and because the $\partial_{k_2}$
derivative can now act on the kernels $\G_\n$. However:
\begin{itemize}
\item[(i)] repeating the same argument used in the case $\m=0$ and using
Eq.(\ref{appe.4}), we see that the sum of the
first and fifth graphs is equal to
\be e\sum_{\n=0}^2(i\bar e_\n)^2\!\!\int \!\!\frac{d\pp}{(2\p)^3}
\hat w^{(\ge h^*)}(\pp)\Big\{\dpr_{k_2}\G_\n(\kk+\pp,-\pp)\hat S_0(\kk+\pp)\G_\n(\kk,\pp)+
\G_\n(\kk+\pp,-\pp)\hat S_0(\kk+\pp)\dpr_{k_2}\G_\n(\kk,\pp)
\Big\}_{\kk=\pp_F^\o}\label{appe.9}\ee
\item[(ii)] combining Eq.(\ref{appe.9}) with the contributions from the second and third
graphs, we get
\bea && e\sum_{\n=0}^2(i\bar e_\n)^2\!\!\int \!\!\frac{d\pp}{(2\p)^3}
\hat w^{(\ge h^*)}(\pp)\Big\{\big[\dpr_{k_2}\G_\n(\kk+\pp,-\pp)+i
\G_{2\n}(\kk+\pp,\V0,-\pp)\big] \hat S_0(\kk+\pp)\G_\n(\kk,\pp)+\nn\\
&&\hskip4.5truecm+\G_\n(\kk+\pp,-\pp)\hat S_0(\kk+\pp)\big[\dpr_{k_2}\G_\n(\kk,\pp)+i
\G_{2\n}(\kk,\V0,\pp)\big]\Big\}_{\kk=\pp_F^\o}\;,\eea
which is zero, simply because the terms in square brackets are identically zero, as one can easily
check.
\item[(iii)] the sum of the fourth and sixth graphs in Fig.\ref{fig1} gives
\be -\frac12 ve^3\sum_{l=1,2}
\int \frac{d\pp}{(2\p)^3}\hat w^{(\ge h^*)}(\pp)\big[\dpr_{k_2}\G_{ll}(\kk,\pp,-\pp)+i
\G_{2ll}(\kk,\V0,\pp,-\pp)\big]_{\pp=\pp_F^\o}\;,\label{vert3} \ee
which is zero, simply because the term in square brackets are identically zero, as one can easily
check.
\end{itemize}
The case $\m=1$ can be obtained from the case $\m=2$ by the discrete rotational symmetry
(4), so this concludes our lowest order check of the Ward identities.


\end{document}